\documentclass{ws-procs9x6}
\begin{document}

\newcommand{\half}{\frac12}
\newcommand{\third}{\frac13}
\newcommand{\eqn}[1]{\label{eq:#1}}
\newcommand{\refeq}[1]{Eq.~(\ref{eq:#1})}
\newcommand{\fig}[1]{\label{fig:#1}}
\newcommand{\reffig}[1]{Fig.~\ref{fig:#1}}
\newcommand{\figgg}[1]{\label{fig:#1}}
\newcommand{\refiggg}[1]{Fig.~\ref{fig:#1}}
\newcommand{\eqsdash}[2]{Eqs.~(\ref{eq:#1}-\ref{eq:#2})}
\newcommand{\eqscomma}[2]{Eqs.~(\ref{eq:#1}, \ref{eq:#2})}
\newcommand{\beq}{\begin{eqnarray}}
\newcommand{\eeq}{\end{eqnarray}}
\newcommand{\nn}{\nonumber}
\def\CO{{\mathcal O}}
\def\CL{{\mathcal L}}
\def\CM{{\mathcal M}}
\def\tr{{\rm\ Tr}}
\def\dim{{\rm dim}\ }
\def\al{\alpha}
\def\bt{\beta}
\def\eps{\epsilon}
\def\mn{{\mu\nu}}
\newcommand{\rep}[1]{{\bf #1}}
\newcommand{\vev}[1]{\langle#1\rangle}
\def\be{\begin{equation}}
\newcommand{\bel}[1]{\be\label{#1}}
\def\ee{\end{equation}}
\newcommand{\eref}[1]{(\ref{#1})}
\newcommand{\Eref}[1]{Eq.~(\ref{#1})}
\newcommand{\mycite}[1]{[\refcite{#1}]}
\newcommand{\rem}[1]{}
\def\tr{{\rm tr}}
\def\half{{1\over 2}}
\def\NN{{\mathcal N}}
\def\hh{y}

\def\nzero{$\NN=0$}
\def\none{$\NN=1$}
\def\ntwo{$\NN=2$}
\def\nntwo{$\NN\approx 2$}
\def\nnfour{$\NN\approx 4$}
\def\nlfour{$\NN\leq 2$}
\def\nfour{$\NN=4$}
\def\neight{$\NN=8$}
\def\Pf{{\rm Pf}\ }
\def\susy{supersymmetry}
\def\susic{supersymmetric}
\def\diag{{\rm diag}}
\def\rarr{\rightarrow}
\def\drarr{\Rightarrow}
\def\betau{\beta_{1/g^2}}
\def\OO{{\mathcal O}}
\def\ZZ{{\bf Z}}
\def\SS{{\bf S}}
\def\TT{{\bf T}}
\def\RR{{\bf R}}
\def\DD{{\tilde D}}
\def\LLam{{\overline{\Lambda}}}
\newcommand{\VV}[1]{{\bf V}_#1}
\newcommand{\II}{{\bf I}}
\def\yy{{\bf \hat y}}
\def\PP{{\mathcal P}}
\def\QQ{{\mathcal Q}}
\def\bQQ{\bar{\mathcal Q}}
\def\ZN{\ZZ_N}
\def\OM{Olive-Montonen}
\def\NO{Nielsen-Olesen}
\def\Lgr{{\mathcal L}}
\def\slz{$SL(2,\ZZ)$}
\def\tim{\tilde m^2}
\newcommand{\dbyd}[2]{{\partial #1\over\partial #2}}
\newcommand{\ddbyd}[2]{{\partial^2 #1\over\partial #2^2}}
\newcommand{\ddbydd}[3]{{\partial^2 #1\over\partial #2\ \partial #3}}
\def\ta{\theta}
\def\bit{\begin{itemize}}
\def\eit{\end{itemize}}
\def\iddx{\int\ d^dx\ }
\def\EXE{{{\underline{\it Exercise:} }}}
\newcommand{\EX}[1]{\vskip 0.2 in {\noindent\underline{\bf Exercise:} #1 \vskip 0.2 in }}
\def\dslash{\partial\ \!\!\!\!\!\slash}
\def\Dslash{D\!\!\!\!\slash\ }

\def\cFthree{{1\over 4\pi^2\alpha'}}
\def\cFfive{\left(\cFthree\right)^2}
\def\cBtwo{\cFthree}

\renewcommand{\theequation}{\thesection.\arabic{equation}}


\setcounter{tocdepth}{3}

\makeatletter

\newcommand\@pnumwidth{1.55em}
\newcommand\@tocrmarg{2.55em}
\newcommand\@dotsep{4.5}
\renewcommand{\theequation}{\thesection.\arabic{equation}}

\newcommand\tableofcontents{%
{\global
\@topnum\z@ 
\@afterindentfalse 
\if@twocolumn
\@restonecoltrue
\onecolumn 
\else 
\@restonecolfalse 
\fi
\vspace*{24pt}
\noindent 
{\small\bf Contents}\par 
\vskip1em 
\nobreak}
{\small
\@starttoc{toc}%
}\if@restonecol
\twocolumn
\fi}

\newcommand*\l@section[2]{%
  \ifnum \c@tocdepth >\z@
    \addpenalty\@secpenalty
    \addvspace{1.0em \@plus\p@}%
    \setlength\@tempdima{1.5em}%
    \begingroup
      \parindent \z@ \rightskip \@pnumwidth
      \parfillskip -\@pnumwidth
      \leavevmode \bfseries
      \advance\leftskip\@tempdima
      \hskip -\leftskip
      #1\nobreak\hfil \nobreak\hb@xt@\@pnumwidth{\hss #2}\par
    \endgroup
  \fi}

  \newcommand*\l@subsection{\@dottedtocline{2}{1.5em}{2.3em}}
  
\makeatother

\def\thefootnote{\arabic{footnote}}

\title{THE DUALITY CASCADE}
\author{MATTHEW\,J.\,STRASSLER}
\address{Department of Physics \\ 
         University of Washington\\
         Box 351560 \\ 
         Seattle, WA 98195 \\
         E-mail: strassler@phys.washington.edu}
\maketitle

\abstracts{ The duality cascade, and its dual description as string
theory on the warped deformed conifold, brings together several
sophisticated topics, some of which are not widely known.  These
lectures, which contain a number of previously unpublished results,
and are intended for experts as well as students, seek to explain the
physics of duality cascades.  Seiberg duality is carefully introduced,
with detailed attention to the physical implications of duality away
from the far infrared.  The conifold is briefly introduced and strings
on the conifold (the Klebanov-Witten model) are discussed.  Next,
fractional branes are introduced.  The duality cascade is then
constructed in field theory and in its dual supergravity description.
Among the newly published results: it is shown why supergravity sees
the cascade as smooth; how the two holomorphic couplings (dilaton and
integrated two-form in supergravity) are related to the three physical
couplings in the gauge theory; that there are actually twice as many
approximate fixed points in the cascade as might be naively expected.
These notes are based on lectures given at TASI 2003 and at the 2003
PIMS Summer School on Strings, Gravity \& Cosmology.}

\tableofcontents
\newpage

\setcounter{equation}{0}
\section{Renormalization Group Flow and Seiberg Duality}

\subsection{Introduction: A Careless Rendering of Seiberg Duality}

What is Seiberg duality?\footnote{I assume that you, the reader,
already know something about quantum field theory, gauge theory,
supersymmetry and how to build supersymmetric gauge theories.  If you
don't, you'll find these lectures too advanced; I recommend you first
read my TASI 2001 lectures, or Ken Intriligator's lectures in this
volume, or any number of suitable texts on supersymmetry.   I also assume you
know some string theory and about D-branes and AdS/CFT; you may
want to consult the lectures by Maldacena and by Kachru in this volume,
or the review article by Aharony et al.} Let us begin with a short summary.

Consider, first, a theory we will call ``SQCD,'' a supersymmetric
version of QCD.  The model involves \none\ \susic\ $SU(N)$ gauge
theory, with matter consisting of $N_f$ flavors of quarks and squarks.
The left-handed quarks $\psi^r$ and their partner squarks $Q^r$,
transforming in the ${\bf N}$ representation of $SU(N)$, are organized
into $N_f$ chiral multiplets, also labelled $Q^r$, where $r=1,\dots,
N_f$.  Note the gauge indices are not shown.  The flavor indices
indicate that the $Q^r$ transform as an ${\bf N_f}$ of an $SU(N_f)$
flavor group, which we will call $SU(N_f)_L$, to distinguish it from
the $SU(N_f)_R$ flavor group under which the left-handed antiquarks
$\tilde \psi$ and their partner antisquarks $\tilde Q$ transform.
These fields form $N_f$ chiral multiplets $\tilde Q_u$, $u=1,\dots,
N_f$, transforming in the ${\bf \bar N}$ of the $SU(N)$ gauge group.
Note that $r$ and $u$ are indices in different groups, and cannot be
contracted to make a flavor singlet; contracting them would break the
flavor symmetry to the diagonal
$SU(N_f)$.  (By convention, we may take $Q^r$ to transform in
the ${\bf N_f}$ of $SU(N_f)_L$ and $\tilde Q_u$ to transform in the
${\bf \overline N_f}$ of $SU(N_f)_R$.)

In analogy to QCD, the theory has a baryon number symmetry $U(1)_B$
under which $Q^r$ has charge $1/N$ and $\tilde Q_u$ has the opposite
charge.  In addition, there is an axial $U(1)$, but this is anomalous,
as in QCD.  However, because of the gluinos, SQCD has a non-anomalous
axial symmetry, a so-called $U(1)_{\mathcal R}$, under which the
gluinos $\lambda$ have charge $1$, the squarks $Q$ and $\tilde Q$ have
charge $1-{N\over N_f}$, and the quarks $\psi$ and $\tilde \psi$ have
charge $-{N\over N_f}$.  The action of SQCD is very simple: it
consists of the kinetic terms for the fields, including
the minimal couplings to the gauge fields, plus the minimal
number of additional
terms required by supersymmetry.  In particular, the
superpotential $W(Q,\tilde Q)$ is zero.

\EX{Verify that instantons, which have $2N$ gluino zero modes and 1
zero mode for each $\psi$ and each $\tilde \psi$, are indeed invariant
under the above-mentioned R-symmetry.  [You may wish
to consult Ken Intriligator's
lectures.]}

Now consider a {\it different} theory, which we will call ``SQCD+M'' (a 
terminology which is not standard, but will prove useful.)  This theory has gauge group
$SU(\tilde N)$, with $\tilde N\equiv N_f-N$.  It also has $N_f$
flavors in the fundamental representation of the gauge group, labelled
$q_r$ and $\tilde q^u$; notice the location of the indices is different,
indicating that $q_r$ transforms as an ${\bf \overline N_f}$ of
$SU(N_f)_L$ and $\tilde q^u$ as an ${\bf N_f}$ of $SU(N_f)_R$.  But
the theory is not quite SQCD again, because it also has another set of
gauge-singlet chiral superfields, labelled $M^r_u$, and coupling to the matter
fields by the superpotential
\bel{magsup} W = y
M^r_u q_r\tilde q^u \ee 
where $y$ is a coupling constant we will discuss later.  Note $M^r_u$
transforms as $({\bf N_f},{\bf \overline N_f})$ of $SU(N_f)_L\times
SU(N_f)_R$.  The baryon number of the $q$ fields is $1/\tilde N$ (the
field $M$ is uncharged).  Finally, there is again an anomaly-free R
symmetry under which both $q$ and $\tilde q$ transform with charge
$1-{ \tilde N\over N_f}$.  The field $M$ transforms with R-charge
$2{\tilde N\over N_f}$, which ensures that the superpotential $W$ has
R-charge 2, as required by supersymmetry.  

In 1994, Seiberg wrote an extraordinary paper \mycite{NAD} arguing that 
SQCD with $N$ colors and $N_f$ flavors
is dual to
SQCD+M with $\tilde N = N_f-N$ colors and $N_f$ flavors. 
What does this mean?

Seiberg explained that these two
theories, which are manifestly different when they are both weakly
coupled, nonetheless 
have the same physics at low momentum (in the ``far infrared''), 
where at least one, if not both, are strongly coupled.  In particular,
the Green's functions (and S-matrices, if they exist) of the two
theories become identical in the limit that the external momenta are
all taken to zero, as long as we match the gauge-invariant operators
of one theory to those of the other. For instance, $Q^r\tilde Q_u$ is
matched to $M^r_u$; the Green functions of $Q^r\tilde Q_u$ in the SQCD
theory approach the Green functions of $M^r_u$ in the SQCD+M theory at
low momenta.

But this is just the beginning.  There are many additional
consequences of duality that Seiberg did not explicitly discuss in his
original paper.  It is these implications --- which Seiberg himself
understood fully, but which are widely misunderstood --- which we will
seek to elucidate in the first section of these lectures.

After we understand Seiberg duality, we'll then discuss continuous
spaces of conformal field theories, then combine this with AdS/CFT
(discussed in Maldacena's lectures), and finally more general gauge
theory/string theory dualities.  Finally, the whole set of tools will
be brought to bear in the context of strings propagating on the
so-called ``warped deformed conifold,'' a space rather more beautiful
than its name implies, and the duality of this theory with a quantum
field theory exhibiting a ``duality cascade,'' a sequence of Seiberg
dualities.

\subsection{Classical and Quantum Beta Functions}

If you want to understand what it means for two theories to become
identical in the infrared, the first thing you must do is ensure your
understanding of renormalization is complete.  What precisely is a
beta function?  What precisely do terms such as ``relevant,''
``irrelevant'' and ``marginal'' mean --- irrelevant to what?  What is
a ``dangerous irrelevant'' operator, and what makes it dangerous?  Why
is the ``renormalization group'' (RG) not a group?  A substantial
discussion of this subject is presented in my  TASI 2001
lectures \mycite{TASIone} and I refer you to that document for details.
Still, there are a few things we should review for use in these
lectures.

\EX{Give an example of a relevant operator in {\it
classical supersymmetric field theory}, and define its beta function
in a way which shows most clearly that it is relevant.  What is the
sign of its beta function?}

Let's consider a massive free complex scalar field.
\bel{massivescalar}
S = \iddx  \left[
\partial_\mu\phi^\dagger \partial^\mu\phi - m^2\phi^\dagger \phi
\right] \ .
\ee
What is the renormalization group flow associated with this theory?
It can't be completely trivial even though the theory is free and
therefore soluble. The number of degrees of freedom is two (one
complex scalar equals two real scalars) in the ultraviolet and zero in
the infrared, so obviously the theory is scale dependent.

\EX{Verify this statement by computing the propagator in position
space; show it is scale invariant at extremely small and extremely
large distances, but is not scale-invariant at distances of order
$m^{-1}$.}

Clearly the mass term is not of any importance for ultraviolet
physics, but is enormously important for infrared physics.  What can
we do to make this intuitively obvious fact precise?  The correct
approach is to define a {\it dimensionless} coupling $\nu^2\equiv
m^2/\mu^2$ where $\mu$ is the renormalization-group scale -- the scale
at which we observe the theory. Then we can think of this theory as
transitioning, as in \reffig{nuflow}, between two even simpler
theories: the scale-invariant theory at $\mu\to\infty$, where the mass
of $\phi$ is negligible and $\nu\to 0$, and the empty though
scale-invariant $\nu\to\infty$ theory in the infrared, where the
scalar does not propagate.

\begin{figure}[th]
\begin{center} 
\centerline{\psfig{figure=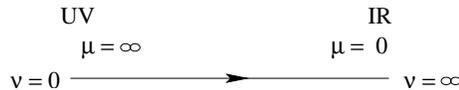,width=6cm,clip=}} 
\caption{\small The effect of a mass term grows in the infrared. }
\fig{nuflow}
\end{center}
\end{figure}

In this way of thinking, scale transformations move the theory from
the $\nu=0$ conformal fixed point to the empty theory at $\nu=\infty$.
The arrows indicate the change in the theory as one considers it at
larger and larger length scales $\mu^{-1}$.

A mass term is known as a ``relevant'' operator, where the relevance
in question is {\it at long distances} (low energies.)  Although the
mass term has no effect in the ultraviolet --- at short distance ---
it dominates the infrared (in this case by removing degrees of
freedom.)  We can see this from the fact that the dimensionless
coupling $\nu$ grows as we scale from the ultraviolet toward the
infrared.  In fact we can define a beta function for $\nu=m/\mu$ as
follows:
\bel{betar} \beta_\nu
\equiv \mu \dbyd{\nu}{\mu} = -\nu  \ .
\ee
That $\nu$ grows in the infrared is indicated by the negative beta
function.  More specifically, the fact that the coefficient is $-1$
indicates that $\nu$ scales like $1/\mu$.  This tells us that the mass
$m$ has dimension 1.

What about a classically irrelevant operator?  A good example would be
given by adding an $h\phi^6$ interaction to the above Lagragian.  In
four dimensions, the field $\phi$ has dimension $1$, so for $\int d^4x
\ h\phi^6$ to be dimensionless, $h$ must have dimension $-2$.  This
operator is classically irrelevant; any amplitude where the external
momenta are of order $\mu$ must, by dimensional analysis, be a
function of $h\mu^2$.  But this tells us the natural coupling constant
here is the dimensionless quantity $\eta=h\mu^2$; and
\bel{betaeta}
 \beta_\eta
\equiv \mu \dbyd{\eta}{\mu} = +2\eta \ .
\ee
Here, we have a positive classical beta function, telling us that
$\eta$ scales as $\mu^2$, and therefore the operator has negligible
effect on amplitudes for which all external momenta are small.  

What about operators with classically dimensionless coefficients, such
as $\lambda\phi^4$?  The coupling $\lambda$ is dimensionless, so its
classical beta function is zero.  Such a coupling is classically just
as important in the infrared as in the ultraviolet, and it is called
``marginal.''  Another example of a marginal coupling is a gauge
coupling in four dimensions.\footnote{Note that the number of
space-time dimensions affects the story!  In three dimensions, the
$\phi^4$ operator is classically relevant, as is the gauge coupling,
while the operator $\phi^6$ is marginal.}  
We can summarize these various cases by noting 
that in all of these classical examples, the beta function for the dimensionless
 quantity corresponding to a coupling $y$
is simply proportional to its dimension (times a minus sign.)

Of course, these classical beta functions will get quantum
corrections, which in perturbation theory will be given as a series
expansion in $\eta$, $\lambda$ and $\nu$.  As long as these couplings
are small, the beta functions will not change much; the relevant
operators will still be relevant, and the irrelevant operators will
still be irrelevant.  But marginal operators will be different!  The
quantum corrections can make them irrelevant (as in QED) or relevant
(as in QCD) or in very rare cases leave them marginal (as in \nfour\
Yang-Mills theory --- more on this below.)  And when coupling
constants are big, all bets are off: classically irrelevant operators
can become relevant, and vice versa.  Such big effects are not visible
in perturbation theory, but we, armed with nonperturbative methods,
will see many examples in what follows.

\subsection{Beta Functions in Supersymmetric Theories}

What happens in \none\ supersymmetric theories?  The story begins with
some confusion, because there are two different types of couplings 
here.  There are holomorphic couplings, 
and there are
physical couplings.  Holomorphic couplings, which appear in the
Wilsonian effective action for these theories in an appropriate
supersymmetric scheme, 
 include the couplings which
appear explicitly in the superpotential, and the gauge coupling which
multiplies $\int d^2\theta \ W_\alpha W^\alpha$ in the action.  There
are nonrenormalization theorems which say (1) couplings in the
superpotential are not renormalized at any order in perturbation
theory; (2) the gauge coupling is not renormalized at any order in
perturbation theory beyond one loop; (3) additional renormalizations
are holomorphic functions and are therefore not functions of $\mu$,
and so do not affect the beta functions as defined above.  Naively,
this seems to say that all quantum corrections to beta functions for
superpotential couplings must vanish, and that
$$\beta_{1/g^2} = {b_0\over 8\pi^2}
$$ 
exactly. (Here $b_0$ is the one-loop beta function coefficient;
$b_0=3N-N_f$ for supersymmetric $SU(N)$ QCD with $N_f$ flavors.)

But of course this is wrong, and the reason is that there is a part of
the theory --- the K\"ahler potential --- which both contributes to
physical Green functions and is renormalized.  In particular, there is
nontrivial renormalization of the coefficients of all kinetic terms
(also known as wave-function renormalization); the kinetic terms
take the form $\int d^4\theta Z_QQ^\dagger Q$, where 
the wave-function coefficient $Z_Q$ of the chiral superfield
$Q$ is a function of the coupling constants.   Physical couplings
are given by canonically normalizing the fields, $Q\to \sqrt{Z_Q} Q$,
This properly
normalizes the pole of a propagator of $Q$, and ensures that the 
couplings appearing in the Lagrangian are related to
the matrix elements for scattering of the $Q$ particles.  These
couplings are not holomorphic, since the $Z_Q$ are real functions
of the couplings.

Without going into
detail (again the reader is referred to my TASI 2001 lectures), the
fact that $Z_Q$ renormalizes physical Green functions, and thereby
enters actual measurements,  implies that there is a
relation between beta functions of physical couplings involving $Q$
and the anomalous dimension $\gamma_Q$
of $Q$.  The latter, 
$$
\gamma_Q = -\dbyd{\ln Z_Q}{\ln \mu} \  ,
$$ 
is a function of all of the coupling constants in the theory, evaluated at
a scale $\mu$ of order the 
momentum scale of the physical process in question.  (Note 
$\dim Q = 1 + \half \gamma_Q$, with a $\half$ because $Z$ multiplies a bilinear
in $Q$.)
In particular, if a coupling $W=h \phi_1\phi_2\phi_3$ appears in the
superpotential, (note in four dimensions this implies $h$ is
classically dimensionless, since $\dim W=3$ and $\dim \phi=1$), then
its beta function is
\bel{betah}
\beta_h = \half h[\gamma_1(h)+\gamma_2(h)+\gamma_3(h)] .
\ee
Note that (as in the classical cases) the full quantum beta function
for $h$ is proportional to its dimension (times a minus sign): 
$$\dim h
= \dim W - \dim \phi_1-\dim \phi_2 -\dim\phi_3 
=-\half(\gamma_1+\gamma_2+\gamma_3)\ .
$$

This relation between quantum beta function and quantum dimension uses
supersymmetry extensively.  In \none\ theories, no singularities arise
when defining composite operators built as a product of chiral fields.
Therefore, since $\dim \phi = 1+\half \gamma$, $\dim
(\phi_1\phi_2\phi_3) = 3 + \half(\gamma_1+\gamma_2+\gamma_3)$, and
$\dim h = 3 - \dim \phi_1 - \dim \phi_2 - \dim \phi_3 = -\half
(\gamma_1+\gamma_2+\gamma_3)$.  And the classical relation $\beta_h =
-h\ \dim h$, which relates to the scaling properties of the coupling,
would be violated by {\it additive} renormalizations in generic
theories; in this case $\beta_h$ could be nonzero even if $h$ were
zero.  Supersymmetry helps to preserve this unusually simple relation
between dimensions and beta functions through its nonrenormalization
theorems.

As for gauge couplings, the physical beta function is given in the
NSVZ form discussed in Intriligator's lectures (and see my TASI 2001
lectures \mycite{TASIone} for a partial derivation)\footnote{Be careful of signs! In an
asymptotically free theory like Yang-Mills theory, $\beta_g <0$ but
$\beta_{8\pi^2/ g^2} > 0$.}
\bel{NSVZ}
\beta_{8\pi^2\over g^2}=  {b_0 + \half \sum_r \gamma_r 
+ \half \sum_u\gamma_u \over
1 -  g^2N/8\pi^2} = -{16\pi^2\over g^3}\beta_g
\ee
where $\gamma_r$ is the anomalous dimension of $Q^r$ and $\gamma_u$
that of $\tilde Q_u$.
In supersymmetric QCD, where in the absence of a superpotential all
charged fields are related by symmetry, and therefore have the same
anomalous dimension $\gamma_0$, we may write
\bel{betaSQCD}
\beta_{8\pi^2\over g^2}=  {3N - N_f[1-\gamma_0] \over
1 -  g^2N/8\pi^2} \ .
\ee
The reason to write the beta function for $8\pi^2/g^2$ instead of $g$
is merely that the former appears naturally in many nonperturbative
contexts; it is also written as ${\rm Im} [2\pi \tau]$, where $\tau$
is the often-defined
\bel{taudef}
\tau \equiv {1\over 2\pi} \left[\theta+ i{8\pi^2\over g^2}\right] \ .
\ee
Here $\theta$ is the usual theta angle for the gauge theory.
Either form will do, as long as one keeps track of the implications.
Another useful form is
\bel{betaSQCDb}
\beta_{g}=  -{g^3\over 16\pi^2}{3N - N_f[1-\gamma_0] \over
1 -  g^2N/8\pi^2}
\ee
which makes it clear that the beta function for $g$ starts at order
$g^3$ even if $\gamma_0\neq 0$.  Defining the 't Hooft coupling
$\lambda\equiv g^2N$ (sometimes normalized with an additional
$1/4\pi$), we have
\bel{betaSQCDc}
\beta_{\lambda}=  -{\lambda^2\over 8\pi^2}{3 - {N_f\over N}[1-\gamma_0] \over
1 -  \lambda/8\pi^2} \ .
\ee
From this it is more evident that perturbation theory is really an
expansion in $\lambda$, not $g^2$; as 't Hooft pointed out,
perturbation theory breaks down when $\lambda\sim 1$, even if $g^2\ll
1$.

Let us now consider the first gauge theory in Seiberg's dual pair:
$SU(N)$ SQCD, with $N_f$ flavors $Q^r,\tilde Q_u$ and $W=0$.  The only
coupling is the gauge coupling, and the beta function is \Eref{betaSQCDb}.
Note that this expression applies {\it assuming} that all expectation
values of scalar fields are zero, so that all fields have zero mass.
If there are any nonzero expectation values ({{\it i.e.}, if we are
not at the origin of ``moduli space,'' the space of supersymmetric
vacua) or there are any explicit mass terms, then the above formula
will be drastically modified below the mass scale of the massive
particles.

When $g$ is small, $\gamma_0$, which is zero for $g=0$, must be small.
(It turns out to be negative.)  Therefore, if $N_f<3N$, then for $g\ll
1$ we have $\beta_g <0$, and thus the coupling $g(\mu)$ will grow as
$\mu$ decreases.  But when $g$ becomes large (more precisely, when
$\lambda\gg 1$) we simply don't know what happens from perturbation
theory alone.  The anomalous dimension $\gamma_0$ is an unknown
function of $g$, and we simply don't know what it does.

An exception occurs when $N_f$ is very close to $3N$.  When $N, N_f\gg
1$ and $b_0 = 3N-N_f \ll N$, it can be shown, working to two-loop
order, that a conformal fixed point exists at $g\sim 1/N$.
The way this happens is this: the function $\gamma_0$ behaves as $-c
g^2N +$ order $(g^2N)^2$, 
where the number $c$ is {\it positive} and of
order one. From \Eref{betaSQCDb}, the beta function for
the gauge coupling looks like
\bel{betaBZ}
\beta_{g}=  
-{g^3\over 16\pi^2}{b_0 + N_f[-cg^2 N+  
{\rm order}\ (g^2 N)^2] \over
1 -  g^2N/8\pi^2} \ 
\ee
which then means the renormalization group (RG) flow looks like
figure \reffig{gflow}, with a zero at 
$$
g=g_* \approx \sqrt{{b_0\over3c}}{1\over N} \ .
$$
This means the 't Hooft coupling is very small: $\lambda_*=g_*^2N\sim 1/N$.  
Since perturbation theory is an expansion in $\lambda$,
all three-loop and higher corrections are suppressed by
higher powers of $1/N$.  Therefore, this two-loop fixed point
(sometimes called a Banks-Zaks fixed point) survives to all orders.

\begin{figure}[th]
\begin{center} 
\centerline{\psfig{figure=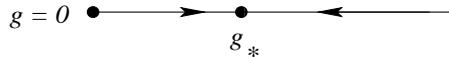,width=6cm,clip=}} 
\caption{\small For $N_f$ in the conformal window, the gauge
coupling has a stable infrared fixed point at $g=g_*$. }
\fig{gflow}
\end{center}
\end{figure}

If the coupling starts at small values in the ultraviolet, it grows to $g=g_*$,
and then stops.  If it starts at larger values, it shrinks to $g_*$,
and stops.  The value $g_*$ is called a ``fixed point'' for obvious
reasons; if the coupling reaches this value, it never leaves.  It is
called an ``infrared fixed point'' because it reaches this value at
small $\mu$ (large distance.)  And it is called stable because no
matter how one perturbs $g$ away from $g_*$, then (if the perturbation
is not too large) $g$ will flow back to $g_*$. (By contrast, the point
$g=0$ represents an unstable ultraviolet fixed point in the context
of this RG flow.)

One of Seiberg's key suggestions ({\it still unproven!})  was that this
sort of fixed point exists even for $3N-N_f\sim N$, in particular,
that it exists for any $N_f$ in the range ${3\over 2}N < N_f < 3N$.
In this ``conformal window,'' the RG flow always takes the form in
\reffig{gflow}, except that $g_*$ is at larger values.  In
particular, $g_*$ grows as $N_f$ shrinks.

We will have more to say later about what happens for $N_f\leq {3\over
2}N$, but for now, let us study these fixed points further.  Using
some important theorems, there are a number of extremely valuable
observations which one can make about these fixed points, even before
considering Seiberg duality.

Here is one crucial theorem.  Near any conformal fixed point
(including a free field theory) all spin-zero gauge-invariant
operators $\OO$ must have dimension greater than or equal to 1 (or
more generally, $(d-2)/2$).  If its dimension is 1 (or more generally,
$(d-2)/2$), then $\partial^2\OO = 0$ (i.e., the operator satisfies the
massless Klein-Gordon equation.)  This is true without any appeal to
supersymmetry!

From this, it follows quickly that if all gauge couplings are zero,
but $\phi$ is not free, then $\gamma_\phi >0$ for any field $\phi$.
This is because if all gauge couplings vanish, then $\phi$ is gauge
invariant, and its dimension $\dim\phi= 1 +\half \gamma_\phi$ must be
greater than one.

Here is another theorem.  At an \none\ \susic\ conformal fixed point, there is
a close relation between the dimensions of many chiral operators and
the R-charges that they carry.  The current of the R-symmetry and the
energy-momentum tensor (of which the scale-changing operator, the
``dilation'' or ``dilatation operator,'' is a moment) are part of
a single supermultiplet of currents.  At a conformal fixed point, the
dilation current and the R current are both conserved
quantities,\footnote{Caution: the conserved R-current at a particular
fixed point may not be a symmetry in the ultraviolet. This happens in
SQCD for $N_f\leq {3\over 2}N$, where the dual theory is free in the
infrared and develops new symmetries, including a new R-charge, in the
infrared limit.}  and the superconformal algebra can then be used to
show that the dimension of a chiral operator is simply $3/2$ times its
R charge:
$$
\dim \OO = {3\over 2}R_\OO \ ,
$$ 
which implies that $\int d^2\theta\ \OO$ is a relevant (irrelevant)
operator at a conformal fixed point when the R-charge of $\OO$ is less
(greater) than 2.

Let's now prove \mycite{NAD} that in SQCD for $N_f\leq
\frac32 N$ there Seiberg fixed points --- specifically, fixed
points at
which no fields are free, and which are located at the origin of
moduli space, where no fields have expectation values ---
cannot occur.  In SQCD a
fixed point requires
$$
\beta_{8\pi^2\over g^2} \propto b_0 + N_f\gamma_0 = 0 \ 
\Rightarrow \gamma_0 = 1-{3N\over N_f} \ .
$$ 
Using the earlier formula that $\dim Q =
1+\half \gamma_0 = {3\over 2}R_Q$, we see that we recover
Seiberg's R-charge assignment,
\bel{dimRQ}
\dim Q = {3\over 2} - {3N\over 2N_f} \ , \ 
R_Q = 1 - {N\over N_f} \ ,
\ee
which he obtained using the fact that this particular R-symmetry is
the unique non-anomalous chiral symmetry of the theory that commutes
with all other symmetries.  However, if $\gamma_0\leq-1$, then the
gauge-invariant operator $Q^r\tilde Q_u$ would have dimension
$2(1+\half\gamma_0)\leq1$.  This is not allowed at a nontrivial fixed
point.  Thus, to have such a fixed point (at least one in the simple
class we have been discussing) it must be that
$$
\gamma_0 >-1 \ \ \Rightarrow\  \ b_0 < N_f\  \ \Rightarrow\  
\ N_f > {\frac32} N \ .
$$
Equivalently, $R_Q>1/3$.

\subsection{Quartic Operators and Their Significance}
\label{subsec:quarticops}

To illustrate the concepts of relevance and irrelevance at the quantum
level, let us turn to a very important operator.  Consider SQCD ---
for reasons that will become abundantly clear, let's call this the
``A'' theory for the following discussion --- and consider adding to
it the superpotential
\bel{QQQQ} W =  h (Q^r\tilde Q_u) (Q^u\tilde Q_r) \ee
with gauge indices contracted inside the parentheses.  (In this and
all subsequent expressions, summation over repeated and contracted
indices is implicit; indices are raised and lowered with a Kronecker delta
function.)  The coupling $h$ has mass dimension $-1$, and
thus the operator above is classically irrelevant.

Crucially, this superpotential, which explicitly breaks part of
the global flavor symmetry (since $r$ and $u$ indices are contracted),
still preserves a diagonal $SU(N_f)$ symmetry and charge conjugation.
This is enough symmetry to ensure that all of the fields still share the
same anomalous dimension $\gamma_0(g,h)$, as was true also for $h=0$.
As always we should study the dimensionless coupling constant
$\eta\equiv h\mu$ and ask how it scales.  Classically it scales like
$\mu$ (and thus has $\beta_\eta = \eta > 0$) but quantum mechanically
it is a different story.\footnote{ Wait a minute.  This theory, whose
potential contains $({\rm scalar})^6$ terms, is nonrenormalizable.
Can we even discuss it quantum mechanically?  Well, nonrenormalizable
simply means that the operator in the superpotential is irrelevant, so
in the ultraviolet regime the effective coupling is blowing up and
perturbative diagrams in the theory don't make sense. But we're
interested in the infrared anyway.  We'll deal with the ultraviolet
later; for now we 
will think of $1/h$ as setting an ultraviolet cutoff on the
theory. Note that we do essentially the same thing with QED in four
dimensions, which is {\it perturbatively renormalizable} but {\it
nonperturbatively nonrenormalizable}, since we cannot take the cutoff
on the theory to infinity without the gauge coupling diverging in the
ultraviolet.  Perturbation theory may not converge, but we are asking
perfectly valid nonperturbative infrared questions which do not depend
on the details of the ultraviolet cutoff.}  In the quantum theory, the
coupling $\eta$ will have a beta function
\bel{etabeta}
\beta_\eta = \eta\left[1+\half(4\gamma_0)\right] = \eta(1+2\gamma_0)\ ,
\ee
where the 1 is the classical scaling and the 4 represents the
contributions of the four chiral superfields in the operator.
The formula for the gauge
coupling is unchanged
\bel{SQCDbeta}
\beta_g \propto - \beta_{8\pi^2\over g^2} \propto -[3N-N_f + N_f\gamma_0]  \ .
\ee

Now, let us recall that $\gamma_0$ is a function of $g$ and $\eta$
with the following properties. (1) If $g=0$ and $\eta\neq 0$, then
$\gamma_0>0$. (2) If $\eta=0$ and $0\neq g\ll 1$, then $\gamma_0<0$.
(3) For $3N>N_f>{3\over 2}N$, there is at least one nontrivial fixed point at
$g=g_*$, $\eta=0$ with $\gamma_0 = 1-{3N\over N_f} \ $.  We can use
these facts to determine the qualitative forms (in particular, the
signs and zeroes) of the beta functions.  For instance, using
\Eref{etabeta}, which changes sign at $\gamma_0=-\half$, we see that
at the nontrivial fixed point on the $\eta=0$ axis, $\eta$ is
irrelevant (as it is classically) if $N_f>2N$, marginal if $N_f=2N$,
and {\it relevant} if $N_f<2N$.

\begin{figure}[th]
\begin{center}
 \centerline{\psfig{figure=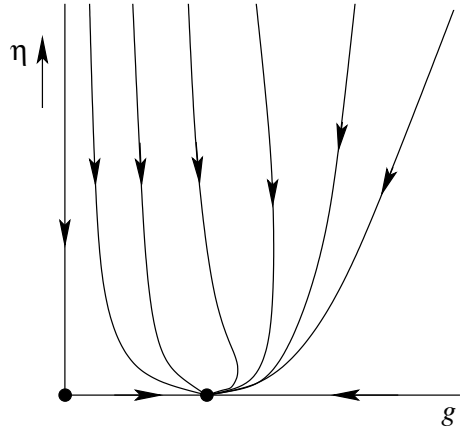,width=6cm,clip=}}
 \caption{\small For $N_f>2N$ the coupling $\eta$ is irrelevant both 
at $g=0$ and at $g=g_*$.}
\fig{piflowNF}
\end{center}
\end{figure}
\noindent 

From these observations we can guess the qualitative features of the
renormalization group flow.  For $N>2N_f$, the qualitative picture is
given in \reffig{piflowNF}.  Even if $\eta\neq 0$, we still end up at
the Seiberg fixed point.  For $N<2N_f$, however, there is a very
different picture, as in \reffig{piflownf}.  Notice that if we start
at weak gauge coupling initially, $\eta$ is irrelevant and flows
toward zero as we would expect classically; but as we flow toward the
infrared, the gauge coupling grows, $\gamma_0$ becomes more negative,
and eventually the coupling $\eta$ turns around and becomes {\it
relevant}.  Although at first it seems as though it will be negligible
in the infrared, it in fact {\it dominates.}  This is called a
``dangerous irrelevant'' operator, since although it is initially
irrelevant it is dangerous to forget about it!  In the infrared it
becomes large, and we must be more precise about what happens when it
gets there\dots on this, more below.

\begin{figure}[th]
\begin{center}
 \centerline{\psfig{figure=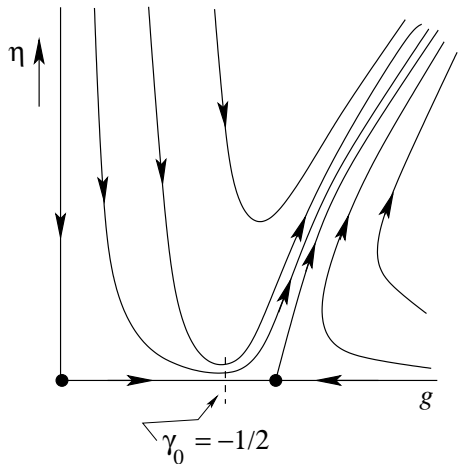,width=6cm,clip=}}
 \caption{\small For $N_f<2N$ the coupling $\eta$ is {\it relevant}
at $g_*$; its initial decrease
is reversed once $g$ is sufficiently large. }
\fig{piflownf}
\end{center}
\end{figure}

What about $N_f=2N$?  In this case, since the operator is marginal at
leading order in $\eta$ (in the same way that the operator $\phi^4$ is
marginal in a theory of a scalar field,) we need to compute more
carefully to find out whether the operator is marginally relevant or
marginally irrelevant.  We'll do this later.  For now, let's conclude
our discussion of this operator by analyzing its description in the
dual theory.

In the dual $SU(\tilde N)$ SQCD+M theory, which we will now refer
to as the ``B'' theory, the operator $(Q^r\tilde Q_u)
(Q^u\tilde Q_r)$ is mapped into the operator $M^r_u M^u_r$; thus the
superpotential is
\bel{dualtoqqqq}
W=yM^r_uq_r\tilde q^u + \hat h M^r_u M^u_r \ .
\ee
I have written $\hat h$ instead of $h$. Although $Q\tilde Q$ is 
mapped to $M$, the former is
classically of dimension 2 and the latter is classically of dimension
1 in the ultraviolet (which is OK, since theory A and theory B are only
the same in the infrared.)  In other words, $h$ is proportional to
$\hat h$ --- if one vanishes, so does the other --- but they
are not equal.  Only $\eta=h\mu$ and $\hat\eta=\hat h/\mu$ could
possibly be equal,
and even then, not until we approach the Seiberg fixed point.

Clearly $\hat h$ has units of mass, because it multiplies a mass term for $M$.
And so we would expect it must be a relevant perturbation, as it is
classically.  Classically, we could choose to look at processes at
scales below the mass of $M$.  We may integrate out $M$ by using its
equation of motion
$$
\bar D^2 (M_r^u)^\dagger = y q_r\tilde q^u + 2\hat h M^u_r  \ .
$$
The left-hand side goes to zero in the infrared, giving 
$$
M^u_r = - {y\over 2\hat h}
\ q_r\tilde q^u \ .
$$
Substituting this into the superpotential gives the low-energy
superpotential
\bel{qqqqsupot}
W_L = - {y^2\over 4\hat h} q_r\tilde q^u q_u\tilde q_r 
\equiv \tilde h\  q_r\tilde q^u q_u\tilde q_r 
\ee
which, remarkably, is of the same form as the superpotential in the
original A theory! except that $\tilde h = -y^2/4\hat h\propto 1/h$.

Of course this was a classical analysis, in which we assumed $\hat h$
was relevant, as it is classically.  But this is surely wrong.  For
$N_f>2N$, it cannot {\it both} be that the $Q\tilde QQ\tilde Q$
operator in the A theory is irrelevant (so that $\eta = h\mu$ flows
to zero in the infrared) {\it and} that the $MM$ operator in the
B theory is relevant
(so that $\hat\eta= \hat h/\mu$ flows to infinity in the infrared, causing
$\tilde \eta = \tilde h\mu$ to flow to zero).  If both $\eta$ and
$\tilde \eta$ flowed to zero, then we would learn that $SU(N)$ SQCD is
dual to $SU(\tilde N)$ SQCD with {\it no} superpotential and no neutral
scalars $M$.  This is simply not true.
Clearly duality must somehow avoid this absurd conclusion.

Of course it does avoid it, and it does so because anomalous
dimensions for operators are so large.  When $3N>N_f>2N$, then $N_f
< 2\tilde N
= 2N_f-2N$.  In the ultraviolet, all scalar fields, being
weakly-coupled, have dimension close to 1; thus $2\approx \dim
[Q\tilde Q] = \dim [q\tilde q] > {3\over 2} > \dim M \approx 1$.  But
as we saw earlier, $R_Q > \half > R_q$ for $N_f>2N$, from which we
learn that near the infrared fixed point $\dim [Q\tilde Q] = \dim M>
{3\over 2} > \dim[q\tilde q]$.  (Note we do not need duality for 
this conclusion; we need only the anomaly-free R-symmetries, which we can 
determine in each of the two theories separately.)
So while $MM$ is classically relevant,
and $q\tilde qq\tilde q$ is classically irrelevant, exactly the
opposite is true in the infrared; $q\tilde qq\tilde q$ is a dangerous
irrelevant operator (becoming relevant in the infrared) and $MM$ is a
{\it harmless relevant} operator --- the mass term for $M$ is actually
infrared irrelevant, and flows to zero!  Thus what happens is this:
$\eta\to 0$, as we would expect, but surprisingly $\tilde \eta \to
\infty$ and $\hat\eta\to 0$ in the infrared.  The A theory flows
to SQCD, and the B theory flows to SQCD+M with $W = yMq\tilde q$, 
consistent with Seiberg duality.

\EX{Show the reverse is true for $N_f<2N$; $Q\tilde QQ\tilde Q$ is
relevant as is the $MM$ mass term.  The B theory therefore loses its
mesons, and its infrared physics is $SU(\tilde N)$ {SQCD}.  Show
the A theory becomes $SU(N)$ {SQCD+M}, by reversing the process of
integrating out.  Following Intriligator and Seiberg \mycite{kins}
and Leigh and
Strassler \mycite{EMOp}, introduce massive gauge-singlet auxiliary fields ${\mathcal
M}_r^u$ to break the $(Q\tilde Q)(Q\tilde Q)$ superpotential into two
terms of the form $\hat y {\mathcal M}Q\tilde Q+ m {\mathcal M}^2$.
Then show that when $N_f<2N$, the mass term for ${\mathcal M}$ is
irrelevant and the resulting A theory is indeed dual to $SU(\tilde N)$
SQCD.  Why is this a legitimate technique?  [Hint: Consider the
large-$N$ treatment of the $O(N)$ model, or the $CP(N)$ model.]}
 
Not surprisingly, the borderline case $N_f=2N$ is the most interesting
of all.  But before turning to it, we have a number of other tasks
ahead of us.

\subsection{Relevant Operators and the Seiberg-Dual Theory}

\def\nnn{n}

Let us consider another important example of a relevant operator by
examining a theory with the same field content as Seiberg's SQCD+M
theory: again $SU(\nnn)$ SQCD, with chiral superfields $q_r$ and
$\tilde q^u$, and gauge-neutral chiral superfields $M$.  However, let
us take $W=0$.  Then the neutral fields $M^i_j$ are free.
They transform under their
own, distinct, $U(N_f)_1\times U(N_f)_2$ symmetry, which is quite
separate from the $SU(N_f)_L\times SU(N_f)_R\times U(1)_B\times
U(1)_{\mathcal R}$ global symmetry group of the $q$, $\tilde q$ and
gauge superfields.  The $M$ fields are completely decoupled from the
physics of the fields charged under $SU(\nnn)$.  The
$SU(n)$ gauge fields and the $q,\tilde q$ multiplets form an SQCD
theory with $\nnn$ colors and $N_f$ flavors, whose
renormalization group flow is the same as for the $SU(N)$ SQCD theory
considered earlier (indeed we have here simply relabelled the SQCD
theory.)  As long as ${3\over 2} \nnn <N_f < 3\nnn$, the
theory will reach a Seiberg fixed point at some $g_*$. The $M$ fields,
meanwhile, are just floating around as a free and independent sector
of the conformal theory.  Since they have dimension $1$, we should
assign them R-charge $2/3$.

But now let us consider what happens if we add the operator $W=
yM^r_uq_r\tilde q^u$ to the action, where $y$ is infinitesimal (and
$g=g_*$).
First, as we can no longer rotate $M$ and $q,\tilde q$ separately,
this breaks the global symmetries down to the usual $SU(N_f)_L\times
SU(N_f)_R\times U(1)_B\times U(1)_{\mathcal R}$.  Second, since the
requirement of anomaly-freedom fixes the R-charge of $q$ and $\tilde
q$ at $1-{\nnn\over N_f}$, we must assign the R-charge of $M$ to be
$2{\nnn\over N_f}$ in order that the R-charge of $W$ be 2.  But this
is not consistent with the dimension of $M$, which we know is 1 +
order($y^2$) --- and this is a sign that nonzero $y$ is inconsistent
with our original $y=0$ fixed point.  This means that for small and
nonzero $y$ the theory cannot be at a fixed point any longer, so a
renormalization group flow is going to ensue.  But will $y$ grow or
shrink?  We can compute the beta function for $y$ at extremely small
$y$, where we know $\dim M= 1$ and $\dim q=\dim\tilde q = {3\over 2}(1
- \nnn/N_f)$:
\begin{eqnarray}\label{betatinyh}
\beta_y& =& -y\ (3 - \dim M - 2\ \dim q) \nonumber \\
&=& -y\ \left(3-1-2 \ \frac{3}{2}\left[1-\nnn/N_f\right]\right) 
= y\ \left(1 - {3\nnn\over N_f}\right)
\end{eqnarray}
and thus for any $N_f< 3\nnn$ this is a relevant operator.  Therefore, $y$
grows, and the Seiberg SQCD fixed point is destabilized.   This is
shown in \reffig{gyflow1}.

\begin{figure}[th]
\begin{center} 
\centerline{\psfig{figure=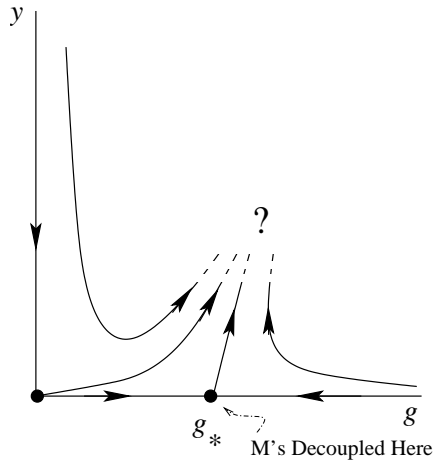,width=6cm,clip=}} 
\caption{\small The $M$ fields are decoupled on the $y=0$ axis;
the theory reaches a Seiberg SQCD fixed point.
If $y\neq0$, the  Seiberg SQCD fixed point is destabilized; 
what happens next?  }
\fig{gyflow1}
\end{center}
\end{figure}

What happens next is anyone's guess; but Seiberg is not just anyone,
and he guessed something very brilliant.  He considered the
possibility that the theory with nonzero $y$ does reach a fixed point,
as illustrated in \reffig{gyflow2}.

\begin{figure}[th]
\begin{center} 
\centerline{\psfig{figure=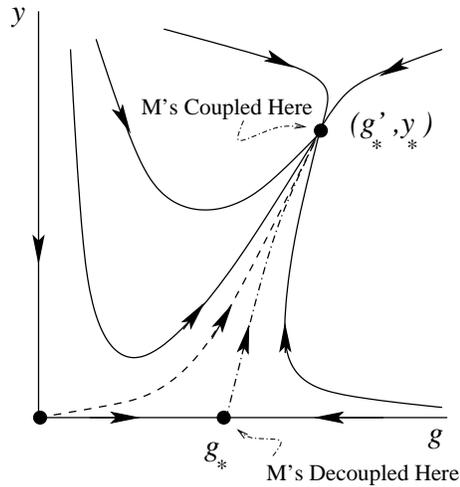,width=6cm,clip=}} 
\caption{\small Seiberg's conjecture: if $y\neq 0$, the flow
ends at a new fixed point $(g,y)= (g'_*,y_*)$.  The dashed curve
is an SQCD+M theory, flowing from a free
theory to its fixed point.  The dot-dashed curve flows between
an SQCD fixed point (with additional free $M$ fields) to an 
SQCD+M fixed point.}
\fig{gyflow2}
\end{center}
\end{figure}

Notice this fixed point is stable; no matter what $g$ and $y$ we start
with, as long as both are nonzero, we will reach the $g'_*, y_*$ fixed
point.  Only if $y$ is zero will we reach the fixed point with
$g=g_*,y=0$; and if only $g$ is zero, the well-known fact that $\phi^4$
has positive beta function in four dimensions implies $y$ will flow to
zero.  (See my TASI 2001 lectures \mycite{TASIone}
for a more convincing proof.)  This
is all consistent with the two beta functions
\bel{bothbetas}
\beta_y = \half y (\gamma_M+2 \gamma_0) \ , \ \beta_g 
\propto -[3\nnn -N_f + N_f\gamma_0 ]
\ee
where now $\gamma_M$ and $\gamma_0$ are functions of $g$ and $y$.
[Note $b_0$ is still $3\nnn-N_f$, independent of $y$, since the
$M$ fields are gauge-singlets and do not affect the one-loop gauge
beta function.] These beta functions both vanish (as they must) when
$\dim q = {3\over 2} R_q$ and $\dim M = {3\over 2} R_M$; you can work
out the $\gamma$'s for yourself.

Note that on this graph the free fixed point at $g,y=0$ is the
ultraviolet fixed point for the SQCD+M theory with $\nnn$ colors,
whereas the $g=g_*', y=y_*$ fixed point is the infrared fixed point
for the SQCD+M theory.  More precisely, any choice of $g,y$ with
$g\neq 0$, $y\neq 0$ at the ultraviolet cutoff will lead to the same
SQCD+M infrared fixed point.  However, if one wishes to define the
SQCD+M theory in the continuum, with no ultraviolet cutoff, one must
take $g,y\to 0$ as the ultraviolet cutoff is removed; this must be
done in a region where $\beta_y<0$.  The dashed line is an example of
a flow from the free fixed point in the infinite ultraviolet to the
SQCD+M fixed point in the far infrared.  Other flows with the same
ultraviolet and infrared fixed points exist between the dashed line
and the $y=0$ axis.  The dot-dashed line represents a different flow,
from the Seiberg SQCD fixed point with $\nnn$ colors, plus a decoupled
set of $M$ mesons, to the infrared fixed point of SQCD+M.

Seiberg then noticed that the dimension and global charges of $M^r_u$
at the $g=g_*', y=y_*$ fixed point --- the infrared fixed point of the
SQCD+M theory --- are now precisely the same as that of the gauge
invariant operator $Q^r\tilde Q_u$ in the $SU(N)$ SQCD gauge theory,
where $\nnn=N_f-N\equiv \tilde N$.  This led him to his duality
proposal --- that the putative fixed point of the $SU(\tilde N)$
theory with the $Mq\tilde q$ superpotential is identical to the
putative fixed point of the $SU(N)$ SQCD theory with $N_f$ flavors
$Q$, $\tilde Q$.  Other checks (see Intriligator's notes)
confirm this idea is consistent.  For instance, the baryon operators
$[Q]^N$ and $[q]^{\tilde N}$ nontrivially match; they have the
same R-charges, and hence the same dimensions, and they have the same
$SU(N_f)_L$ transformation properties (here we see how important it is
that $Q^r$ and $q_r$ are in conjugate representations of $SU(N_f)_L$.)
\reffig{Seidual1} illustrates this suggestion.

\begin{figure}[th]
\begin{center} 
\centerline{\psfig{figure=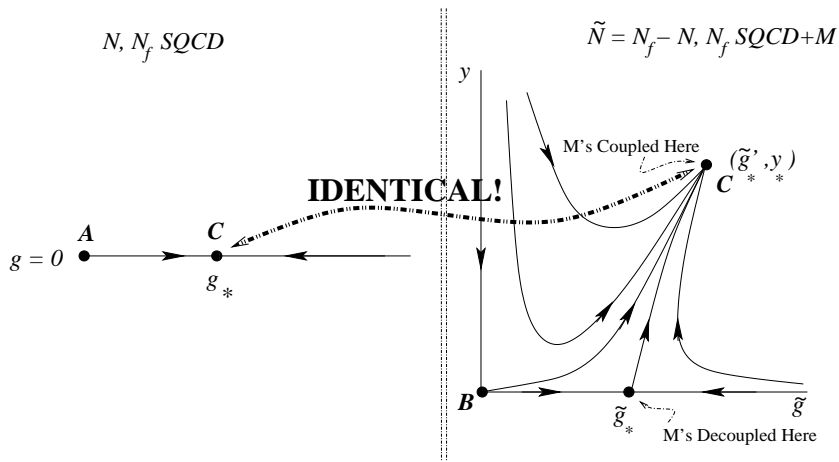,width=11cm,clip=}} 
\caption{\small Seiberg's conjecture: the two fixed points marked
$C$ are actually identical.  The $SU(N)$ SQCD and $SU(\tilde N)$ SQCD+M
theories,
which approach points $A$ and $B$ in the ultraviolet,
give two different ways to reach this one fixed point marked $C$,
and thus provide two different choices of variables for 
describing it.}
\fig{Seidual1}
\end{center}
\end{figure}

\subsection{Seiberg Duality, Precisely}

We'll come back and discuss this further a bit later on.  First,
we'd better ask ourselves ``what is duality''?  Let's examine what
Seiberg is telling us.

In the regime of ${3\over 2}N<N_f< 3N$, Seiberg has given us the
following picture of duality.  There are two flows, each from a free
theory, which approach the same nontrivial infrared conformal fixed
point.  Let us call the original weakly-coupled ultraviolet theory A,
the dual weakly-coupled ultraviolet theory B, and the infrared fixed
point which the theories share C.  The flow is as shown in
\reffig{Seidual2}.  The theories are different in the ultraviolet, and
match only in the infrared, approaching each other as the distance
scale is taken longer and longer.\footnote{The flow into an infrared
fixed point is always controlled by the least irrelevant operator
allowed by symmetries, and thus both flows must approach the fixed
point from the direction associated with this operator.  However, the
two flows might enter the fixed point with a different sign, or phase,
for the least irrelevant coupling. Thus the theories may only match in
the extreme infrared.  With some additional work it is possible to
match them more precisely in some circumstances.  See for example
Kapustin and Strassler \mycite{akms}.  This issue will not arise in the
context of ``exact Seiberg duality,'' which we will discuss in great
detail below.}

\begin{figure}[th]
\begin{center} 
\centerline{\psfig{figure=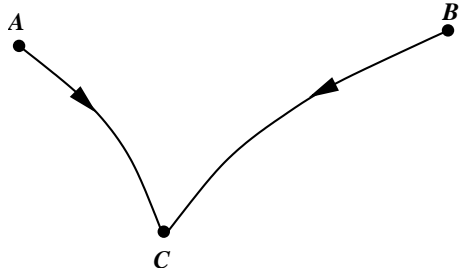,width=6cm,clip=}} 
\caption{\small Fixed points A and B can each be perturbed so
that they flow to fixed point C.}
\fig{Seidual2}
\end{center}
\end{figure}

This is not the only case for which Seiberg duality applies, however.
As you saw in Intriligator's lectures, the cases of $N_f=N$ and
$N_f=N+1$ are ones in which there is a dual theory of massless mesons
and baryons; for $N_f=N$ there is chiral symmetry breaking on the
entire moduli space, while for $N_f=N+1$ all chiral symmetries are
preserved at the origin of the moduli space.  In both of these cases
the low-energy theory is infrared-free, and is not well-defined at
arbitrarily short distance.  Moreover, because the superpotential
interactions $W=X\det M+\cdots$ or $W=\det M+\cdots$ are of degree
$N+1$, these are perturbatively nonrenormalizable theories for $N>2$,
so we cannot even do reliable perturbative calculations.  Instead, the
meson-baryon theories must be defined with an ultraviolet cutoff
$\Lambda_{UV}$ if we are to make any sense of them; and consequently
the RG flow now takes the form in \reffig{ChiralL}.
Here, theory C is a free theory --- yet another conformal
fixed point ---
of massless mesons and baryons (since
all couplings flow to zero in the extreme infrared.)

\begin{figure}[th]
\begin{center} 
\centerline{\psfig{figure=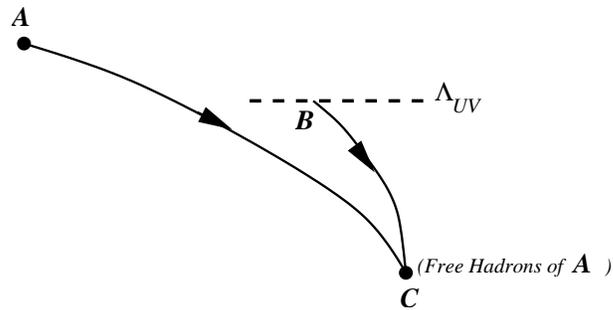,width=8cm,clip=}} 
\caption{\small Think of theory A as QCD with massless
quarks, theory B as the 
chiral Lagrangian with its ultraviolet cutoff, and theory C as
the infrared theory of free massless pions; so it is in the case
of SQCD with $N_f=N$ or $N+1$. }
\fig{ChiralL}
\end{center}
\end{figure}

This should look familiar.  This is precisely the diagram which one
would use for real-world QCD, in the limit where the three light
quarks are massless.  The theory of massless quarks and gluons becomes
strongly coupled at some scale $\Lambda$.  A separate theory, known as
the linear sigma model, is used to describe mesons\footnote{and
baryons, as solitons called ``skrymions''} as far as their
low-momentum interactions are concerned.  This sigma model is non-renormalizable and must be defined
with a cutoff, $\Lambda_{UV}$; the theory is ambiguous near this
scale, and does not exist above it.  Now why do both theories appear
in the same textbook?  {\it It is because they are believed to have
the same infrared physics!}  The sigma model is known as an
``effective theory'' for QCD because it provides a simpler and more
analytically tractable model for the infrared collective behavior of
quarks and gluons.  But we wouldn't introduce it if we thought that a
nonperturbative calculation in QCD, using lattice gauge theory, would
give different answers for physical questions. The statement that one
theory is an effective theory for another is the statement of duality
in a particular context: namely, that Theory A and Theory B approach
each other in the limit where all external momenta are small, and that
the theory they approach (a weakly coupled theory of pions, in this
case) has two descriptions, one using the variables of quarks and
gluons (and computable using lattice gauge theory) and a second using
the variables of pions themselves (computable in  sigma-model perturbation theory.)\footnote{It should perhaps be remarked
that some members of our community still see this viewpoint --- that
the sigma model is ``dual'' to QCD --- as merely
giving a fancy name
to something very simple, namely effective field theory.
But an essential point of these lectures 
is that effective field theory, when it employs
quasi-particles or bound states as effective
degrees of freedom, is a {\it special
case} of something much more general, namely duality.}

Seiberg duality also contains one more spectacular generalization of
these ideas, known as the free magnetic phase, illustrated
in \reffig{FreeMag}.  For $N+1 <N_f \leq
{3\over 2}N $, the dual theory of $SU(N)$ is of the same form as
described before --- it has gauge group $SU(\tilde N)$, fields
$q,\tilde q, M$, etc. --- but now there is a big difference.  The
$SU(\tilde N)$ gauge group has a {\it positive} beta function.
Consequently, just as with QED with massless electrons, and just as
with the sigma model, the coupling constants of the theory run
to zero in the far infrared, and the theory must be defined with an
ultraviolet cutoff.  In some ways, this case is very similar to those
with $N_f=N$ or $N+1$, and to real-world QCD.  But the big
difference is that the low-energy effective theory is itself a gauge
theory, and the dual degrees of freedom are not hadrons of the
original theory, but fractional quasiparticles.  For instance, in the
particular case of $SU(5)$ with 8 flavors, the baryon operator
$Q^1Q^2Q^3Q^4Q^5$ is not a particle in the low-energy theory; instead,
this five-particle state of the high-energy theory falls apart into a
three-quasiparticle state $q_6q_7q_8$ in the low-energy description as an
$SU(3)$ gauge theory, with its 8 flavors of $q$ and $\tilde q$
and 64 $M^u_r$ singlets.

\begin{figure}[th]
\begin{center} 
\centerline{\psfig{figure=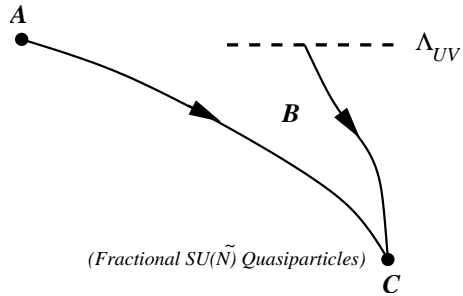,width=6cm,clip=}} 
\caption{\small In the free magnetic phase, theory B is
a gauge theory with a Landau pole, so it requires a cutoff;
theory C is a free theory of the gauge bosons and quarks of
theory B, which are quasiparticles of fractional
charge in theory A. }
\fig{FreeMag}
\end{center}
\end{figure}

Thus Seiberg duality, in its original manifestation, incorporates
three different types of relationships, as illustrated in the three
figures above.  In the
conformal window, two asymptotically free gauge theories flow to a
single nontrivial conformal field theory.  For smaller
$N_f$, below the conformal window,
an asymptotically-free gauge theory and an infrared-free gauge theory
with a cutoff flow to the same infrared physics, described best using
the variables of the
infrared-free gauge theory.  And for $N_f=N$ and $N+1$, the second
theory is an infrared-free theory of chiral superfields --- hadrons of
the original theory --- and again the latter theory well-describes the
infrared physics.  Note that in each of these cases the far-infrared
physics is conformally invariant, although in the last two the fixed
point is free (in the dual variables.)\footnote{Note the
fixed point is approached rapidly in the conformal window, while it is
approached logarithmically when the low-energy theories have an
infrared-free gauge coupling.}

\subsection{Seiberg Duality versus Electric-Magnetic Duality}

The alert student will no doubt be wondering, however, what this type
of duality has to do with electric-magnetic duality, and in
particular, the duality present in \nfour\ \susic\ Yang-Mills theory.
In such dualities, there is a coupling constant $g$, or more precisely
$\tau$, defined in \Eref{taudef}, which is truly constant.  Unlike the
Seiberg conformal fixed points, which are isolated points $g=g_*$
within the space of coupling constants, the \nfour\ theory has a
conformal fixed point for every value of $\tau$.  And unlike Seiberg's
duality, which is an infrared duality, the electric-magnetic duality
of \nfour\ is supposed to be an exact duality between the theory with
$SU(N)$ gauge group and coupling $\tau$ and the theory with
$SU(N)/\ZZ_N$ gauge group and coupling $-1/\tau$.  In fact,
an entire $SL(2.\ZZ)$ acts on $\tau$, generated by shifts of the
theta angle by $2\pi$ $(\tau\to\tau+1)$ and by the electric-magnetic duality
transformation $\tau\to -1/\tau$.

One of the apparent distinctions between these dualities is no
distinction at all.  Although we phrased Seiberg duality as an
``infrared duality,'' it is nonetheless true that Seiberg duality is exact
when applied to the Seiberg conformal fixed points directly.  The
reason for this is trivial: fixed points are the same in the
ultraviolet as they are in the infrared, so if two fixed
points match in the
infrared, they are the same at all scales.  That is, if we take the
$SU(N)$ theory with $g=g_*$, it is identical to the $SU(\tilde N)$
theory with $\tilde g=\tilde g_*$, $y =  y_*$; both are
describing the exactly conformal theory C.  So if we look only at the
conformal theory C, then Seiberg duality provides two descriptions of
a single theory, just as $SU(N)$ \nfour\ is a single theory with
multiple descriptions.  I'll have a bit more to say about this later.

On the other hand, the Seiberg fixed point is an isolated point inside
the space of coupling constants, whereas there really are an infinite
number of \nfour\ theories, each indexed by a different value of
$\tau$.  (The $SU(N)$ theory with $\tau = .2 i$ is equivalent to the
$SU(N)/\ZZ_N$ theory with $\tau = -(.2i)^{-1}=5 i$,
in that there exists a map between operators of the two theories such
that their Green functions are identical; however, the $SU(N)$ theory with
$\tau = .3 i$ is a distinct theory from that with $\tau = .2 i$, in
that there exists no such map.)  As we will soon see, \none\ \susic\
theories also sometimes have continuously-infinite spaces of
conformal field theories, so this kind of phenomenon is not limited to
extended supersymmetry.  But still, duality is applied to a finite set
of theories in SQCD, and to an infinite set of theories in \nfour.

Nonetheless, this distinction is not very important.  What is
important for us here is that in both cases duality acts as a {\it
transformation on a space of theories}, whatever that space may happen
to be.\footnote{To understand how electric-magnetic duality manifests
itself as a change of variables within a path integral, the reader
should study the second section of Seiberg and Witten \mycite{nsewone},
or the pedagogical presentation in my TASI 2001 and Trieste 2001
lectures \mycite{TASIone,Trieste}.}  In \none\ theories with continuous
spaces of conformal field theories, there are sometimes dualities
which are both of Seiberg type {\it and} of electromagnetic type, and
in other cases one finds dualities which generalize both Seiberg
duality and electric-magnetic duality simultaneously.  So the two
types of dualities are really just different manifestations of a
single phenomenon.

\subsection{Why Seiberg Duality can be Exact, and When}
\label{subsec:exactdual}

Let us now see that Seiberg duality is an exact duality, not merely an
infrared duality, even in some contexts beyond the purely conformal
field theories. This will be very important for us.  

We spent some time earlier investigating the precise relationship
between the $SU(N)$ theory with $N_f$ flavors and its $SU(\tilde N)$
dual; we saw that the weakly-coupled versions of these theories, which
we called A and B, flow to an interacting conformal field theory C.
Of course, if this is true for $N, N_f, \tilde N=N_f-N$, then it is
also true for $N, N_f-1, \tilde N-1 = (N_f-1)-N$.  Let us refer to the
$SU(N)$ theory with $N_f-1$ flavors as A', and its dual as B'; their
infrared fixed point we can call C'.  Then A' and B' both flow to C',
just as A and B both flow to C.  

We can check Seiberg duality by
considering flows from A to A', which induce flows from C to C'.  In
particular, let us reduce $N_f$ by one by adding a mass term
$W=mQ^1\tilde Q_1$.  In the dual theory $W= yMq\tilde q + \hat m
M_1^1$, which leads\footnote{Here $m\propto \hat m$, but they have different
dimensions, so they cannot be viewed as equal.} 
to the condition $0 = \partial W/\partial
M_1^1 = y q_1 \tilde q^1 + \hat m$.  Therefore $\vev{q_1\tilde q^1}\neq 0$,
breaking the $SU(\tilde N)$ gauge
group to $SU(\tilde N-1)$.
The massive gauge bosons absorb the fields $q_1$ and $\tilde q_1$,
leaving $N_f-1$ flavors.
Thus theory B indeed flows to theory B', the dual of A'.  

But now we have found an important consequence, illustrated in
\reffig{Seidual3}.  If $m$ is very large, much larger than the strong
coupling scale $\Lambda_A$, then the flow of A is of the form A $\to$
A' $\to$ C'.  Similarly, in this regime B flows to C' by way of B'.
But suppose instead that $m$ is extremely small.  Then the flow is of
the form A $\to$ C $\to$ C', while that of B is B $\to$ C $\to$ C'.
What we obtain in this case, then, is {\it two descriptions of a
single flow} from C to C'.  We can now dispense with the ultraviolet
starting points A and B, by taking the strong-coupling scale
$\Lambda_A$ to infinity, and similarly for $\Lambda_B$, while holding
$m$ fixed.  This leaves us, in this limit, with a single flow from C
to C', described from two different points of view.  And thus we see
that although we began with an ``infrared duality,'' which was between
theories A and B, and involved two different theories flowing to the
same infrared fixed point, the existence and consistency of this
duality implies inevitably that {\it we must also have ``exact
dualities'' as well, in which we have {\bf two} descriptions,
identical at all length scales, of {\bf one and only one} nontrivial
renormalization group flow.}

\begin{figure}[th]
\begin{center} 
\centerline{\psfig{figure=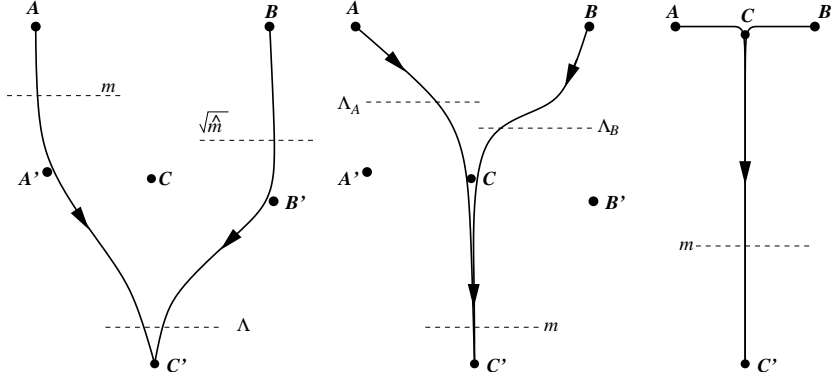,width=11cm,clip=}} 
\caption{\small Theories $A$ and $B$ both flow to $C$ in the
infrared; if perturbed by $m$ to $A'$
and $B'$, they both flow to $C'$.  If $m$ is large
the theories flow near to $A'$ and $B'$; if $m$ is small
they flow close to $C$.  By scaling $\Lambda_A\to\infty$
and $\Lambda_B\to\infty$, holding $m$ fixed, we obtain two
descriptions of the flow from $C$ to $C'$ generated when
$C$ is perturbed by $m$.  }
\fig{Seidual3}
\end{center}
\end{figure}

So if anyone ever tells you that ``duality is only an infrared effect
in field theory, while in string theory duality is exact,'' tell them
to read this chapter!  (And to read a paper by Kapustin and
myself \mycite{akms}, from 1999, in which we worked out more complex and
complete three-dimensional examples in great detail.)  As we will see,
this point undergirds the duality cascade.

Let us see how this comes into our discussion of quartic operators,
using \reffig{CCprime}.
We saw that if ${3\over 2}N<N_f<2N$ then the Seiberg fixed point of
SQCD has quartic $Q\tilde Q Q \tilde Q$ operators which are relevant.
In these cases, the coupling $\eta$ grows in the infrared.  Similarly,
the meson mass term is a relevant coupling $MM$ at the dual SQCD+M fixed
point, which means the quartic operator $q\tilde q q \tilde q$ is {\it
irrelevant}: $\tilde \eta$ shrinks in the infrared.  If we now
consider a flow in which $\eta$ is initially very small, then the
SQCD theory will flow from A (a free theory) to C (the Seiberg fixed point
perturbed very slightly by $\eta$) to C' (a fixed point where $\eta\to
\infty$.)  This is shown in the left side of \reffig{CCprime}.
On the right side of the figure, we see
the dual theory will flow from B (a different free theory)
to C (the same intermediate fixed point, though described in
$SU(\tilde N)$ variables, slightly perturbed by a meson mass term, or
equivalently large $\tilde \eta$) to C' (the same final fixed point,
in dual variables, where $\tilde\eta\to 0$.)  This is important.  {\it
The flow from the Seiberg fixed point of $SU(N)$ SQCD and $N_f$
flavors with $\eta\ll1$ to the theory of $\eta\gg 1$ is {\bf exactly}
Seiberg dual to the flow from the $SU(\tilde N)$ SQCD+M with small
meson mass ($\tilde\eta\gg 1$) to the $SU(\tilde N)$ SQCD fixed point
with $\tilde\eta\ll1$.}  
As illustrated in \reffig{CCprime},
we thereby obtain two descriptions of one physical
theory, the theory which flows from an ultraviolet fixed point
at  C to an infrared fixed point at C'.

\begin{figure}[th]
\begin{center} 
\centerline{\psfig{figure=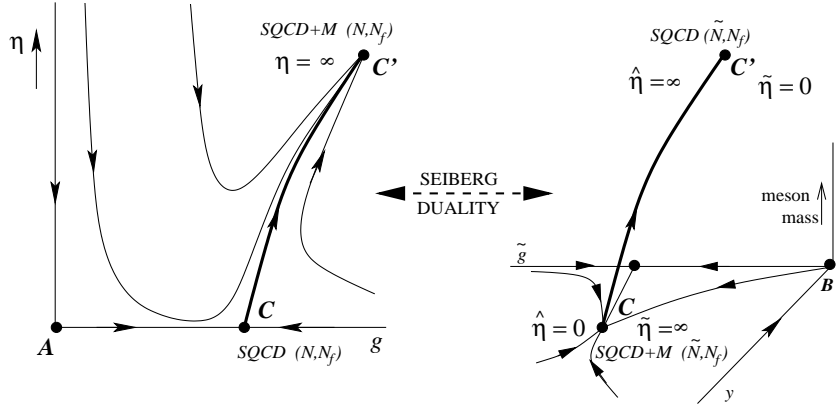,width=11cm,clip=}} 
\caption{\small The flow on the left, in which the quartic
coupling $\eta$ flows from
$0$ to $\infty$, is exactly Seiberg-dual to the flow on the 
right, in which the meson mass coupling $\hat\eta$ also flows
from $0$ to $\infty$, or equivalently the dual quartic
coupling $\tilde\eta$ flows from $\infty$ to $0$.  Theory
$C$ can be reached from either theory $A$ or theory $B$,
but only along the solid line from $C$ to $C'$ is Seiberg-duality
exact.}
\fig{CCprime}
\end{center}
\end{figure}

Before continuing,
please make sure you understand this point, and \reffig{CCprime}, fully. 
It takes some serious thought, but it is absolutely
crucial for an understanding of why the duality
cascade is {\it exact} Seiberg duality in action.

\subsection{Exactly Marginal Operators}
\label{subsec:emop}

We now turn to marginal couplings, and the
special circumstances in supersymmetric
theories that sometimes make them exactly marginal.

Let's argue that \nfour\
Yang-Mills is finite.  
Consider an \none\ gauge theory with three
chiral superfields $\Phi_i$ in the adjoint representation,
and a superpotential $W = \sqrt{2}\hh\ \tr\
\Phi_1[\Phi_2,\Phi_3]$.  I will use canonical normalization here for
the $\Phi_n$, so $\hh=g$ gives the \nfour\ supersymmetric theory.  But
let's not assume that $\hh=g$.  For any $g,\hh$, the symmetry relating the
three fields ensures they all have the same anomalous dimension
$\gamma_0$, which is a single function of two couplings.  The beta
functions for the couplings are
$$
\beta_\hh = {\frac32}\hh\gamma_0 \ ; \ \beta_{g}= 
{-g^3\over 16\pi^2}{3N\gamma_0\over 1 -  g^2N/8\pi^2}
$$ 
These are proportional to one another, so the conditions for a fixed
point ($\beta_\hh=0$ and $\beta_g=0$) reduce to a single equation,
$\gamma_0(g,\hh)=0$.  But this is one equation on two variables, {\it so
if a solution exists, it will (generically) 
be a part of a one-dimensional space of
such solutions}.

Now, does a solution exist?  For small $g,\hh$, we know that
$\gamma_0(g,\hh=0)<0$ and that
$\gamma_0(g=0,\hh)>0$; so yes, by continuity, there must be a curve,
passing through $g=\hh=0$, along which $\gamma_0=0$ and
$\beta_\hh=\beta_g=0$ (and thus perturbation theory has {\it no}
infinities along this line.)  The renormalization group flow must look
like the graph in \reffig{finitetheory}.  Both the theory with $\hh=0$
and the theory with $g=0$ are infrared free; yet a set of nontrivial
field theories lies between.  Notice that we do not know the
precise position of the curve $\gamma_0=0$; in particular, we have
not shown that $g=\hh$ gives $\gamma_0=0$.  However, the {\it
existence} of a finite theory (and one which is renormalization-group
infrared stable against perturbations) only requires arguments using
\none\ supersymmetry.  Of course, since the theory at $g=\hh$ has more
symmetry (namely \nfour) it is natural to expect $g=\hh$ to be the
solution to $\gamma_0(g,\hh)=0$.

\begin{figure}[th]
\begin{center}
 \centerline{\psfig{figure=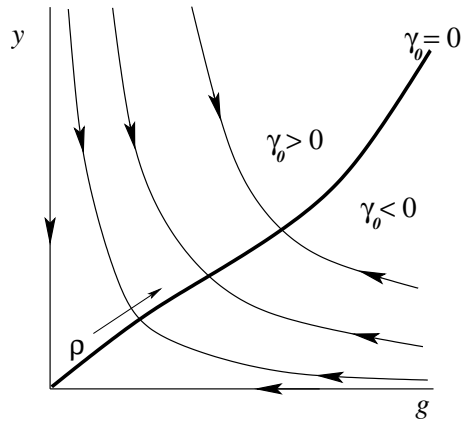,width=6cm,clip=}}
 \caption{\small In some \none\ theories one can argue for
a line of fixed points indexed
by an exactly marginal coupling $\rho$; perturbation theory
has no divergences on this line. Only in \nfour\ is the
equation for this line  $g=\hh$.}
\fig{finitetheory}
\end{center}
\end{figure}

The motivation for introducing this \none--based reasoning is there are
many \none\ field theories which are also finite, as one can show
using similar arguments.  For example, replace the \nfour\
superpotential with $W=\hh\ \tr \Phi_1\{\Phi_2,\Phi_3\}$; the discussion
is almost unchanged, except that $g=\hh$ is not the solution to
$\gamma_0=0$.  Another simple example is $SU(3)$ with $N_f=9$ and the
superpotential
$$
W = \hh[Q^1Q^2Q^3 + Q^4Q^5Q^6+Q^7Q^8Q^9 + \ (Q\to\tilde Q)\ ]
$$
for which
$$
\beta_\hh = {3\over 2} \hh\gamma_Q \ ; \ \beta_g \propto (9-9)+9\gamma_Q \ .
$$ 
The existence of these theories was discovered in the 1980s; the
slick proof presented above is in Leigh and Strassler (1995).

The coupling which parametrizes the line of conformal fixed points
(which is actually a complex line, since the couplings are complex) is
called an ``exactly marginal coupling,'' and the operator which it
multiplies is called an ``exactly marginal operator.''  Let's call
this complex coupling $\rho$.  (In the \nfour\ case we can identify
$\rho$ as equal to the gauge coupling $i/\tau$, but in a more general
\none\ finite theory these will not be simply related.)  Unlike
$\lambda$ in $\lambda\phi^4$, which is marginal at $\lambda=0$ but
irrelevant at $\lambda\neq 0$, $\rho$ is marginal at $\rho=0$, and
remains marginal for any value of $\rho$.  Thus $\rho$ is a truly
dimensionless coupling, indexing a continuous class of scale-invariant
theories.  It can be shown\footnote{In theories with exactly marginal
operators, one can always find a dimensionless holomorphic invariant
which is neutral under all continuous global
symmetries \mycite{kins,EMOp,argyresetal,nelsonstras}.} that there is always a
holomorphic version of $\rho$. Moreover, it is very common for such
classes of theories to be acted upon by duality transformations.  In
fact, for \nfour, electric-magnetic duality (S-duality) acts on this
coupling $\rho$ as in \reffig{Sduality}, identifying those theories at
large $\rho$ with those at small $\rho$.  (Strictly speaking, the
duality relates the $SU(N)$ theory with the $SU(N)/\ZZ_N$ theory;
these have the same local dynamics, but have slightly different
behavior for subtle questions beyond the scope of these lectures.)

\begin{figure}[th]
\begin{center}
 \centerline{\psfig{figure=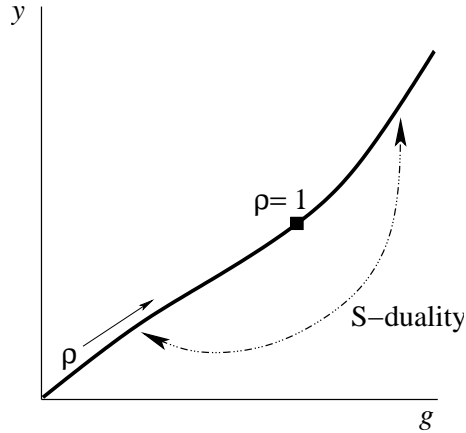,width=6cm,clip=}}
 \caption{\small The action of S-duality on the line of fixed points. }
\fig{Sduality}
\end{center}
\end{figure}

One more example of a finite theory: \ntwo\ $SU(N)$ Yang-Mills with
$N_f=2N$.  This theory is just like \none\ $N_f=2N$ SQCD except
for the presence of a chiral superfield $\Phi$ in the adjoint, and
the superpotential 
\bel{ntwosup}W=f\tilde Q_r \Phi Q^r\ .
\ee
  The $SU(N_f)$ global
symmetry and charge conjugation symmetry
assure all $Q, \tilde Q$ fields have the same $\gamma$.  Because
$$
\beta_f = \half f [\gamma_\Phi+2\gamma_0 ] \ , \  \beta_g\propto -[3N - N(1-\gamma_\Phi)
- 2N(1-\gamma_0)] = -N[\gamma_\Phi+2\gamma_0 ] 
$$ 
we again have a finite theory with a marginal coupling
$\rho=i/\tau$ passing through $g=0,f=0$.  (As in \nfour, we know this
coupling is acted on by a duality transformation that takes $\tau \to
-1/\tau$.)

\EX{Suppose only $k$ of the fields are coupled to $\Phi$.  In this
case, those $Q$'s which are coupled to $\Phi$ have one anomalous dimension, and those
that aren't have a different one.  Show that there is no marginal
operator in this theory.}

\EX{Suppose that $k$ of the fields have
one coupling $f$ and $N-k$ have a nonzero coupling $f'$; this breaks
\ntwo\ \susy\ to \none.  Argue
that $f-f'$ flows to zero and \ntwo\ is restored at low energy.}

Now let us return to our discussion of Sec.~\ref{subsec:quarticops}.
We consider the theory of \none\ SQCD with
$N_f=2N$, and add the quartic superpotential \eref{QQQQ}
$$
W = h (Q^r\tilde Q_u)(Q^u\tilde Q_r) \ .
$$ 
The symmetries ensure that all the $Q$ and $\tilde Q$ have the same
anomalous dimension $\gamma_0$, which is a function of the gauge
coupling $g$ and the coupling $\eta=h\mu$.  Classically, the gauge
coupling is scale-invariant, but recall that $\beta_\eta = +\eta$,
since the quartic superpotential is classically irrelevant.  Quantum
mechanically, the beta functions have the form
$$
\beta_\eta = \eta[1+2\gamma_0] \ ; \ 
\beta_{8\pi^2/g^2}\propto 3N - 2N(1-\gamma_0) = N + 2N\gamma_0
$$ 
and thus we see that $\beta_\eta\propto \beta_g$.  This means that,
as before, the conditions for a fixed point to exist, namely
$\beta_\eta=0=\beta_g$, reduce (for nonzero $g$ and $\eta$)
to a single condition,
$$
1+2\gamma_0 = 0 \ \Rightarrow \ \gamma_0=-\half \ .
$$
Again, this is one condition on two couplings, {\it so any solution
will generically be part of a one-dimensional space of solutions.}  Seiberg has
already convinced us that it is likely that a solution {\it does}
exist for $h=0$; if $W=0$, this theory is SQCD in the conformal
window, and we expect it has a fixed point $g=g_*$ where
$\gamma_0(g_*) = -\half$.  Even though we know nothing
about the curve $\gamma_0(g,h)=-\half$, the very existence of the
Seiberg fixed point then
implies that a continuous space of fixed points emanates from this
point, as in \reffig{pimarginal}.\footnote{The curve $\gamma_0=-\half$
drawn here, with the property that $dh/dg>0$, is
an educated guess.  The guess is based on the fact that 
superpotential couplings tend to
give positive contributions to anomalous dimensions, therefore
requiring a larger gauge coupling to cancel them and bring $\gamma_0$ back to
$-\half$.  We will use this guess
throughout in our schematic discussions.
However, if this guess is wrong, the only
effect is to require a minor redraw of the figures.  The main
results of these lectures
are all independent of the precise shape of this
curve.}

\begin{figure}[th]
\begin{center}
 \centerline{\psfig{figure=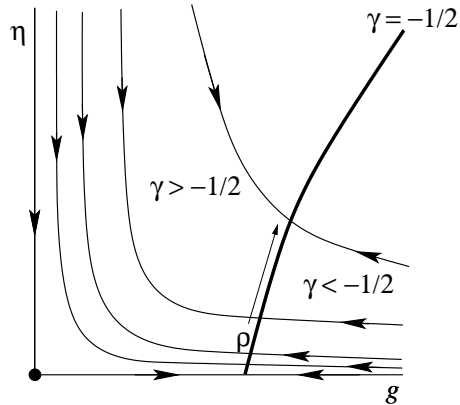,width=6cm,clip=}}
 \caption{\small In the \none\ theory of $SU(N)$ with
$2N_f$ flavors and a quartic superpotential, $\eta$ is marginal at
$g=g_*$ and there is a line of fixed points emerging from
the fixed point at $(g,\eta)=(g_*,0)$.}
\fig{pimarginal}
\end{center}
\end{figure}

So here the classically irrelevant operator has been converted into an
exactly marginal one!  The coupling $\rho$ which parametrizes the line
of fixed points, and on which duality symmetries might act, now has
nothing to do with the gauge coupling.  
  In fact, $\rho=0$ corresponds
to $g=g_*$, not $g=0$.  Granted, the choice of coordinate $\rho$ is not
unique.
But no matter what coordinate we choose, we see that
nowhere are these conformal field theories
near $g=\eta=0$.  We therefore have no hope of seeing them appear
in any perturbation
expansion.  (Nonetheless, we will be able to 
see them in another weakly-coupled
limit, namely that of supergravity.)

But now consider how Seiberg duality acts on these theories.  The
SQCD+M dual has $\tilde N = N$ colors, but it is not the same as
the original theory; it has
$W=yMq\tilde q$ when $h=0$. When $h\neq 0$ one must add $\hat h
MM$ to the superpotential, as before, but now $\hat h$ is marginal.  The
beta functions near the Seiberg fixed point with $(\tilde g, y,
\hat h) = (\tilde g'_*,y_*,0)$ take the form
$$
\beta_{\hat h} = \hat h[-1+\gamma_M] \ , \ 
\beta_y = \half y[\gamma_M + 2\gamma_0] \ , \
\beta_{\tilde g}\propto - N[1+2\gamma_0] \ .
$$
Thus a fixed point requires $\gamma_M=1$ and $\gamma_0=-\half$,
namely two conditions on
three couplings, meaning that the generic solution is
a one-dimensional manifold of fixed points.  
Since the conditions are satisfied at the $\hat h=0$ fixed point,  
a line of conformal field theories 
extends from that point.  But of course
it {\it must}!  The $N_f=2N$ SQCD and $N_f=2N$ SQCD+M theories
are dual, so in the infrared they have the same
physics.  If a line of conformal fixed points ends on the fixed point
of one, a similar line must end on the fixed point of
the other.  This is just as in \reffig{CCprime}, except in this
case the theories marked $C$ and $C'$ are connected by a
line of conformal fixed points instead of a renormalization-group
flow.

In the presence of a marginal mass term, it is not obvious whether we
should or should not integrate out $M$, but suppose we do.  Then, by
the analysis we did earlier for $N_f \neq 2N$, we find the SQCD+M
theory, having lost its $M$ fields, becomes SQCD with a quartic
superpotential $q\tilde q q \tilde q$, with a physical coupling
$\tilde \eta$.  Thus we have a duality between two theories of
identical form, with the same gauge group and the same matter content
and the same quartic superpotentials.  The only difference between
them is that the exactly marginal coupling of one is roughly the
inverse of the other.  Moving away from the $\eta=0$ fixed point, with
$\eta$ increasing from zero, is the same as moving away from the dual
fixed point with the dual coupling $\tilde \eta$ decreasing from
infinity.  Conversely, as we increase $\eta$ to infinity, we expect
$\tilde\eta$ will decrease to zero.  In fact, we can guess that there
should be an exactly marginal coupling $\rho$, whose exact relation to
$g,\eta$ we do not know (and which, indeed, is not uniquely specified)
but which has the property that $\rho=0 $ corresponds to $\eta=0$,
$\rho=\infty$ corresponds to $\tilde\eta=0$, and {\it under $\rho \to
1/\rho$, the theory is self-dual \mycite{kins,EMOp}!} The Seiberg fixed
points sit at $\rho=0,\infty$, and Seiberg-duality has been
generalized to something that more closely resembles the
electric-magnetic duality of \nfour\ Yang-Mills.

\begin{figure}[th]
\begin{center}
 \centerline{\psfig{figure=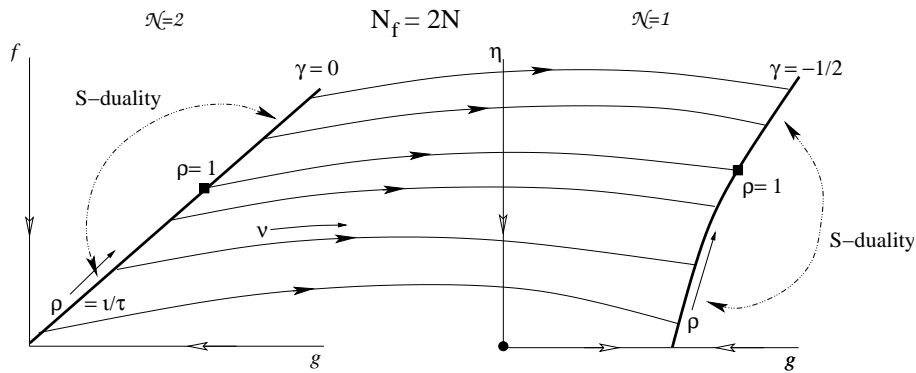,width=12cm,clip=}}
 \caption{\small The \none\ theory with $N_f=2N$ inherits S-duality
from the finite \ntwo\ theory with $N_f=2N$; when the latter is
perturbed by a mass $m_\Phi$ for the adjoint $\Phi$,
the ensuing flow along $\nu=m_\Phi/\mu$
carries it to the former in the infrared. 
Note the S-duality of the low-energy theories {\it is}
Seiberg-duality in this case. }
\fig{thewholeshebang}
\end{center}
\end{figure}

\noindent

Now, since the theory is nonrenormalizable, we probably should say
something about the ultraviolet.  We know a perfectly good ultraviolet
theory into which we can embed this theory, namely the finite \ntwo\
gauge theory with $N_f=2N$ discussed above, perturbed by a mass term
$m_\Phi\Phi^2$ for the adjoint chiral superfield.  When we integrate
out $\Phi$, we obtain\footnote{I am glossing over a subtlety: when the
quartic operator is rewritten as a product of two gauge-invariant
bilinears $Q\tilde Q$, the precise flavor structure of the operator is
very slightly different from the one given in \Eref{QQQQ}.  We could
fix this if there were time, but it would require too much of a
detour.  Moreover, the subtlety is not present for the conifold
theory, which we will study in the next section.  The interested
reader can examine the last section of Leigh and Strassler \mycite{EMOp}
where the precise operator is obtained, and can adjust the discussion
of this section accordingly.}  a quartic operator $(Q\tilde Q)(Q\tilde
Q)$ with coefficient $h = -f^2/4m_\Phi$.  Thus the endpoint of the
flow is one of the \none\ $N_f=2N$ conformal theories we were just
discussing.  It is natural to expect that each \ntwo\ field theory
flows to a unique \none\ conformal field theory, along a flow which
looks schematically like \reffig{thewholeshebang}.  In the figure, the
flow is parameterized by $\nu=m_\Phi/\mu$.  We thereby obtain a
one-to-one map between finite \ntwo\ theories, indexed by the \ntwo\
gauge coupling $\tau$, and our \none\ conformal theories, indexed by a
holomorphic coupling $\rho$.  Since $\rho$ is an arbitrary coordinate,
we can replace it by $i/\tau$, if we wish --- but be careful to
remember that this ``$\tau$,'' while it is the inverse of the gauge
coupling in the \ntwo\ theory, is {\it not} the inverse gauge coupling of the
\none\ theory, as is clear from the graph!

Note one more nice thing about this picture.  The \ntwo\ theories have
an electromagnetic duality $\tau\to -1/\tau$.  The manifold of \none\
fixed points is acted on by Seiberg duality.  Clearly, in this case,
the electromagnetic duality of the high-energy theory descends to
Seiberg duality of the low-energy theory!

\subsection{Seiberg Duality with a Quartic Operator}
\label{subsec:SeiDualQuartOp}

In summary, in the presence of quartic operators Seiberg duality 
becomes even more elegant than before.  We have seen that
$SU(N)$ SQCD with $N_f$ flavors near its Seiberg
fixed point and with a quartic coupling $\eta$
is dual to a theory {\it of exactly the same form}, except with $\tilde N$
colors and with a quartic coupling $\tilde \eta\sim1/\eta$.
Usually $\eta$ flows either to zero or to infinity, and
$\tilde\eta$ flows to the reverse.    
For $N_f>2N$, $\eta$ is irrelevant and $\eta=0$ is stable.
But, importantly for these lectures,
when $N_f<2N$, $\eta$ grows in the infrared --- and as in \reffig{CCprime},
the flow from $\eta=0$ to $\eta=\infty$ is {\it exactly} Seiberg dual
to the flow from $\tilde\eta=\infty$ to $\tilde\eta=0$.

Finally, the case $N_f=2N$ is very special.  The theory is self-dual,
$\eta$ is exactly marginal, and 
Seiberg duality acts to invert $\eta$, or more precisely 
to invert some suitable
holomorphic coupling $\rho(\eta)$ (\reffig{noneselfdual}.)

\begin{figure}[th]
\begin{center} 
\centerline{\psfig{figure=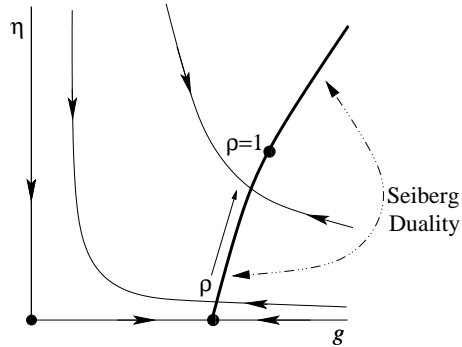,width=6cm,clip=}} 
\caption{\small Seiberg duality acts on the space of fixed points of
\reffig{pimarginal}. }
\fig{noneselfdual}
\end{center}
\end{figure}

Let us conclude with an analogy that helps us to summarize these
results.  We may attempt to think about the beta functions, and the
figures which describe how the theory moves within the space of
coupling constants, in terms of a model dynamical system.  With the
caveat that it has not been proven that this is legitimate, let us
attempt to think metaphorically of the flow of a theory to the
infrared in terms of a ball rolling in a potential, subject to high
friction.  The friction is necessary because the beta function
equations are first-order differential equations, not second-order.
Then the renormalization group flows in
Figs.~\ref{fig:finitetheory}, \ref{fig:pimarginal} and
\ref{fig:noneselfdual} are associated with a ball rolling in a
potential which has an absolutely-flat valley bottom (the manifold of
conformal fixed points, where all beta functions vanish) surrounded by
steep rising walls on either side (which drive all flows into the
valley bottom.)  If you put the ball anywhere in the space, it will
flow (subject to high friction) into the valley; once there, it feels
no potential gradient, and comes to a stop.  By contrast, consider
\reffig{piflownf}, or the left-hand diagram in
\reffig{CCprime}, for SQCD with $N_f<2N$.  Here, the valley (the line
connecting C and C') is still surrounded by steep hillsides (since all
flows approach this line) but the valley also has a tilt (which drives
all flows away from C and toward C').  All generic flows therefore end
at C', even if they initially flow toward C.  We will find this
metaphor of the ball rolling in a landscape very useful below, so keep
it in mind.

\setcounter{equation}{0}
\section{The Conifold Theory}

We now turn to a simple generalization of the model with $N_f=2N$ and
a quartic superpotential. We will gauge an $SU(N)$ subgroup of the
$SU(2N)$ flavor symmetry, and we will also alter the operator slightly
to enhance the global symmetries.  It is this theory which appears
when we study D3-branes on the conifold.  The theory first arose in
the context of Type IIA and M-Theory brane constructions,
but it was first investigated in the AdS/CFT context, and thereby
popularized, by Klebanov and Witten \mycite{ikew}.  We will consequently
refer to it as the ``Klebanov-Witten model.''

\subsection{The Field Theory}

The model of interest has $SU(N)\times SU(N)$ gauge group, two chiral
superfields $A_1, A_2$ in the $({\bf N},{\bf \overline N})$
representation (called the ``bifundamental representation'') of the gauge group, and two chiral superfields $B_1,
B_2$ in the $({\bf\overline N},{\bf N})$ ``anti-bifundamental''
representation.  It has an
$SU(2)_L\times SU(2)_R\times U(1)_B$ symmetry, where the $A_r$
transform under the first factor, the $B_u$ under the second, and both
transform under baryon number with charge $\pm 1/N$.  Finally, we will
add a superpotential which preserves the full $SU(2)\times SU(2)\times
U(1)$:
\bel{conifsup}
W = h\ \tr\det_{r,u} (A_rB_u)
\ee
where the trace is necessary now because $A$ and $B$ both carry {\it
two} gauge indices, one for each group, and thus should be viewed as
matrices.  Explicitly, letting $\alpha$ be a group index in the first
$SU(N)$ gauge group and $a$ be a group index in the second, this can
be written as
\bel{conifsupB}
W = h\left[(A_1)^\alpha_a(B_1)^a_\beta(A_2)^\beta_b(B_2)^b_\alpha - 
(A_1)^\alpha_a(B_2)^a_\beta(A_2)^\beta_b(B_1)^b_\alpha 
\right] \ .
\ee

This theory has not one but {\it two} exactly marginal coupling
constants.\footnote{Actually there are more if one allows a smaller
global symmetry.}  For each of the two gauge groups,
the number of ``flavors'' contributing to the one-loop 
beta function is $2N$ 
($N$ from $A_1$ and $B_1$ and $N$ from $A_2$ and $B_2$) so
$b_0=3N-2N=N$ for both groups.
The beta functions for the two gauge couplings $g_1, g_2$
(and the corresponding 't Hooft couplings)
and for the coupling $\eta=h\mu$ are
\begin{eqnarray}\label{kwbetas}
\beta_{g_1} &=& -{g_1^3\over 16\pi^2}{N+2N\gamma_0
\over 1-{g_1^2N\over 8\pi^2}}  
\ \ \Rightarrow \
\beta_{\lambda_1} = -{\lambda_1^2\over 8\pi^2}{1+2\gamma_0
\over 1-{\lambda_1\over 8\pi^2}} \  , \nonumber \\
\beta_{g_2} &=& -{g_2^3\over 16\pi^2}{N+2N\gamma_0
\over 1-{g_2^2N\over 8\pi^2}} 
\ \ \Rightarrow \
\beta_{\lambda_2} = -{\lambda_2^2\over 8\pi^2}{1+2\gamma_0
\over 1-{\lambda_2\over 8\pi^2}} \  , \nonumber \\
\beta_\eta &=& \eta(1+2\gamma_0) \ ,
\end{eqnarray}
where all four charged fields have the same anomalous dimension
$\gamma_0(g_1,g_2,h)$ because of the global symmetries (including
charge-conjugation.)  Since the
condition for a fixed point clearly puts one condition on {\it three}
couplings (namely $\gamma_0=-\half$), any nontrivial fixed point is
part of a two-dimensional complex manifold of fixed points,
parametrized by two exactly marginal couplings.  And we know solutions
to these conditions exist, since when $g_2=0$ and $h=0$ the theory
becomes $SU(N)$ SQCD with $N_f=2N$, which has a Seiberg fixed point with
$\gamma_0=-\half$.

It is useful to draw a picture representing where this two-complex-dimensional
space of fixed points sits within the space of three complex couplings.
Suppressing the phase of $h$ and the imaginary part of the $g_i$,
we can draw the manifold, somewhat 
schematically (since we don't know its
exact equation), as in \reffig{KWCFTs2}.

\begin{figure}[th]
\begin{center} 
\centerline{\psfig{figure=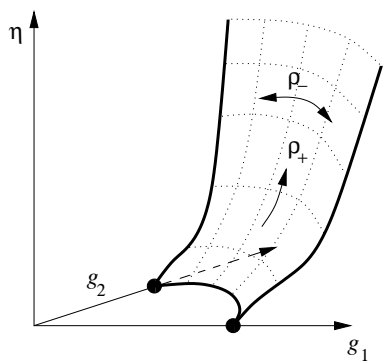,width=5cm,clip=}} 
\caption{\small The two-dimensional space of \none\ conformal 
theories inside the three-dimensional space of couplings.  The exact
shape of the manifold is unknown, but inessential.  }
\fig{KWCFTs2}
\end{center}
\end{figure}

Of course the manifold is symmetric under $g_1\leftrightarrow g_2$.  
Note that $\beta_{g_1}\propto \beta_{g_2}$,
which means there is a line of interesting conformal fixed points lying
in the $h=0$ plane and connecting the two Seiberg fixed points
at $(g_1,g_2)=(g_*,0)$ and $(g_1,g_2)=(0,g_*)$.
When $g_2=0$ (or $g_1=0$), the theory becomes an 
$N_f=2N$ SQCD model with a quartic operator, of the sort we
studied earlier (though the quartic operator is very slightly
different).  For this reason, the intersection of the manifold
of fixed points with the $g_2=0$ plane looks very 
similar to \reffig{pimarginal}.

 \reffig{pimarginal} also reminds us that 
this manifold is infrared stable.  The signs of the beta functions
are such that from any point in coupling constant space that is
off the manifold of fixed points, the flow into the infrared will
bring us onto the manifold at some particular fixed point.  Our
analogy of a ball rolling in a potential applies here too; the
valley is now two-dimensional, but otherwise is similar to what
we discussed at the end of Sec.~\ref{subsec:SeiDualQuartOp}.
We have an valley, with a perfectly flat floor, 
surrounded by high, steeply rising walls.  The ball will
always flow down to the valley floor, and once there, it will 
come to a stop.

\begin{figure}[th]
\begin{center} 
\centerline{\psfig{figure=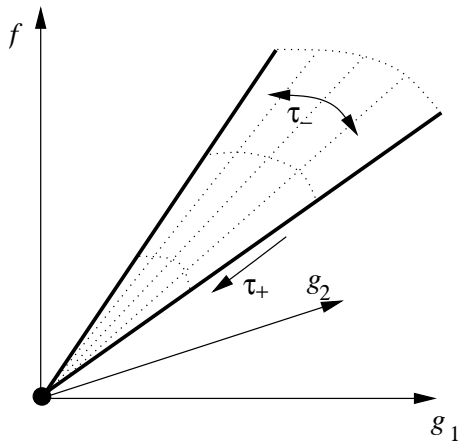,width=6cm,clip=}} 
\caption{\small The two-dimensional space of \ntwo\ conformal 
theories inside the three-dimensional space of couplings.
The dark lines lie in the $g_2=0$ and $g_1=0$ planes. }
\fig{orbiCFTs}
\end{center}
\end{figure}

It is very natural to parameterize this manifold of fixed points using
two complex parameters $\rho_+$ and $\rho_-$ that are symmetric and
antisymmetric in $g_1\leftrightarrow g_2$.  We can do this as we did
in the $N_f=2N$ SQCD model, by finding a finite \ntwo\ field theory,
with its own dualities, to serve as an ultraviolet definition of the
theory. There is indeed such a model, given by adding to the Klebanov-Witten
theory the superfields $\Phi$ in the adjoint of the first $SU(N)$ and
$\phi$ in the adjoint of the second.  The \ntwo\ superpotential, a
simple generalization of \eref{ntwosup}, is $W= f\ \tr [A_r \Phi B^r +
B_r\phi A^r]$.  Notice the flavor group contains only a single $SU(2)$,
since the flavor indices of $A$ and $B$ are contracted (although
there is a second hidden $SU(2)$, an R-symmetry associated with 
the \ntwo\ supersymmetry.)  This theory
--- which corresponds to D3-branes sitting on a $\ZZ_2$ orbifold, by
the way --- is finite and has {\it two} marginal couplings $\tau_1$
and $\tau_2$, corresponding to its gauge couplings --- or at least,
that is the correspondence at large imaginary $\tau_i$.  The manifold
of fixed points is shown in \reffig{orbiCFTs}.  As we did before, we
can choose $\tau_1$ and $\tau_2$ to parametrize the manifold, or even
better, we can use $\tau_\pm=\tau_1\pm\tau_2$.  Although it was known
from the finite \ntwo\ theories with a single gauge group that this
\ntwo\ theory is acted on by duality transformations of $\tau_+$, such
as $\tau_+\to -1/\tau_+$, it was not until Witten constructed
this theory using branes \mycite{ewbranes} 
that it was demonstrated that the dualities
include also $\tau_-\to \tau_- + 2\tau_+$. 

\EX{Show this \ntwo\ 
theory has two marginal couplings, using the beta function
arguments given earlier.}

\begin{figure}[th]
\begin{center} 
\centerline{\psfig{figure=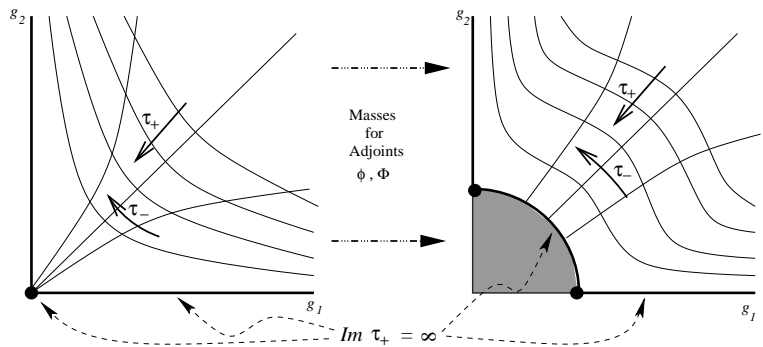,width=10cm,clip=}} 
\caption{\small On the left the two-dimensional space of \ntwo\ conformal 
theories projected onto the $g_1,g_2$ plane; on the right, the
same for the \none\ Klebanov-Witten model.  
The space of theories
on the left, naturally parameterized by $\tau_\pm$, flows
under perturbation by masses to the space of theories on the
right.  This flow allows the use of $\tau_\pm$ as coordinates on
the space of Klebanov-Witten conformal theories.  At the thick
solid boundaries, $\tau_+\to i\infty$.  Note there are no
Klebanov-Witten fixed points near $g_1=g_2=0$. 
}
\fig{tauflows}
\end{center}
\end{figure}

We proceed as we did in Sec.~\ref{subsec:emop}: by adding masses for
$\Phi$ and for $\phi$ {\it with opposite sign} ($m_\Phi=-m_\phi)$, we
can obtain the Klebanov-Witten theory in the infrared.  The two-dimensional
manifold of \ntwo\ fixed points shown in \reffig{orbiCFTs} flows down,
in a simple generalization of \reffig{thewholeshebang}, to the
two-dimensional manifold of \none\ fixed points shown in
\reffig{KWCFTs2}.  We can choose $\tau_\pm$ as the coordinates of the
latter, in lieu of $\rho_\pm$.  This is indicated schematically in
\reffig{tauflows}.  In the left graph, the manifold of fixed points in
\reffig{orbiCFTs} has been projected down onto that $g_1,g_2$ plane,
with the coordinate $\eta$ suppressed; the coordinates
$\tau_\pm=4\pi/g_1^2\pm 4\pi/g_2^2$ here.  In the right graph, to which the
left graph flows (as in \reffig{thewholeshebang}) when the adjoint
fields of the \ntwo\ theory are given masses, the manifold of fixed
points in \reffig{KWCFTs2} has similarly been projected down onto its
own $g_1,g_2$ plane, with the coordinate $\eta$ suppressed.  In this
case $\tau_\pm$ are not simply related to the gauge couplings of the
\none\ gauge theory.  Instead, the \none\ theory on the right with
labels $\tau_+,\tau_-$ is the low-energy limit of the \ntwo\ theory on
the left with gauge couplings $\tau_+,\tau_-$ and adjoint mass terms.

In the Klebanov-Witten model, $\tau_-$ still has some connection with
the difference between the two \none\ gauge couplings, since it
vanishes when $g_1=g_2$; thus for small $g_1-g_2$ we can expect
$\tau_- \propto {4\pi i\over g_1^2}-{4\pi i\over g_2^2}$.  But this is
not true when the difference of the gauge couplings is large, as is
clear from Figs.~\ref{fig:KWCFTs2} and \ref{fig:tauflows}; in some
places $\tau_-$ varies along with $\eta$.  Meanwhile $\tau_+$ in
general has little to do with the sum of gauge couplings; in
some places it reaches
$i\infty$ at finite values of ${1\over g_1^2}+{1\over g_2^2}$.  This
is to be expected when a complicated two-dimensional manifold is
embedded in three dimensions; the coordinates $\tau_\pm$ natural on
the manifold need not have any simple relation to the coordinates
$g_1,g_2,\eta$ of the space in which it is embedded.

\EX{Using a classical analysis,
show that when $\Phi$ and $\phi$ have masses of opposite
sign, the Klebanov-Witten theory emerges in the infrared.  What
happens if the masses are not in the ratio $-1$?}

Finally, and crucially,
we can expect the \ntwo\ duality transformations on
$\tau_\pm$ will be inherited by the Klebanov-Witten theory,
generalizing \reffig{thewholeshebang}.  Indeed Witten's brane-based
arguments \mycite{ewbranes},
properly supplemented, show that they are.  The $SL(2,\ZZ)$ duality
$\tau_+\to-1/\tau_+$ acts on both gauge groups symmetrically.
On the other hand, the transformation $\tau_-\to\tau_-+2\tau_+$
does something stranger.  To see what it is, we need to compare
\reffig{thewholeshebang} to \reffig{tauflows}, and
\reffig{noneselfdual} to \reffig{KWCFTs2}.  In both cases, the
first figure represents the second in the $g_2=0$ plane.  (The
slight difference in the shapes is not meant to be meaningful;
we don't know the shapes exactly anyway.)  What does the transformation
$\tau_-\to\tau_-+2\tau+$ mean near the $g_2=0$ plane?  Comparing the 
$g_2=0$ axis in the right-hand diagram of \reffig{tauflows} to the
picture in \reffig{noneselfdual}, we see that a shift of $\tau_-$
is related to changing $\eta$ from small to large --- in short,
to motion from one region of the manifold
of fixed points to a second region related to the first
by Seiberg duality of the first gauge group.  And indeed,
that's what this duality transformation is: Seiberg duality of one
gauge group with the other gauge group fixed.  The theory can 
easily be seen to be invariant under this transformation.

\EX{Show that the Klebanov-Witten model is invariant in form if you
perform Seiberg duality on one of the two gauge groups.  You cannot
prove $\tau_-\to\tau_-+2\tau_+$ using the methods I've given you; 
instead, you should read
Witten's paper on the matter \mycite{ewbranes}}.

\subsection{D3-branes on the Conifold}

What makes this particular set of conformal field theories so
interesting is that they are easily embedded into string theory, yet
(unlike \nfour\ Yang-Mills) they do not contain a free theory, and as
such do not share many properties with free theories.  In particular,
in contrast to \nfour\ Yang-Mills, where all chiral superfields have
vanishing anomalous dimensions, the chiral superfields here have
$\gamma_0=-\half$.  How does a weakly-coupled string theory manage to
incorporate such a theory?

To answer this question, we need to study the singular
complex manifold known
as the ``conifold.''  As we will see, just as
\nfour\ Yang-Mills theory is realized as the low-energy
dynamics of D3-branes placed in flat space, the Klebanov-Witten theory 
can be
similarly found on D3-branes placed on the conifold.  
In particular, the branes will sit in a
ten-dimensional space consisting of four-dimensional Minkowski space
$x^0, x^1, x^2, x^3$, to which they lie parallel, and a conifold
in the remaining directions.

The conifold, a six-real-dimensional singular space, can be partially
defined by embedding it as a three-complex-dimensional space inside
four complex dimensions.  We choose four complex coordinates
$z_1,z_2,z_3,z_4$, and define the three-dimensional space using a
single complex condition
\bel{conidef}
z_1^2+z_2^2+z_3^2+z_4^2= 0 \ .
\ee
Notice it is not $|z_i|^2$ which appears here!  Alternatively, we may
define the space by using coordinates 
$$
\tilde z_{ru} = z_i(\sigma^i)_{ru} + iz_4 ({\bf 1}_{ru})
$$
where the $\sigma^i$ are the Pauli matrices and ${\bf 1}$ is a two-by-two
unit matrix; then the conifold is defined as 
\bel{conifolddefn}
\det \tilde z =0 \ .
\ee
The symmetries on this space
are $SO(4)\approx SU(2)\times SU(2)$ (which act on the $i$ and $r,u$
indices respectively) along with a $U(1)$ which rotates all of the
$z_i$ by the same phase.  The space is singular
when all $z_i=0$.  This is shown schematically in \reffig{T11}.

\begin{figure}[th]
\begin{center} 
\centerline{\psfig{figure=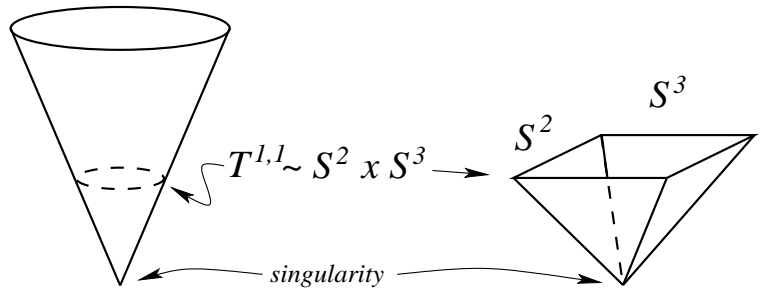,width=10cm,clip=}} 
\caption{\small Two schematic views of the conifold
as a cone with the space $T^{1,1}$ as its sections and with 
a singular point at the origin.  The diagram at left
highlights the smoothness of the space away
from the origin; that at right emphasizes
that $T^{1,1}$ is topologically $S^2\times S^3$,
with the radii of both spheres shrinking to zero at
the origin.  }
\fig{T11}
\end{center}
\end{figure}

I have not fully defined the space yet, because I have given only its
complex structure and not its metric; for now I will leave the metric
unspecified.  But let us see what we can learn about its topological
and algebraic 
features.  First, the defining equation
\eref{conidef} for the space has an obvious rescaling symmetry $z_i\to
\alpha z_i$, where $\alpha$ is any positive definite constant.  Let us
therefore consider the intersection of this space with the additional
condition $\sum_i|z_i|^2=r^2$, $r$ a constant, which removes this
symmetry.  What does the resulting five-real-dimensional space, which
goes by the name $T^{11}$, look like?  It is shown in Candelas and de
la Ossa \mycite{CdlO} 
that this space is topologically $S^2\times S^3$.  We can see
this fairly easily.  Let $z_i = x_i+iy_i$; then our space is defined
by
$$\sum_i x_i y_i = 0\ ; \sum x_i^2 = \sum y_i^2 = r^2/2 \  .
$$
The solution to this equation is to take $x_i$ to be a
four-dimensional real vector of length $r/\sqrt2$, which parametrizes
a 3-sphere, and then to take $y_i$ to be a four-dimensional real
vector of length $r/\sqrt2$ lying orthogonal to $x_i$, which
parametrizes a 2-sphere.  Clearly the nature of the 2-sphere is
$x_i$-dependent, which makes this an $S^2$ fibration over an $S^3$;
however the fibration is trivial in the end, as Candelas and de la
Ossa demonstrate.  These authors also show how to define $T^{11}$ as
the coset space $[SU(2)\times SU(2)]/U(1)$ (note each $SU(2)$ is an
$S^3$ and $U(1)$ is an $S^1$, so the coset is indeed
five-real-dimensional) and how to write an appropriate
metric on this space which preserves these symmetries.\footnote{Do not get 
confused between the two
$SU(2)$ factors which are {\it symmetries} of the space and the two
$SU(2)$ factors which can be used to {\it define} the space!  An
analogy is a two-sphere: the symmetries which act on it are
$SO(3)\approx SU(2)/\ZZ_2$ but the space itself is the coset
$SO(3)/SO(2)\equiv SU(2)/U(1)$.}


A standard technique in
string theory is to ``probe'' a  manifold by
placing a D-brane on it, as in \reffig{probes}, and examining the
low-energy field theory of the modes localized on the brane.  This
``D-brane gauge theory'' has a moduli space of vacua which corresponds
to the configuration space of the D-brane on the manifold.  In
the cases where the
D-brane can be positioned, stably, at any point on the manifold, the
moduli space of the gauge theory and the manifold
should precisely be equal.
And if $k$ identical D-branes are placed on the manifold, 
then the D-brane gauge theory's moduli space
should consist of $k$ copies of the manifold, up to identifications
due to the fact that the $k$ branes are identical.

The simplest example of this is to probe flat space: four-dimensional
Minkowski space times a six-dimensional Euclidean space $\RR^6$.  The
D-brane gauge theory for a single D3-brane placed in flat space is a
four-dimensional $U(1)$ \nfour\ supersymmetric gauge theory.  This
theory has six real scalar fields, which have no potential energy
constraining their expectation values.  Consequently the moduli space
is indeed $\RR^6$.  If $k$ D3-branes are placed on the flat space,
then the resulting $U(k)$ \nfour\ theory has six scalars which are
$k\times k$ matrices.  The D-term and F-term conditions of the gauge
theory force these matrices to be diagonal, which means that only the
$6k$ eigenvalues of these matrices can be nonzero.  There are no other
constraints, so the moduli space is almost $\RR^{6k}$, the
configuration space of $k$ points in $\RR^6$.  However, the D3-branes
are identical, so the moduli space is reduced by permutations of one
D3-brane with another; similarly, in the gauge theory, the eigenvalues
can be permuted by Weyl transformations in the $U(k)$ gauge group.

  If we place a single D3-brane on the conifold, the moduli
space of the D3-brane gauge theory should be the conifold.
What gauge theory would give this result?
The solution turns out to be
a $U(1)\times U(1)$ gauge
theory with four charged fields, $A_1$ and $A_2$ of charge $(1,-1)$
and $B_1$ and $B_2$ of charge $(-1,1)$.  The superpotential is zero.
We will now confirm that this theory's moduli space is the conifold.

What is the space of vacua of this theory?  There
are no F-term conditions, since the superpotential vanishes.
The D-term conditions of
the two $U(1)$ factors are both of the form $|A_1|^2 + |A_2|^2 =
|B_1|^2+ |B_2|^2$.  We may also use gauge invariance to set the phase
of any one of the four fields to zero, making it real, or to set, say,
$A_1$ and $B_1$ to have the same phase.  This leaves a total of six
independent degrees of freedom out of the original eight real scalars,
so we should have a six-real-dimensional moduli space.  How do we
characterize it?  

We can appeal to a slick argument, that in any supersymmetric gauge
theory the solution to the D-term equations (subject to gauge
equivalence) is always given by expressing the moduli space in terms
of {\it expectation values of holomorphic gauge-invariant operators.}
The gauge-invariant objects which are functions of $A_r, B_u$, and not
of $A^\dagger, B^\dagger$, are the four complex bilinears $Z_{ru} =
A_rB_u$; there are no others.  But these four bilinears are not
independent.  In this abelian theory, the chiral superfields are not
matrices; they commute, and thus $(A_1B_1)(A_2B_2)=(A_1B_2)(A_2B_1)$.
This condition may be written $\det Z_{ru}=0$.  And this is precisely
the defining condition \eref{conifolddefn} of the conifold, written as
a three-dimensional complex space embedded in the
four-complex-dimensional space of the $Z_{ru}$.  Since we have already
checked that the moduli space should have six real parameters, there
cannot be any additional constraints on the moduli space.  Thus our
demonstration is complete: this theory's moduli space is a conifold.

\begin{figure}[th]
\begin{center} 
\centerline{\psfig{figure=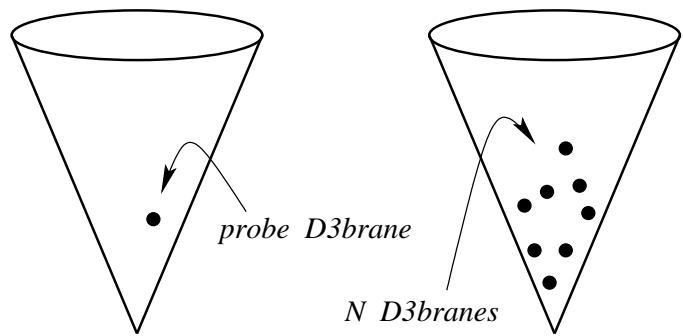,width=9cm,clip=}} 
\caption{\small At left, a single D3-brane placed on the
conifold serves to probe its structure;
at right, $N$ D3-branes are placed on the conifold.  }
\fig{probes}
\end{center}
\end{figure}

If we place $N$ D3-branes on the conifold, the generalization is a bit
more involved.  The gauge group becomes $U(N)\times U(N)$; the fields
$A_r$ and $B_u$ are now in the bifundamental and anti-bifundamental
representations; and
the superpotential is nonzero and takes the by now familiar
form
$$
W = h\ \tr\det_{r,u} (A_rB_u)  \  .
$$
This is the theory discussed in the previous section (except that
its gauge group was $SU(N)\times SU(N)$, a difference we will
address shortly.)
The conditions for supersymmetric vacua can be written
$$
0 = B_2A_rB_1-B_1A_rB_2 \ , \ 0 = A_2B_uA_1-A_1B_uA_2 \ .
$$ 
Notice these are $N\times N$ complex matrix conditions.
Combined with the D-terms, it can be shown (with a small
amount of work) that these equations are
solved if and only if $A_r,B_u$ can be simultaneously diagonalized,
$$
A_r = {\rm diag}\left[a_r^{(1)},a_r^{(2)},\dots,
a_r^{(N)} \right] \ ,
\ B_u = {\rm diag}\left[b_u^{(1)},b_u^{(2)},\dots,
b_u^{(N)} \right] \ .
$$
Since it is a trivial identity that
$$
\det_{r,u} a_r^{(\sigma)}
b_u^{(\sigma)} = a_1^{(\sigma)} b_1^{(\sigma)} a_2^{(\sigma)} b_2^{(\sigma)}
 - a_1^{(\sigma)} b_2^{(\sigma)} a_2^{(\sigma)} b_1^{(\sigma)}
=0
$$ 
valid for all $\sigma=1,\dots,N$, it follows that $\det_{r,u} A_r
B_u=0$, and that the moduli space of the $4N$ complex eigenvalues is
$N$ copies of the conifold, up to permutations of the eigenvalues
which are gauge-equivalent.  This is the same as the configuration
space for $N$ indistinguishable D3-branes on the conifold.

The analysis just performed was purely classical.  However, the $U(1)$
factors are quantum mechanically problematic.  The diagonal $U(1)$
factor is actually decoupled from everything else in the low-energy
limit; it has no renormalizable interactions.  For this reason, we can
truly remove it from our discussions; it makes no contribution to the
interesting physics.  The other linear combination of the $U(1)$
factors, under which $A$ and $B$ are oppositely charged, 
does couple, but its beta function is positive and of order
$N$, from loops of $A$ and $B$ fields.  Consequently
it seems it will have a Landau pole in the ultraviolet ---
unless, of course, its coupling constant is actually {\it zero}.
The correct interpretation is not immediately obvious, but it
does turn out to be true that the coupling constant is zero.  As a result,
this $U(1)$ actually is not gauged in the quantum 
theory.  Instead, this $U(1)$ is a {\it global}
symmetry --- a true symmetry naturally called ``baryon number'' --- in the
quantum theory.

Now, that's fine as far as the moduli space is concerned; if all the
branes are away from the origin of the moduli space, then the gauge
group is broken to a smaller subgroup (or set of subgroups) and in
each subgroup the gauge theory is \nfour\ Yang-Mills.  For instance,
suppose we just have one D3-brane and we allow $A_1B_1$ to have an
expectation value, so that the D-brane sits at some point away from
the singular point of the conifold.  Then the gauge group is broken to
$U(1)$, and six scalars remain massless --- the six possible
translations of the D3-brane away from its initial point --- exactly
the number needed to fill out an \nfour\ $U(1)$ vector multiplet.  I
leave the $U(N)\times U(N)$ case to you as an exercise.

\EX{Show that if all $N$ D3-branes sit at the same nonsingular point, the
low-energy theory is $SU(N)$ \nfour\ Yang-Mills theory.  Then show that
if the branes are all moved slightly apart, one obtains $N-1$ copies
of $U(1)$ \nfour\ Yang-Mills.}  

But what happens if the branes all sit at the singularity?  Then the
details of the singularity are important.  It may appear as though we
can analyze the singularity gradually by getting closer and closer to
it on the moduli space.  But this is misleading, and doesn't work.
Away from the origin of moduli space, at least one scalar field has an
expectation value --- call it $v$ --- so the lightest massive
particles have physical masses $gv$, as shown in \reffig{moduCFT}.
The analysis on the moduli space is valid at energies low compared with
all of these masses:  $E\ll gv$.  But at the origin of
moduli space $v\to 0$, where the singularity lies, $gv$ is also going
to zero, so all of the physics lies {\it above} the scale $gv$ in this
limit.  In other words, to understand the physics at the origin of
moduli space, we actually want to know what is happening for energies
$E\gg gv$ as we take the $v\to0$ limit.  For this purpose, all the
low-energy information about the moduli space is useless.  In other
words, the order of limits matters; the limit $E\to 0$ does not
commute with the limit $v \to 0$.

\begin{figure}[th]
\begin{center} 
\centerline{\psfig{figure=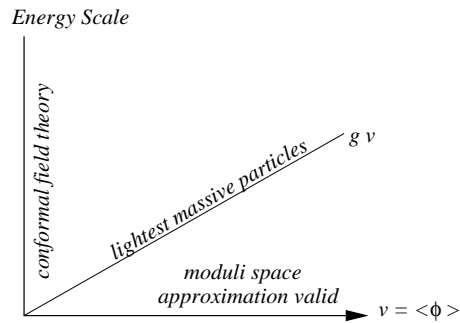,width=6cm,clip=}} 
\caption{\small For given $v$, nonzero particle masses are at least
$gv$. The conformal field theory is found at all energies when $v\to 0$,
while the moduli space involves the massless fields at low
energies for fixed nonzero $v$. }
\fig{moduCFT}
\end{center}
\end{figure}

That's actually fortunate, because there is a profound difference between
the field theory on the moduli space and the field theory at the
singularity.  If, as in the exercise above, all $N$ D3-branes are
sitting away from the origin, the low-energy theory is \nfour\
Yang-Mills, a theory which has no anomalous dimensions ($\gamma_0=0$)
for its chiral superfields.  But when all branes sit at the origin, we
expect the conformal field theory that we studied earlier, for which
the anomalous dimensions for $A_r, B_u$ are $\gamma_0=-\half$.  So
these two regimes are really very far apart, and there is no way to
interpolate between them easily.

\EX{Argue that one can use the theory at the singularity to study the
low-energy theory on the moduli space, but not vice versa; explain the
connection with the irreversibility of renormalization ``group''
transformations.}

How does the theory of D3-branes sitting on the conifold's singularity
know that the superfields $A_r$ and $B_u$ have anomalous dimensions?
The metric of the space, when expressed in terms of $A_r$ and $B_u$,
requires it; we will see this in a moment when we write down the
Maldacena limit.  Another way to see it is to look carefully at
the definition of the R-charge.  We won't explore this carefully
for lack of time.

\subsection{The Maldacena Limit}

Now we want to obtain the string theoretic dual to this gauge theory,
in a regime where supergravity is a good approximation.  So let us
take $N\gg 1$, with the string coupling $g_s\ll 1$, holding $g_sN\gg
1$ fixed, while accounting for the back-reaction of the branes on the
space.  We will take the low-energy limit of the theory on the
world-volume of the branes \mycite{klebAdS}, and take the corresponding
limit in the ambient spacetime by rescaling the radial coordinate $r$,
the distance from the conifold singularity, in the correct
way \mycite{maldaAdS,MAGOO}. In the case of D3-branes in flat space, or
on orbifolds of flat space, the effect of this limit was (1) to leave
the five angular directions transverse to the D3-branes unchanged,
with the same metric they had before, but with a fixed radius $R=(4\pi
g_sN)^{1/4}$, and (2) to combine the Minkowski spacetime directions
with the radial direction $r$ into a single $AdS_5$ space of radius
$R$.  (The $AdS_5$ is associated with the fact that the theory is dual
to a conformal field theory, since its $SO(4,2)$ isometry group is
also the conformal group in four spacetime dimensions.)  Klebanov and
Witten pointed out \mycite{ikew} that we could do the same here, as shown
in \reffig{Maldalim}.  In flat space, a section of ${\bf R}^6$ of
fixed radius was a five-sphere; here, as we discussed earlier, the
corresponding section of the conifold is the space $T^{11}$, a space
of $S^2\times S^3$ topology and $SU(2)\times SU(2)\times U(1)$
symmetry but with a nontrivial metric.  So they proposed that the
conformal field theory on the D3-branes at the conifold singularity
has a stringy description as Type IIB string theory on the space
$AdS_5\times T^{11}$. Thus the metric on the Poincare' patch (which we
will work with exclusively, since our goal is eventually to study
nonconformal field theory on Minkowski spacetime) is
$$
ds^2 = R^2 ds_{AdS_5}^2 + R^2 ds_{T^{11}}^2
={r^2\over R^2}(dx^\mu)^2 + {R^2\over r^2} (dr^2 + r^2 ds^2_{T^{11}}) .
$$
Note the final expression in parentheses is the metric on the conifold,
which is warped here by the $R^2/r^2$ factor.
Also nonzero are the dilaton and axion, which combine together
into the complex string coupling $\tau_{IIB}$
of type IIB string theory, and 
the self-dual 5-form, which as always in AdS/CFT
tells us the number of D3-branes via Gauss's law:
$$
\cFfive\int_{T^{11}} \ F_5 = N \ .
$$

\begin{figure}[th]
\begin{center} 
\centerline{\psfig{figure=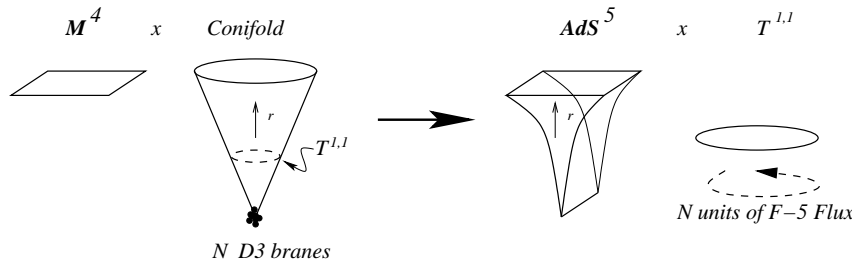,width=11.2cm,clip=}} 
\caption{\small In the Maldacena limit, the D3-branes on Minkowski
space times the conifold are replaced with an $AdS_5\times T^{1,1}$
space with $F_5$ flux on it.}
\fig{Maldalim}
\end{center}
\end{figure}

Actually, this doesn't give us the full set of solutions. The string
theory has two tunable parameters.  One of these is the Type IIB
string coupling $\tau_{IIB}$.  But clearly $\tau_{IIB}$, a parameter
on the space of conformal field theories, cannot be equal to the
Yang-Mills gauge coupling. Instead it must correspond to an exactly
marginal coupling in the gauge theory. In particular, it must be
related to the $\tau_+$ that we decided to use as a coordinate on the
space of conformal fixed points, since both are invariant under
$g_1\leftrightarrow g_2$.
 
The other tunable parameter, which presumably is something like
$\tau_-$, is associated with the presence of a non-trivial two-cycle
in the space $T^{11}$, and a corresponding volume-form $\omega_2$.  We
do no harm to the supergravity equations if we turn on a constant
complex two-form $B_2+iC_2 \propto \omega_2$, where $B_2$ ($C_2$) is
the Neveu-Schwarz (Ramond-Ramond) two-form of IIB string theory.
(Note $H_3, F_3$ are zero since $d\omega_2=0$.)  In fact, a little
work on the string theory side --- for instance, see Hanany and Uranga
1998 --- shows that (when all theta angles are zero, for simplicity),
\bel{Bint}
{\rm Im}(\tau_-) = {\rm Im}(\tau_1) - {\rm Im}(\tau_2) \sim 
2{\rm Im}(\tau_{IIB})
\left[\left(\cBtwo\int_{S^2} B_2 \ - \half\right)\ {\rm mod}\ 1\right]
\ .
\ee
Note that $B_2$ is actually not zero when the gauge couplings are
equal.  We are naturally led to identify $\cBtwo\int B_2 - \half$ with
$\tau_-/2\tau_+$, which is also, via a duality, periodic with period
1.  As we discussed earlier, this means that a shift of $\cBtwo\int
B_2$ by 1 is related to a Seiberg duality transformation on one of the
two gauge groups.  More precisely, this shift moves the theory 
from one region in the manifold of fixed points over to a second 
that is related to
the first by Seiberg duality.

%

An important check on this proposal is that the supergravity agrees
that operator dimensions are quantized in units of ${3/4}$ and carry
appropriate $SU(2)_L\times SU(2)_R\times U(1)_{\mathcal R}$ quantum
numbers.  The gauge invariant operators of the conformal field theory
should be $\tr[A_rB_u]$, $\tr[A_rB_uA_sB_v]$, $\tr[ABABAB]$, etc.,
(with indices in each $SU(2)$-flavor group symmetrized and with traces
removed), and these should have dimension $3/2$, $3$, $9/2$, etc.
Supergravity on $AdS_5\times T^{11}$, when reduced to five-dimensional
supergravity on $AdS_5$, does indeed have five-dimensional scalar
fields with the corresponding charges and five-dimensional masses.  I
won't cover this straightforward calculation here; the interested
reader can read the papers by Gubser \mycite{GubserT11} and by Ceresole
et al. \mycite{CeresoleT11}

Yet another check, due to Gubser and Klebanov \mycite{sgik},
involves the presence of ``dibaryon operators,''
$$
{\mathcal D} = \epsilon^{a_1a_2a_3\dots a_N}
\epsilon_{\alpha_1\alpha_2\alpha_3\dots \alpha_N}
(A_{r_1}) _{a_1}^{\alpha_1}(A_{r_2})_{a_2}^{\alpha_2}
\dots (A_{r_N})_{a_N}^{\alpha_N}
$$ 
where two epsilon tensors create gauge-invariants of the two
$SU(N)$ gauge group factors.  Note the $r_i$ are automatically
symmetrized.  There is a similar di-anti-baryon operator, of course.
These operators have dimension $3N/4$ in the gauge theory.  Their
appearance in the gravity theory is very interesting.  A D3-brane can
be {\it wrapped} nontrivially on $T^{11}$, since $T^{11}$ contains
(topologically) an $S^3$. (More precisely, $\pi_3(T^{11})=\ZZ$.)  From
the perspective of $AdS_5$, this D-brane, with its three
world-volume spatial directions wrapped on a compact subspace, looks
like a particle --- something created by an ordinary field in $AdS_5$.  
It therefore corresponds to a local operator in the field
theory.  Its topological charge corresponds to an additive charge in
the gauge theory, which we are naturally led to identify with baryon
number.  And the object of baryon number 1 with lowest dimension is
the D3-brane of lowest mass with wrapping number 1; it has mass of
order $R^4/g_s\propto N$, and a complete calculation shows that indeed
it corresponds to an object of dimension $3N/4$.  Again the reader is
referred to the original papers on the subject.

Gubser and Klebanov also raised another question.  Since
$\pi_3(T^{11})=\ZZ$, we could wrap D3-branes on $T^{11}$ and get
dibaryons; but $\pi_2(T^{11})=\ZZ$ also.  What can we use this for?
They argued \mycite{sgik} that we could {\it change the theory} --- in
particular, change the gauge group from $SU(N)\times SU(N)$ to
$SU(N+M)\times SU(N)$ --- by wrapping $M$ D5-branes on the $S^2$ of
$T^{11}$.  Each wrapped D5-brane has two world-volume directions on
the $S^2$, so from the $AdS_5$ directions it looks like a 3-brane.
There are many contexts in which D$p+2$ branes wrapped on an $S^2$
give a {\it fractional} Dp brane --- a brane with $p+1$ large
world-volume dimensions but with fractional charge.  This is a subject
all its own, and there's no time for it here, but I refer you to the
references in Gubser and Klebanov's paper. If we do wrap a D5-brane of this
type on an $S^2$, its tension grows with $r$, so it prefers to sit at
$r=0$. Thus, unlike an integer D3-brane, which is free to move around
on the conifold, fractional D3-branes must be placed at the
singularity of the conifold, at $r=0$.  This is shown in
\reffig{frcD3s}.

\begin{figure}[th]
\begin{center} 
\centerline{\psfig{figure=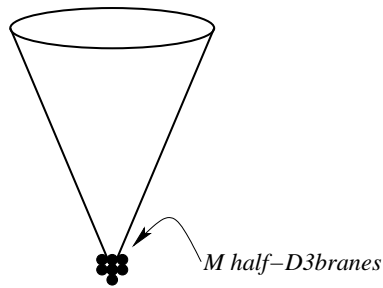,width=5cm,clip=}} 
\caption{\small Fractional D3-branes must sit at the singularity,
unlike full D3-branes which are free to roam anywhere on the conifold.
 }
\fig{frcD3s}
\end{center}
\end{figure}

The presence of D5-branes breaks the S-duality of the IIB-string
theory, thereby removing the $\tau_+\to -1/\tau_+$ duality of the
field theory.  The D5-branes are a source for the 6-form potential
$C_6$, and generate a corresponding 7-form electric flux $F_7$.  The
indices of the nonzero component of $F_7$ span the world-volume
directions of the D5-brane and the radial direction $r$.  We can
Poincare-dualize $F_7$ to a 3-form flux $F_3= *F_7$, which will lie
perpendicular to the volume of the D5-branes, and to the coordinate
$r$.  In short, $F_3$ will be proportional to the volume form of the
$S^3$, which we will call $\omega_3$, and will satisfy $\cFthree\int
F_3 = M$, counting the number of D5-branes.

The presence of $F_3$ will now act as a source for the metric,
and for $F_5$ (and potentially for the dilaton, although we will
see later that the dilaton is not in the end affected.)  Moreover,
$B_2$ will no longer be constant, and consequently $H_3=dB_2$ will be nonzero.
Therefore, once the effect
of the $M$ fractional branes is accounted for, the supergravity metric
can no longer be $AdS_5\times T^{11}$.  
This is consistent with the gauge theory: once 
the two gauge groups no longer have the same number of
colors, the theory is no longer conformal.  But what then is the metric in 
the presence of fractional branes,
and, if we can find the answer, what field theory physics 
does it correspond to?

\setcounter{equation}{0}
\section{The Cascade: Preliminaries}

The answer, of course, is the duality cascade.
I am not going to tell you the detective story which led to its
discovery.  That would take too long, and would
require us to study issues which are not essential for the end result.
Instead, for pedagogical reasons, we will build the cascade
not from the top down, as was done historically, but from the bottom
up.  

\subsection{The Base of the Cascade}

What would happen if we had only fractional D3-branes and no integer
D3-branes?  Our field theory would then be \none\ $SU(M)\times SU(0)$
gauge theory, and would have matter fields in the $({\bf M}, {\bf 0})$
representation --- in short, no matter fields at all.  Thus the theory
would be pure $SU(M)$ \none\ Yang-Mills, whose properties you know.
Classically, it has gluons and gluinos, and a $U(1)$ R-symmetry which
rotates the gluinos by $\lambda \to \lambda e^{i\alpha}$.  It has no
moduli space; there are no scalar fields, corresponding to the fact
that the fractional branes must sit at the origin of the conifold and
cannot move off of it.  At one loop, it has a beta function with
$b_0=3M$, and a holomorphic strong-coupling scale satisfying
$$
\Lambda_0^{3M} = \mu^{3M}\exp[-8\pi^2/g^2(\mu)] \ .
$$ 
It also has an anomaly which breaks the $U(1)_{\mathcal R}$
 to a $\ZZ_{2M}$
R-symmetry, under which the gluinos rotate as above but with $\alpha =
\pi k/M$, $k = 0,\dots,2M-1$.  Nonperturbatively the theory has many
interesting features, some analogous to pure nonsupersymmetric
Yang-Mills, some analogous to QCD, and some different from both.
These include
\begin{itemize}
\item A discrete spectrum, with a mass gap; 
\item Confinement (and associated strings carrying
electric flux);
\item Chiral symmetry breaking (a $\vev{\lambda\lambda}$ condensate
spontaneously breaks $\ZZ_{2M}$ to $\ZZ_2$, since now the vacuum
is invariant only under chiral transformations with
$\alpha=0,\pi$); and its two consequences:
\item $M$ identical isolated degenerate vacua, with the gluino
condensate taking values
$\vev{\lambda\lambda}=e^{2\pi i n/M} (\Lambda_0^{3M})^{1/M}$,
$n=0,\dots,M-1$ (since as with any spontaneously broken symmetry $G\to
H$, the vacua must form a representation of $G/H \approx
\ZZ_{2M}/\ZZ_{2}\approx\ZZ_M$); and
\item Domain walls which separate one vacuum
from the next --- and it turns out these are BPS saturated
(Dvali and Shifman, 1997).
\end{itemize}

But this is very interesting already. We know that the conifold has a
continuous $U(1)_{\mathcal R}$ symmetry.  If we put $M$ fractional
branes on it, it must somehow happen that (a) the continuous symmetry is broken by an
anomaly to $\ZZ_{2M}$, and (b) the remaining symmetry is spontaneously
broken to $\ZZ_2$ by the dynamics of the theory.  How is this going to
play out?  We won't discuss the anomaly here (see Ouyang, Klebanov and
Witten \mycite{OKW})  but we will carefully investigate the spontaneous
breaking of chiral symmetry.

\subsection{The Base with an Extra D3}
\label{subsec:extraD3}

What if we had $M$ fractional branes and {\it one} integer D3-brane,
as in \reffig{Mplus1}?
This is a particularly interesting case.  The D3-brane is free to move
all around the conifold, unlike the fractional branes.  As such, it
makes an excellent {\it probe} of the space, and what has happened to
it.  We will see in a moment that it serves our purpose admirably.

\begin{figure}[th]
\begin{center} 
\centerline{\psfig{figure=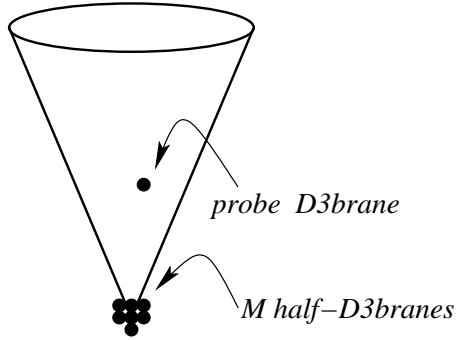,width=6cm,clip=}} 
\caption{\small $M$ fractional D3-branes stuck at the origin
of the conifold, with one full D3-brane serving as a probe. }
\fig{Mplus1}
\end{center}
\end{figure}

The field theory on the D3-branes is $SU(M+1)\times SU(1)$ with fields
$A_r$ in the $({\bf M+1},{\bf 1})$ and $B_u$ in the $({\bf
\overline{M+1}},{\bf 1})$ representations.  But $SU(1)$ is no group at
all, so this is simply SQCD with $M+1$ colors and 2 flavors.  However,
it isn't quite that trivial, because the flavors will have the
conifold superpotential, which classically preserves the full $SU(2)_L
\times SU(2)_R\times U(1)_{\mathcal R}\times U(1)_B$ flavor symmetry, namely
\bel{nftwosup}
W = h\left[(A_1 B_1)(A_2 B_2) 
- (A_1 B_2)(A_2 B_1)\right] = h \det_{r,u} (A_rB_u) \ .
\ee
Classically, our single
D3-brane knows that it is moving on the conifold.  The moduli space
must be described by the four gauge invariant operators
$Z_{11}=A_1B_1$, $Z_{22}=A_1B_2$, $Z_{21}=A_2B_1$, $Z_{22}=A_2B_2$.
But, as we have seen before, the classical equations for a
supersymmetric vacuum include
$$
0=
{\partial W\over \partial A_1}  = B_1(A_2B_2)- B_2(A_2B_1)
$$ which gives (multiplying by $A_1$ and contracting the hanging
$SU(M+1)$ indices)
$$
0 = (A_1B_1)(A_2B_2)-(A_1B_2)(A_2B_1) = \det Z_{ru}
$$
which is exactly the equation for the conifold.

At one loop, an anomaly removes most of the $U(1)_{\mathcal R}$ symmetry.
However, there is a discrete R-symmetry which survives.  The
superpotential, which must have R-charge 2, requires the $A$s and
$B$s have R-charge $\half$, and their fermions carry charge $-1/2$. 
The gluinos carry R-charge 1. An instanton in this theory
has one fermion zero mode for each of $A_1, A_2, B_1, B_2$,
and $2(M+1)$ for the gluinos.  If the gluinos are rotated
by a phase $e^{i\alpha}$, then the $A,B$ zero modes
rotate  by a
phase $e^{-i\alpha/2}$.  Thus the whole instanton rotates by a phase
$e^{i(2(M+1)\alpha-2\alpha)}=e^{i2M\alpha}$.  For the instanton to be
invariant (meaning that the R-charge in question is anomaly-free,) it
must be that $(2M)\alpha$ is a multiple of $2\pi$, or $\alpha = \pi
k/M$, $k=0,\dots, 2M-1$.\footnote{We only include $k$ up to
$2M-1$ because the apparently additional symmetry
under which $\alpha=2\pi$, the gluinos are invariant, and $A$ and
$B$ change sign is already contained in $U(1)_B$ as a rotation
by $\pi$; thus it is already included in our list of symmetries.}
This $\ZZ_{2M}$ non-anomalous discrete R-symmetry is a chiral symmetry,
since it rotates $A$ and $B$ in the same direction.
Remarkably, this calculation suggests
that our theory has a discrete R-symmetry {\it
which only depends on the number of fractional D3-branes}, and not on
the number of integer D3-branes.

What does this theory do nonperturbatively?  When $h\to 0$ this is
just SQCD again, and we know that for $N_f=2$ the theory is best
described using the gauge-invariant variables $Z_{ru}=A_r B_u$ (as
long as all fields have small expectation values.)  And we know (from
Affleck, Dine and Seiberg \mycite{ADS}, as described in Intriligator's
lectures) that the theory generates a dynamical superpotential of the
form
\bel{ADSsup}
W = (M-1)\left[{2\Lambda^{3(M+1)-2}\over\det Z}\right]^{1/([M+1]-2)} \ .
\ee
where $\Lambda$ is the holomorphic strong coupling scale of the
$SU(M+1)$ gauge group.  The condition for a supersymmetric vacuum becomes
\bel{dWdM}
0={\partial W\over \partial Z^r_u} = 
-(2\Lambda^{3M+1})^{1/(M-1)}(\det Z)^{-M/(M-1)}\epsilon_{rs}\epsilon^{uv}Z^s_v 
\ee
and so this theory has no vacuum except at $\det Z\to\infty$. On the
other hand, when $h$ is present but small, the effective
superpotential is of the form
\bel{ADSsupp}
W = (M-1)\left[{2\Lambda^{3M+1}\over\det Z}\right]^{1/(M-1)}  + 
h\det Z + {\rm order}\ (h^2)\ .
\ee
Let us assume that the higher order terms in $h$ vanish (and one can
show, using nonperturbative methods, that they do in this case.)  Then
the conditions for a supersymmetric vacuum take the form
\bel{dWdMp}
0={\partial W\over \partial Z^r_u} = \left[-
(2\Lambda^{3M+1})^{1/(M-1)}(\det Z)^{-M/(M-1)}+h\right]
\epsilon_{rs}\epsilon^{uv}Z^s_v \ .
\ee
Despite appearances, there is no solution at $Z^s_v=0$; if you
take the determinant of \eref{dWdMp}, you will see this is the case.
The only solutions are to have the quantity in brackets
vanish:
\bel{solnAdSp}
(\det Z)^{M} = \left[2\Lambda^{3M+1}/h^{M-1}\right]  
\Rightarrow \det Z = \left[2\Lambda^{3M+1}/h^{M-1}\right]^{1/M}  \ .
\ee

This is a remarkable result, in a wide variety of ways.  First, we see
that the probe D3-brane is not moving on the conifold anymore!
Instead, it moves on a space with a different complex structure, given
by the equation $\det \tilde z_{ru}= \epsilon$, where $\epsilon$ is a
nonzero complex constant.  This space is also well-known and
well-studied \mycite{CdlO}: it is called the ``deformed
conifold.''\footnote{The ``resolved conifold'' is yet another space, with
the same complex structure as the conifold but with other aspects
changed; I will not discuss it here.}

Moreover, since \Eref{solnAdSp} has $M$ solutions for $\det Z$,
the constant $\epsilon$ can have any one of $M$ different phases, so the moduli
space consists not of one copy of this space but of $M$ copies.  This
is reminiscent of the fact that there are $M$ different vacua in
$SU(M)$ Yang-Mills theory.  Indeed, we might guess that what has
happened is that (in analogy to $SU(M)$ Yang-Mills) the expectation
values for $Z_{ru}$ have broken the $\ZZ_{2M}$ symmetry down to
$\ZZ_2$.  This guess is correct!  Let's ignore the anomaly for the
moment, and see that the classical $U(1)_{\mathcal R}$
continuous symmetry is broken down to $\ZZ_2$.  When $\epsilon=0$ we
can rotate the $\tilde z_{ru}$ by any phase without changing the
equation $\det \tilde z=0$, but when $\epsilon\neq 0$ only $\tilde
z_{ru}\to -\tilde z_{ru}$ for all $r,u$ leaves $\det\tilde z
=\epsilon$ invariant.  The anomaly does not affect this remaining
$\ZZ_2$ symmetry.  Thus the $U(1)_{\mathcal R}$ symmetry behaves just
as in pure Yang-Mills; the classical $U(1)$ is broken explicitly by
the anomaly to $\ZZ_{2M}$, and then it is broken spontaneously to
$\ZZ_{2}$, with the result that the moduli space of the theory
consists of $M$ identical branches, with domain walls separating them.

However, unlike \none\ Yang-Mills, this theory has massless scalar
fields, six to be precise, corresponding to the six coordinates of the
probe D3-brane.  And the chiral $SU(2)_L\times SU(2)_R$ symmetry,
which is not present in the pure Yang-Mills case, is {\it also} broken
--- to the diagonal $SU(2)_{L+R}$ subgroup if
$Z_{11}=Z_{22}=\sqrt{\det Z}$ and $Z_{12}=Z_{21}=0$ (or any flavor
rotation of this vacuum), and further to $U(1)_{L+R}$ otherwise.  Thus
the theory has Nambu-Goldstone bosons, one for each broken generator
of the group, along with other scalar and fermion superpartners.  Note
these three (or five) ``pions'' are simply a subset of the six
coordinates of the probe brane.

\begin{figure}[th]
\begin{center} 
\centerline{\psfig{figure=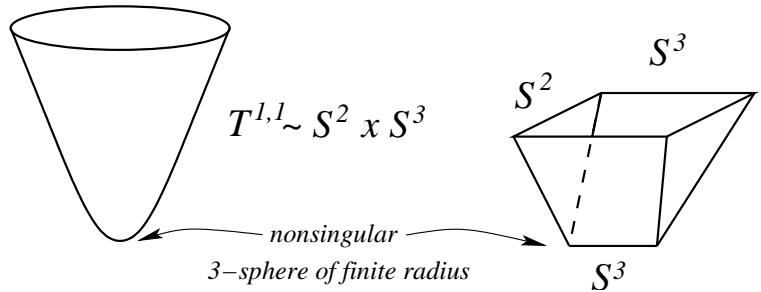,width=10cm,clip=}} 
\caption{\small The conifold is deformed by the $M$ fractional
D3-branes; the singularity is gone, because the $S^3$ remains
of finite size as the radius of the $S^2$ shrinks to zero.
Compare with \reffig{T11}.}
\fig{defcon}
\end{center}
\end{figure}

Also remarkable is that the moduli space is now nonsingular.
Recall that any section of the conifold at a fixed radial coordinate
$r$ is of the form $T^{11}$, which has topology $S^2\times S^3$, and
that the radius of these spheres goes to zero in the limit $r\to 0$,
leading to a singularity, as shown in \reffig{T11}.  
This is not true of the deformed conifold.
Again let $z_i = x_i+iy_i$; and take
$\epsilon$ to be real and positive. (If it is not, then redefine
$z_i=e^{i(\arg \epsilon)/2}(x+iy)$, and proceed in the same
way.)  Then the deformed conifold is
defined by
$$\sum_i x_i y_i = 0\ ; \sum x_i^2 = r^2/2 + \epsilon/2 \ ; \
\sum y_i^2 = r^2/2 - \epsilon/2 \ .
$$ 
Since we chose $\epsilon>0$, and $r^2>0$, we can always take $x_i$
to be a four-dimensional real vector of length
$\sqrt{r^2+\epsilon}/\sqrt2$, which parametrizes a 3-sphere.  Now
we must take $y_i$ to be a four-dimensional real vector, lying
orthogonal to $x_i$, of length $\sqrt{r^2-\epsilon}/\sqrt2$.  This
gives us a 2-sphere if $r^2>\epsilon$, but for $r=\sqrt{\epsilon}$ we
see this 2-sphere degenerates to a point.  Note there is still a
nontrivial 3-sphere at this radius.  Clearly there are no solutions
for $r<\sqrt{\epsilon}$. Thus our space has the property that the
2-sphere shrinks faster than the 3-sphere, and the space ends entirely
when the 2-sphere disappears.  Locally, near $r=\sqrt{\epsilon}$, the
6-dimensional space looks topologically like ${\bf R}^3$ fibered over
a finite-radius $S^3$, which is a nonsingular fibration.  This
is illustrated in \reffig{defcon}, to be compared with \reffig{T11}.

Now, what is the physics that has caused this to happen?  Does it have
a simple description?  If $Z_{11}\sim Z_{22}\sim \sqrt{\det Z}\leq
\Lambda$, then it doesn't; the theory is more or less perturbative
down to the scale of order $\Lambda$, and below that scale the physics
is strongly coupled.  However, we can study the theory in a different
regime, where it is more tractable.  With a fixed $\det Z$, we can
take, for instance, $Z_{11}$ large, while sending $Z_{22}$ very small.
In this limit, the gauge group is broken at high energy, by one unit,
to $SU(M)$.  The superpotential, together with the expectation value
for $A_1$ and $B_1$, gives $A_2$ and $B_2$ a mass $h\vev{A_1B_1}$.
Below this scale the theory is a pure $SU(M)$ \none\ Yang-Mills
theory, with a scale $\Lambda_0$ satisfying $\Lambda_0^{3M} =
h\Lambda^{3M+1}$.  (Note this scale, and much of the ensuing low-energy
physics, is independent of $Z_{11}/Z_{22}$.  This is a common feature
of the holomorphic subsector of
supersymmetric theories; see Argyres et al. \mycite{argyresetal})  
And so gluino
condensation occurs, generating the low-energy superpotential,
confinement, breaking of the $\ZZ_{2M}$ symmetry, and all of the other
details we expect.  The only thing new is that we can do
$SU(2)_L\times SU(2)_R$ rotations on each of these vacua, generating a
larger moduli space for each choice of gluino condensate; this is, in
the end, the only effect of the probe brane on the $M$ fractional
branes.

The implications for supergravity are clear.  The fractional branes
must somehow turn the conifold into the deformed conifold.  The probe
brane will have little effect on the whole structure (which justifies
thinking of it as a probe) and in supergravity it is likely to play
almost no role (except for the nontrivial fact
that the matter it adds can break the confining
flux tubes of the theory.)

\subsection{More Integer D-branes: $SU(N+M)\times SU(N)$}

We could also consider the effect of two D3-branes.  This is
covered partially in the appendix of Klebanov and Strassler \mycite{ikms},
and we do
not need it here; the results are similar to the case just discussed.
More interesting is when the number of D3-branes is of order $M$, or even
greater.  

Let us see what we can say classically and semiclassically about the
theory with $N$ D3-branes and $M$ fractional D3-branes.  The gauge
group is $SU(N+M)\times SU(N)$, and the matter and superpotential are
as usual for the conifold.  The $SU(2)\times SU(2)$ and baryon
symmetries are as usual.  Again the $U(1)_{\mathcal R}$ is broken by
anomalies.  We must assign $R_A=R_B=\half$ to be consistent with the
superpotential (although this does not imply $\dim A=\dim B={3\over 4}$,
since we are not necessarily going to reach a conformal fixed point!)
Let us consider an instanton in the $SU(N+M)$ gauge fields.  Such an
instanton has $2(N+M)$ gluino zero modes. Meanwhile, a fermion in the
fundamental representation of $SU(N+M)$ provides one zero mode; there
are $2N$ such fermions, $N$ each from $A_1$ and $A_2$.  Similarly,
there are $2N$ zero modes from $B_1$ and $B_2$.  The $A$ and $B$
fermions have R-charge $-1/2$, and the gluinos have R-charge 1, so
under a $U(1)_{\mathcal R}$ rotation by a phase $e^{i\alpha}$ the
instanton rotates by a phase $\exp\{i[2(N+M) - 2N]\alpha\}=
e^{i(2M)\alpha}$.  To avoid an anomaly, then, we must have $\alpha =
\pi k/M$, $k =0,1\dots, 2M-1$ --- again a $\ZZ_{2M}$ symmetry.  

What happens if we apply this $\ZZ_{2M}$ symmetry to an instanton of
the $SU(N)$ gauge group?  We simply switch $N$ and $N+M$ in the above
calculation, and find the instanton rotates by $e^{-i(2M)\alpha} =1$.
So the  same $\ZZ_{2M}$ symmetry is anomaly-free under {\it both}
groups.  A remarkable result --- the anomaly-free discrete R-symmetry
of the $SU(N+M)\times SU(N)$ theory is a $\ZZ_{2M}$,
independent of $N$!

Now, what about the RG flow of the theory?  The one-loop beta
functions for the theory are given by $b_0^{(N+M)} = 3 (N+M) - 2N =
N+3M$ and $b_0^{(N)}= 3N-2(N+M) = N-2M$; thus it appears both gauge
groups are asymptotically free, but the $SU(N+M)$ group, which has
less matter, flows faster.  This conclusion is partially correct,
but the reasoning is basically wrong.  We need to be more careful.

\subsection{The Lowest Step of the Cascade}
\label{subsec:onestep}

Let's study in detail the case of $N=M$, that is, $SU(2M)\times
SU(M)$, with $A_r$ in the $({\bf 2M},{\bf \overline M})$
representation and $B_u$ in the conjugate representation.  The
superpotential is the usual one, \Eref{conifsup}.  In this case
$b_0^{(2M)} = 4M$ while $b_0^{(M)} = -M$, so the second group is (even
naively) infrared free.  The low-energy dynamics is therefore
dominated by $SU(2M)$, with $SU(M)$ acting as an unimportant
``spectator.''  The $SU(2M)$ group has $2M$ flavors ($M$ from $A_1$
and $B_1$, and $M$ from $A_2$ and $B_2$).  Such a theory is an 
example of SQCD with an equal number of flavors and colors.  It is
well-described by its mesons $A_rB_u$ and by its baryons ${\mathcal B}
= [A]^{2M}$ and $\overline{{\mathcal B}}= [B]^{2M}$, which we
recognize as related to (but not equal to) the dibaryons of
$SU(N)\times SU(N)$ discussed earlier.  But we must take our time
here.  The mesons $(Z_{ru})^a_b =(A_r)^a_\alpha(B_u)^\alpha_b$ (which
are in the $({\bf 2},{\bf 2})$ representation of $SU(2)_L\times
SU(2)_R$) have their $SU(2M)$ indices contracted, but not their
$SU(M)$ indices.  To form irreducible representations of $SU(M)$, we
must define
$$
Z^0_{ru} = (Z_{ru})^a_a
\ , \ (Z^{adj}_{ru})^a_b = (Z_{ru})^a_b-{1\over M} Z^0_{ru}\delta^a_b
$$
which are in the singlet and in the adjoint representation,
respectively, of $SU(M)$ .  Meanwhile, the baryon ${\mathcal B}$,
an $SU(2M)$ gauge invariant operator, is also (by Bose-statistics) both 
an $SU(M)$ and an $SU(2)_L$ singlet.
The same is true for $\overline{{\mathcal B}}$.

SQCD with $2M$ flavors and $2M$ colors has a ``deformed moduli
space,'' as shown by Seiberg \mycite{powerholo}.  
This can be implemented using a
Lagrange multiplier superfield $X$ in the effective superpotential,
which takes the form
$$
W = h\ \tr_{a,b}[Z_{11}Z_{22}-Z_{12}Z_{21}] 
- X(\det_{r,u,a,b} Z - 
{\mathcal B}\overline{{\mathcal B}} - \Lambda_{2M}^{4M})\ .
$$
Notice the determinant over $r,u,a,b$ treats $Z$ as a $2M\times 2M$ matrix,
not as a $2\times 2$ matrix.  There are multiple solutions to this
equation, and I don't want to go exploring them here.
But the simplest solution is
$${\mathcal B} = \overline{{\mathcal B}}=i\Lambda^{2M}_{2M} \ ,
 \ Z^{adj}=0\ , \  Z_{ru}^0 = 0 \ .
$$
Since ${\mathcal B}, \overline{\mathcal B}$ are singlets of $SU(M)$,
they leave the $SU(M)$ gauge group unbroken.  Meanwhile, the
$Z$ fields are massive.  The
theory at energies below their
masses consists simply of an unbroken
$SU(M)$ gauge group with no massless matter: 
a pure \none\ Yang-Mills theory, as illustrated in \reffig{laststep}.  
Again we
have confinement, chiral symmetry breaking, and all of the other
phenomena we have discussed up to now.

\begin{figure}[th]
\begin{center} 
\centerline{\psfig{figure=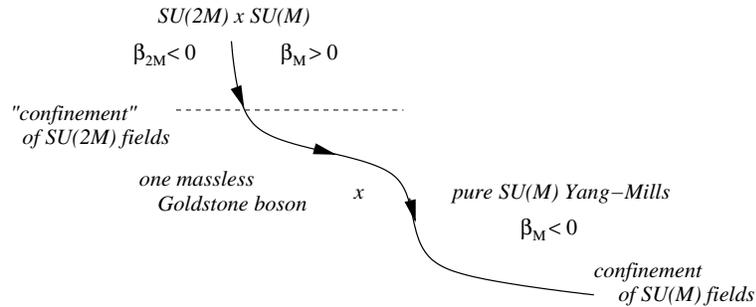,width=10cm,clip=}} 
\caption{\small At one scale, $SU(2M)$ confines, leaving behind a pure
$SU(M)$ \none\ Yang-Mills theory plus a single massless goldstone
supermultiplet (from the spontaneously broken baryon number).  The
goldstone supermultiplet couples only through irrelevant operators to the
$SU(M)$ sector, and does not affect its infrared dynamics.}
\fig{laststep}
\end{center}
\end{figure}

This is almost true.  The symmetry $U(1)_B$ is broken by the
expectation values for ${\mathcal B}$ and $\overline{\mathcal B}$.
(Notice that no other symmetries are broken; the $\ZZ_{2M}$ R-symmetry
and the $SU(2)_L\times SU(2)_R$ global symmetries are retained.)
Consequently, there must also be a massless Goldstone boson which
rotates the phases of ${\mathcal B}$ and $\overline{\mathcal B}$ in
opposite directions.  (Here it is crucial that $U(1)_B$ is not gauged
in the quantum theory; the subtle issues involved were partly
explained by Aharony \mycite{oferbaryon} and finally clarified by
Gubser, Herzog and Klebanov \mycite{barygold}.)  Thus there isn't
actually a mass gap in this theory, though there is still a discrete
spectrum.  The Goldstone mode, its scalar superpartner (which changes
the ratio $|{\mathcal B}/\overline{\mathcal B}|$) and a partner
fermion form a massless chiral multiplet.  It is important to note,
however, that the extra massless chiral multiplet has very minor
effects upon the interesting dynamics of the confining $SU(M)$ gauge
theory.  Because it is a composite of the $SU(2M)$ dynamics, contains
$2M\gg 1$ constituents, is neutral under $SU(M)$, and couples to the
$SU(M)$ sector only via irrelevant operators, it does not participate
in the $SU(M)$ dynamics.  Even with this massless multiplet present, the
$SU(M)$ sector confines in the usual way, and generates a mass gap as
usual in its own sector.

In summary, when the $SU(2M)$ sector confines, (1) it makes neutral
scalar baryons, whose expectation values break baryon number and give
rise to a single massless composite chiral multiplet, and also (2) it
makes scalar mesons charged under $SU(M)$, but these are massive
because of the superpotential, leaving a pure Yang-Mills theory in the
$SU(M)$ sector.  At the scale of the confining $SU(M)$ dynamics, and
at small 't Hooft coupling, the $SU(M)$ sector and the sector
containing the massless chiral multiplet are exponentially-weakly
coupled.  This is because all couplings of the Goldstone mode are
suppressed by $\Lambda_{2M}$, which is exponentially larger than
$\Lambda_M$.  At strong 't Hooft coupling, the situation is more
subtle, and we will return to it only once our understanding of the
cascade is complete.  Still, we will see that even then the presence
of the Goldstone supermultiplet has limited effects on the dynamics, at
least when $|{\mathcal B}|=|\overline{\mathcal
B}|$.

\subsection{Another Step}

Let's try one more example before climbing higher on the cascade.
What if $N=2M$?  Then we have $SU(3M)\times SU(2M)$ to start with.
The $SU(2M)$ theory has $b_0=0$ and is again infrared-free.  Let's
initially ignore it, as we did for $N=M$ (this will be justified more
fully later.)  Then the $SU(3M)$ theory becomes strongly coupled at
some scale $\Lambda_{3M}$, where its coupling $g_{3}$ becomes
large, and we must ask what it does there.  In
this regime the theory is effectively SQCD with $3M$ colors, with
$2(2M)=4M$ flavors, plus the quartic superpotential $W=h\
\tr\det[A_rB_u]$.  This theory is in the free-magnetic phase: below
$\Lambda_{3M}$ 
(momentarily ignoring the spectator $SU(2M)$
group and the small superpotential) the $SU(3M)$
sector is better described as $SU(M)$ SQCD+M with $4M$ flavors $a^u,
b^r$, which is infrared free, along with $SU(M)$-gauge-singlet
operators $(Y_{ru})^\alpha_\beta = A_u^\alpha B_{r\beta}$, where
$\alpha,\beta$ are indices in the $SU(2M)$ spectator group.  The
superpotential is
$$
W = \hat h\ \tr_{\alpha}\ \det_{r,u}\ Y + y\ Y_{ru}a^rb^u
$$
($SU(2M)$ and $SU(M)$ 
indices contracted implicitly.)  Integrating out $Y$ we obtain
$$
W = \tilde h\ \tr\ \det_{r,u}\ (a^rb^u)
$$
where $\tilde h \sim -{y^2/ 4\hat h}$. Indeed this is just the
sort of duality transformation with quartic
operators that we studied in Sec.~\ref{subsec:quarticops};
see Eqs.~\eref{dualtoqqqq}--\eref{qqqqsupot}.

\EX{Verify the previous paragraph!  You'll need to fully grasp
it in order to understand the self-similarity of the duality cascade.}

Now we recall that the coupling of the $SU(2M)$ group is not quite
zero, even if it is small.  Below $\Lambda_{3M}$, where we usefully
change variables from an $SU(3M)$ description to an $SU(M)$
description, we now have an $SU(2M)\times SU(M)$ theory of exactly the
form that we discussed  in Sec.~\ref{subsec:onestep}: one with $M$, not
$2M$, integer D3-branes (in addition to $M$ fractional branes) on the
conifold.  We already know what happens in this theory.  The $SU(2M)$
group now has a {\it negative} beta function ($b_0=+4M$).  Although it
remained weakly-coupled above the $SU(3M)$ strong-coupling transition,
it now flows to strong coupling.  When its coupling $g_{2}$ becomes
large, we again change variables to a hadronic description using the
mesons and baryons constructed from $SU(2M)$ gauge-invariants.  This
leaves only a massless Goldstone mode
and an $SU(M)$ gauge group with pure \none\ Yang-Mills, which
confines, breaks chiral symmetries, etc., etc.

Thus, as in \reffig{twostep}, 
we now have three separate strong-coupling transitions chained
together, the first of which is best described
by making a Seiberg duality transformation (in fact all
three transitions can be viewed this way) and the first hints of
what a duality cascade is going to be. What do these dualities imply?
The three theories that we have discussed, with $N=2M$, $N=M$, and
$N=0$ are generally different theories, whose extreme infrared physics
is the same.  But to say only this underestimates Seiberg's duality.
As can be seen from \reffig{twostep}, the physics of the $N=2M$
and $N=M$ theories become equal far above the extreme infrared.
{\it Below the scale where their two flows merge, exact Seiberg duality is
in operation.}  The $N=2M$ and $N=M$ theories give two equivalent
descriptions of the RG flow all the way from the confining scale of
$SU(2M)$ down to zero momentum.  We will see that this generalizes to
larger $N$.

\begin{figure}[th]
\begin{center} 
\centerline{\psfig{figure=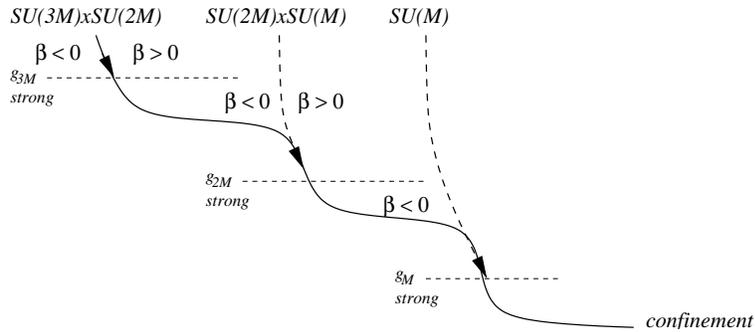,width=10cm,clip=}} 
\caption{\small Three strong-coupling transitions; the bottom
steps of the duality cascade. }
\fig{twostep}
\end{center}
\end{figure}

\setcounter{equation}{0}
\section{The Cascade: Descending the Grand Staircase}

Finally, it is time to move beyond individual cases to study the
full duality cascade.  We will first investigate the very high steps in
the cascade, then understand their supergravity dual description.
Next we will make some observations concerning
the supergravity dual of the bottom steps.  We will
conclude these lectures by considering applications to QCD and to
beyond-the-standard-model physics.

\subsection{The Upper Reaches of the Cascade}
\label{subsec:upreach}

Having investigated
$k=$1, 2 and 3,
our next task should be to study $(k-1)M$ integer D3-branes
on the conifold, with $k\gg 1$ an integer.  The supergravity description
was gradually
elucidated in papers of Klebanov with Nekrasov \mycite{ikNekr},
and then with Tseytlin \mycite{ikat},
and finally with yours truly \mycite{ikms}, where we also understood the field
theory.  
We will investigate the
$SU(kM)\times SU([k-1]M)$ theory by using methods similar to those
above, but paying special attention to an expansion in $1/k$.  The
reason for this should be clear: as $k\to \infty$, the theory comes
closer and closer to the Klebanov-Witten model, and our understanding
of the latter should prove conceptually and technically useful.

Naively speaking, the $SU(kM)$ and $SU([k-1]M)$ theories, which have
$b_0 = 3kM-2[k-1]M = (k+2)M$ and $3[k-1]M-2kM = (k-3)M$ respectively,
are both asymptotically free, although the $SU(kM)$ theory is running
faster toward strong coupling.  If the $SU(kM)$ group should be
replaced by its Seiberg-dual description at some strong-coupling
transition, then since $2[k-1]M-kM = [k-2]M$, we will find ourselves
(through the same arguments as above) with an $SU([k-1]M)\times
SU([k-2]M)$ gauge theory of exactly the same form.  Again both gauge
groups are (naively) asymptotically free; but the $SU([k-1]M)$ beta
function has increased ($b_0 = (k+1)M$ now) so it now flows faster to
stronger coupling, and perhaps we should dualize it, obtaining
$SU([k-2]M)\times SU([k-3]M)$, again of the same form.  And so we can
imagine cascading downward through $k$ separate strong-coupling
transitions, at each of which it is wise and convenient to replace the
more-strongly-coupled gauge group with its dual description, and after
each of which the theory looks the same as before except with $M$
fewer colors in both groups.  This continues until we reach pure
$SU(M)$ Yang-Mills theory, whereupon we get confinement, chiral
symmetry breaking, etc.  This is illustrated in \reffig{kstep}.

\begin{figure}[th]
\begin{center} 
\centerline{\psfig{figure=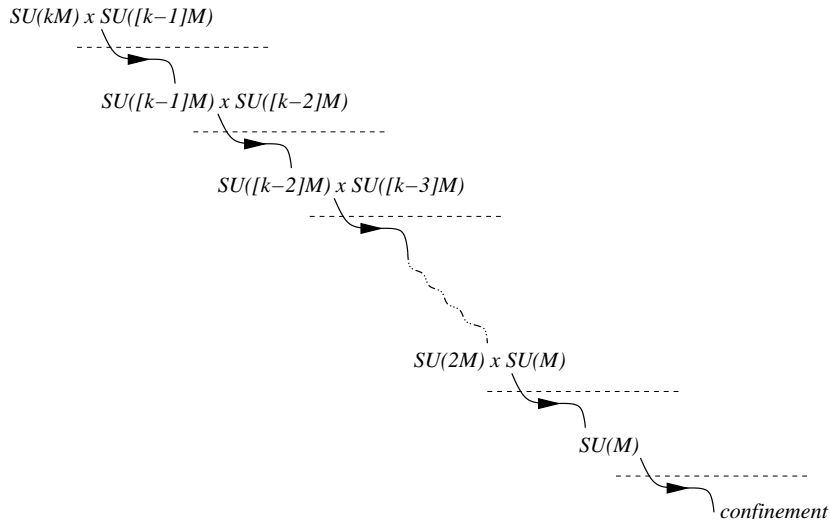,width=11cm,clip=}} 
\caption{\small The duality cascade; a naive view, to be corrected
later.  }
\fig{kstep}
\end{center}
\end{figure}

But while this is correct in its broadest outlines, almost all of the
real physics is wrong.  Our goal here is to say precisely what is
happening, and in doing so bring together all of the field-theory
techniques that we have gradually been assembling.

To build up an understanding of a field theory of this type takes some
time, and we'll start slow.  Let's first imagine that we have, in the
far ultraviolet, a weakly-coupled theory of $SU(kM)\times SU([k-1]M)$,
with gauge couplings $g_k,g_{k-1}$, the usual $A$ and $B$ fields, and
the conifold superpotential, with coupling $h$ and dimensionless
coupling $\eta=h\mu$.  Since all fields have the same anomalous
dimension $\gamma_0$, we have
\begin{eqnarray}\label{ksbetas}
\beta_{g_k} &=& -F(g_k)
[(k+2)M+2[k-1]M\gamma_0]\ \nonumber \\
 & & \Rightarrow \ 
\beta_{\lambda_k}
= - G(\lambda_k)\left[\left(1+{2\over k}\right)
+2\left(1-{1\over k}\right)\gamma_0\right]
\ , \nonumber \\ \nonumber \\
\beta_{g_{k-1}} &=& -F(g_{k-1})
[(k-3)M+2kM\gamma_0]\ \nonumber \\ 
& &\Rightarrow \
\beta_{\lambda_{k-1}}
= - G(\lambda_{k-1})\left[\left(1-{2\over k-1}\right)
+2\left(1+{1\over k-1}\right)\gamma_0\right]
 , \nonumber \\ \nonumber \\
\beta_\eta &=& \eta(1+2\gamma_0) \ ,
\end{eqnarray}
where 
$$
F(g_p)={g_p^3\over 16\pi^2}{1\over 1-{[g_p^2(pM)/ 8\pi^2]}} \ ,
\
G(\lambda_p)={\lambda_p^2\over 8\pi^2}{1\over 1-{[\lambda_p/ 8\pi^2]}} \ .
$$ 
and $\lambda_p = g_p^2 pM$.
When the couplings are small and $|\gamma_0|\ll 1$, we see that
both gauge couplings are asymptotically free, while $\beta_\eta>0$.
However, the signs of the beta functions are all reversed if
$\gamma_0\ll-1/2$.  This is clear from the fact that, at large $k$,
all three beta functions become proportional to $1+2\gamma_0$.  In
fact, comparison with \Eref{kwbetas} shows that at large $k$ the beta
functions of this theory become equal, up to $1/k$ corrections, to
those of the Klebanov-Witten model.  

Let us recall the key features of the Klebanov-Witten theory: (1) it
has a continuous two-complex-dimensional manifold of conformal fixed
points, shown in \reffig{KWCFTs2}, defined by the condition
$\gamma_0=-1/2$ which causes all three beta functions \eref{kwbetas}
of the Klebanov-Witten model to simultaneously vanish; (2) the
manifold is infrared-stable, in the sense that all renormalization
group flows end, in the infrared, on the manifold, in analogy to a
ball rolling with friction into a flat valley bottom; and (3) duality
symmetries act on this manifold, including a blend of 
electric-magnetic dualities and Seiberg dualities.

However, the $1/k$ differences between \eref{ksbetas} and
\eref{kwbetas} are critically important.  It is easily seen in
\eref{ksbetas} that for finite $k$ there is no value of $\gamma_0$ for
which even two, much less three, of the beta functions simultaneously
vanish.  Thus, unlike the Klebanov-Witten model, this theory does not
have continuous manifolds of conformal fixed points.  Still, the
theory with fractional D-branes inherits modified versions of each of
the three key features of the Klebanov-Witten theory: (1) it has a
region, a thin ``slab'' inside the three-dimensional space of
couplings, defined by the condition $\gamma_0\approx -1/2$, where all
three beta functions \eref{ksbetas} are very small; (2) this slab is
infrared-stable, in the same sense as in the Klebanov-Witten model, in
that all renormalization group flows reach this slab in the infrared;
and (3) although $SL(2,\ZZ)$ duality is lost (since fractional
D3-branes are not invariant under $SL(2,\ZZ)$ transformations),
Seiberg duality continues to act on this slab, as we will soon see.

Since \eref{ksbetas} and \eref{kwbetas} are
equal as $k\to\infty$, it is not surprising
that, for large $k$ and $\gamma_0\approx-\half$,
the beta functions are all of order $1/k$ compared
to their ``natural size.''  To be more precise,
let us write $\gamma_0=-\half + \delta_0$; then
$$
\beta_{g_k} = -F(g_k)M
[3+2\left(k-1\right)\delta_0]
\ , \
$$
$$
\beta_{g_{k-1}} = -F(g_{k-1})M
[-3+2k\delta_0]\ , \
$$ 
\bel{ksbetasB} \beta_\eta = 2\eta\delta_0 \ .  
\ee 
For
$|\delta_0|\sim\half$, as will be the case for a generic point in the
space of the coupling constants $(g_1,g_2,\eta)$, the beta functions
for the gauge couplings are of order $g^3kM$, while
$\beta_\eta\sim\eta$.  The resulting flow, as in the Klebanov-Witten
model, pushes the theory toward the region where
$\delta_0(g_1,g_2,\eta)$ is small.  Once within the narrow slab of the
coupling-constant space where $|\delta_0(g_1,g_2,\eta)|\sim 1/k$, the
couplings all slow to a crawl: the two gauge couplings have beta
functions of order $g^3M$, and $\beta_\eta\sim \eta/k$, each a factor
of $1/k$ smaller in magnitude than in the rest of the coupling space.
Thus the region $|\delta_0|\sim 1/k$ is an infrared-stable slow-motion
slab; once the couplings reach it, they remain within it, their values
drifting slowly.  Clearly, by the fact that this theory matches the
Klebanov-Witten theory in the large-$k$ limit, the location of this
slab is very nearly the location of the manifold of Klebanov-Witten
fixed points.

Let's return to our metaphor of the ball rolling in the
potential. In the Klebanov-Witten model the
renormalization group can be thought of as a ball rolling (with
strong friction)
in a landscape which has steep walls
surrounding a deep, perfectly flat valley, representing the manifold
of conformal fixed points.  The ball rolls quickly into the valley,
and stops.  Once we add fractional branes, the only change is that the
valley, instead of being perfectly flat, acquires a small tilt of
order $1/k$.  The
ball rolls quickly into the valley, comes nearly to a stop, then
gradually drifts down the valley toward the regions of lower
elevation and smaller $k$.

Our task, now, is to understand what happens during this drift.  What
does the theory do as its coupling constants slowly change?  We will
see that the answer is intricate and remarkable.

\subsection{Near the Boundary of the Valley}

Although there is no manifold of fixed 
points for this theory, there are {\it isolated} fixed
points when two of the three couplings are {\it zero}.  For instance, if
$g_k=h=0$, then the theory is simply SQCD with $[k-1]M$ colors and
$2kM$ flavors.  This is in the conformal window (for $k>3$) and has a
Seiberg fixed point at which the fields $A_r, B_u$ have dimension
$$
\dim A_r = \dim B_r = {3\over 2}\left(1 - {[k-1]M\over 2kM}\right) 
= {3\over 4}\left({k+1\over k}\right) 
$$
or equivalently 
$$
\delta_0 = {3\over 2k}  \ .
$$
Then, from \eref{ksbetasB}, when $0<g_k,\eta\ll 1$ we have
\bel{firstCFT}
\beta_{g_k} = -F(g_k)[6M] < 0\ , \
\beta_{g_{k-1}} = 0
\ , \
\beta_\eta = {3\eta\over k} >0 \ 
\ee
to leading order in $1/k$.

If instead $g_{k-1}=h=0$, then the theory
is $SU(kM)$ SQCD with $2(k-1)M$ flavors; this is in the conformal
window and 
its fixed point has
$$
\dim A_r = \dim B_r = {3\over 2}\left(1 - {kM\over 2(k-1)M}\right) 
= {3\over 4}\left({k-2\over k-1}\right) 
$$
or equivalently 
$$
\delta_0 = -{3\over 2(k-1)}\approx -{3\over 2k} \ .
$$ 
Then we find, at this fixed point,
\bel{secondCFT}
\beta_{g_k} = 0
\ , \
\beta_{g_{k-1}} \approx -F(g_{k-1})[-6M] >0
\ , \
\beta_\eta \approx -{3\eta\over k} <0\ .
\ee
So the signs are all reversed from the previous case.

\EX{Determine the conditions for a fixed point and verify these statements.}

Let's now encode this information on a graph of the space of
couplings, shown in \reffig{KSbdy1}.  Suppose $g_k$ is zero and we are
at the $SU(k[M-1])$ Seiberg fixed point, the upper dot on the graph.
From \Eref{firstCFT}, if $\eta$ is nonzero, then $\beta_\eta>0$; the
theory flows back to the fixed point.  But if $g_k$ is nonzero (and
$\eta=0$) then the theory flows along the edge of the valley to the
$SU(kM)$ fixed point, where $g_{k-1}=0$, the lower dot on the graph.
Here \Eref{secondCFT} applies; the fact that $\eta$ is relevant at
this fixed point is familiar, as it corresponds to $SU(kM)$ SQCD with
$N_f=2(k-1)M<2kM$, a case we studied in Sec.~\ref{subsec:quarticops}.

\begin{figure}[th]
\begin{center} 
\centerline{\psfig{figure=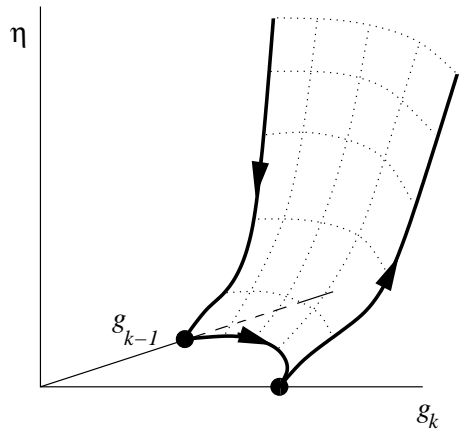,width=6cm,clip=}} 
\caption{\small The drift along the boundary of the valley for
nonzero $M$ and large $k$;
compare with the space of conformal field theories
in \reffig{KWCFTs2}. Each dot represents a Seiberg SQCD fixed point. }
\fig{KSbdy1}
\end{center}
\end{figure}

What happens once $\eta$ is nonzero?  As we saw in 
Sec.~\ref{subsec:exactdual} and is shown in \reffig{CCprime}, the
flow from small to large $\eta$ has an exactly-dual description in which we
replace $SU(kM)$ with $SU([k-2]M)$ and $\eta$ with $\tilde\eta\sim
1/\eta$.
The endpoint of
this flow is the Seiberg fixed point of $SU([k-2]M)$ SQCD with
$2[k-1]M$ flavors.
But at this fixed point $g_{k-1}$ is relevant, so
if it is nonzero we obtain a flow to the fixed point of $SU([k-1]M)$
with $2(k-2)M$ flavors, as shown in \reffig{KSbdy2}.
We can see that the process will repeat until $k\sim 1$.

\begin{figure}[th]
\begin{center} 
\centerline{\psfig{figure=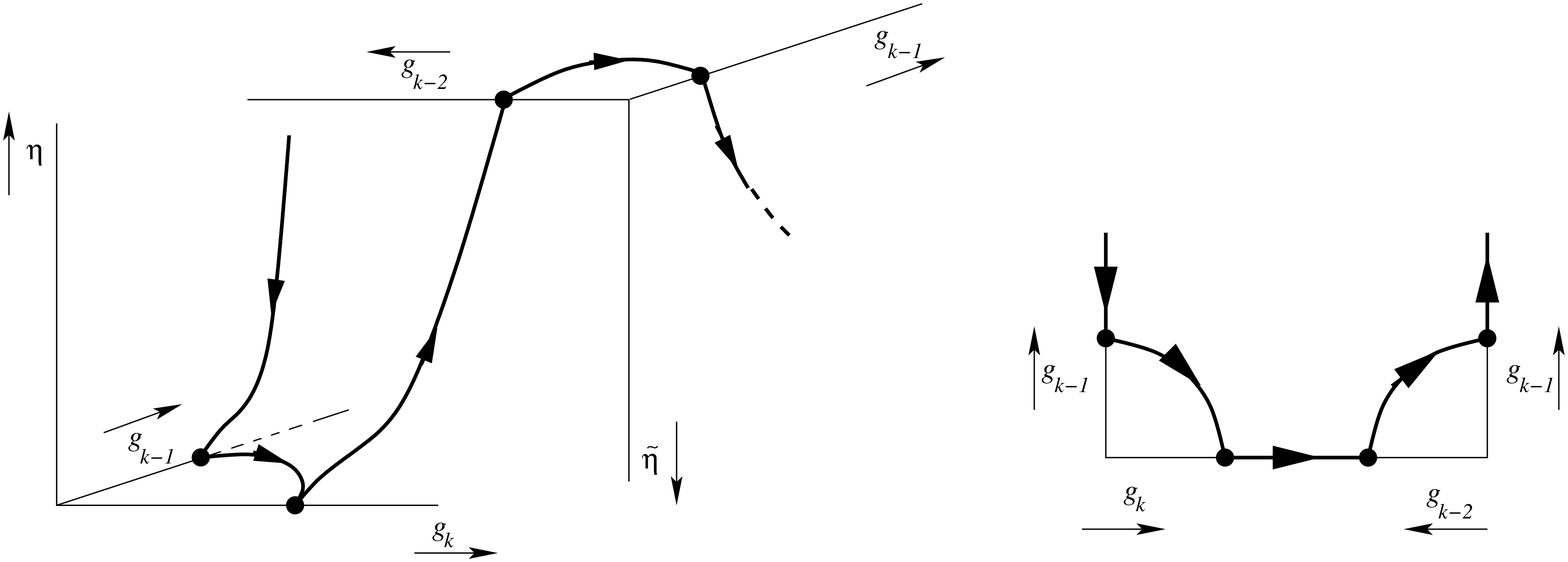,width=11cm,clip=}} 
\caption{\small On the left, the flow along the boundary of
the valley, extending \reffig{KSbdy1} through the region
appropriate for a Seiberg-duality
transformation, as in  \reffig{CCprime}.  
Each dot represents a Seiberg SQCD fixed point.
Initially, the set
of coordinates $g_{k}, g_{k-1},\eta$ is appropriate; after the
transformation, new coordinates $g_{k-2}, g_{k-1},\tilde \eta$
should be used.  On the right, the same image is shown looking
down from above, as in \reffig{tauflows},
so that the $\eta$ (and $\tilde\eta$) coordinate
is suppressed, and only the gauge couplings are retained. }
\fig{KSbdy2}
\end{center}
\end{figure}

The curves we've just discussed simply form the boundary of the
valley.  
In a real flow, the gauge couplings and $\eta$ are never
strictly zero, so the flow takes place within the valley, but away
from its boundary, in a fashion better illustrated by the two curves
added to \reffig{KSbdy3}.  One of these curves (the dashed line) 
lies very close to the
boundary, and passes very close to the Seiberg fixed points.  When it
reaches the vicinity of such a fixed point --- $g_{k-1}\sim g_{k-1*}$ and
$\eta,g_{k}\ll 1$ --- the motion along the curve might slow to a
near-stop. To find out, let's check the beta functions.  Near the
fixed point, $\beta_{g_{k-1}}$ is of order $g_{k-1}-g_{k-1*}$ (since it must
vanish at $g_{k-1*}$) but it also gets corrections from $\eta$ and
$g_{k}$; by simple one-loop perturbation theory estimates, these
will be of order $\eta^2/8\pi^2$ and $g_{k}^2 kM/8\pi^2 =
\lambda_{k}/8\pi^2$.  Meanwhile, from \Eref{firstCFT},
the beta function for $\eta$ is of
order $\eta/k$, while that for $g_{k}$ is of order $-6M F(g_{k-1})$;
both of these are $1/k$ smaller than their typical size.  For the
couplings $g_{k}$ and $\eta$ to remain small, and for $g_{k-1}$ to
remain near its fixed value, for an extended range of energy, it must
be that $\lambda_{k} = g_{k}^2 kM$ --- the 't Hooft coupling
for the weakly-coupled $SU(kM)$ gauge group --- is much less than
1 in this regime.  Also, $\eta$ must be small.

\begin{figure}[th]
\begin{center} 
\centerline{\psfig{figure=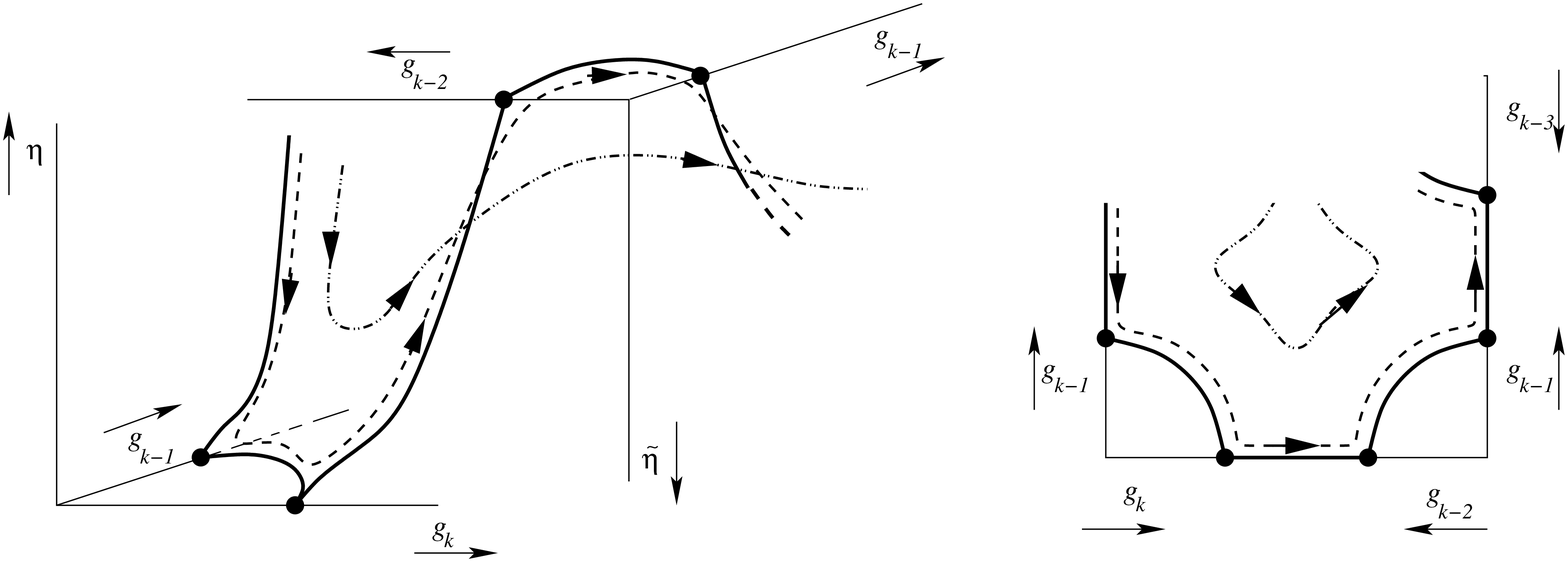,width=11cm,clip=}} 
\caption{\small Two flows inside the valley, close to (dashed) and far
from (dot-dashed) the boundary (solid) of
the valley.  The notation of the two figures is the same as for
\reffig{KSbdy2}.}
\fig{KSbdy3}
\end{center}
\end{figure}

So, if the curve corresponding to a particular flow dips down to a
regime in which the 't Hooft coupling
$\lambda_{k}=g^2_{k}kM\ll 1$ and $\eta\ll 1$, then there
really is a range of energy scales in which the theory can be
described as a conformal Seiberg fixed point of $SU([k-1]M)$ SQCD along
with a weakly-coupled $SU(kM)$ gauge group and a small quartic
coupling $\eta$.  Conformal field theory for the $SU(kM)$ sector
combined with conformal perturbation theory in $\lambda_{k-1}$ and
$\eta$ would be an effective method for doing calculations in this
range.

Eventually, $g_k$ will increase to large values, with $g_{k-1}$
shrinking to zero; more precisely, $\lambda_{k-1}$
will become much less than 1 as the flow approaches
the Seiberg fixed point at $g_k\to g_{k*}$.
Near  this fixed point, $\beta_\eta<0$, so after a long range
of energy where the beta functions \eref{secondCFT} are all very
small, 
$\eta$ will begin to grow
and will increase to large values.  

But then, as we
can see from the curve, we will find yet another range of slow motion
in which the theory can be described as approximately conformal
$SU([k-2]M)$ with weakly coupled $SU([k-1]M)$, two bifundamental
fields, and small $\tilde\eta$.  Logarithmic running of
$\lambda_{k-1}$ would eventually destabilize this quasi-fixed point,
leading us into yet another quasi-fixed point, that of $SU([k-1]M)$
with weakly coupled $SU([k-2]M)$ and small $\tilde\eta$.  And this
repeats, over and over.  For this flow, there is really a
``cascade,'' with a sequence of steps.  The ball rolling down
the valley nearly stops, then rolls to a new location and nearly
stops, then rolls again to a new location, {\it etc.}  The cascade
from one fixed point to the next is represented by the curve in
\reffig{cascade}, and by the dashed curve in \reffig{KSbdy3} and
\reffig{cascade2}.

\begin{figure}[th]
\begin{center} 
\centerline{\psfig{figure=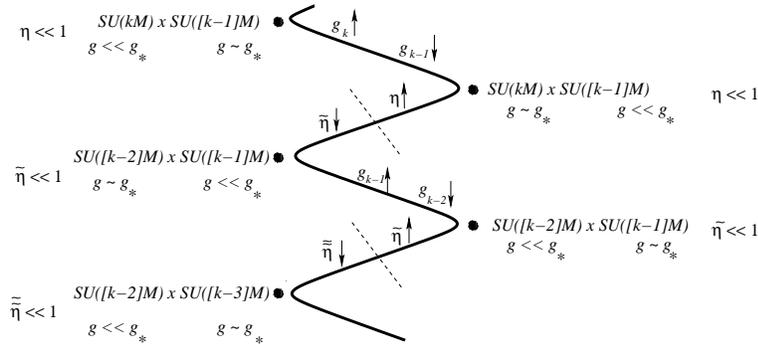,width=10cm,clip=}} 
\caption{\small Near the boundary of the valley, the
theory cascades
from one Seiberg fixed point to another.  First a flow
occurs in which one gauge coupling shrinks from its
Seiberg-fixed-point value to zero, while
the other grows from zero to its Seiberg-fixed-point value.
Next, $\eta$ grows from zero to infinity, necessitating a
change of variables using an exact
Seiberg-duality transformation; this is shown by a dashed line.
From the dual point of view, $\tilde \eta$ shrinks to zero.
The process then repeats.}
\fig{cascade}
\end{center}
\end{figure}

Does Seiberg duality apply exactly or approximately on this curve?
The answer is that it applies exactly. To see this, we must bring
two of our earlier arguments together.
(1) If Seiberg duality
is exact in the ultraviolet, it remains exact forever into the infrared.  
This is clear from \reffig{Seidual3}.  Once the two flows shown
in \reffig{Seidual3} flow together, or more precisely, are brought together
by taking a scaling limit, they simply aren't two flows anymore: they
become a single flow with two descriptions.  A single flow cannot bifurcate
in the infrared; that would make the renormalization group indeterminate.
(2) The Klebanov-Witten model has exact Seiberg duality acting on
its manifold of fixed points.  The theory with fractional
branes differs
from the Klebanov-Witten model only by $1/k$ corrections, and 
the duality cascade lies
within the valley where exact Seiberg duality applies in the
Klebanov-Witten model.  Therefore Seiberg duality must apply up to at
most $1/k$ corrections.  But in the ultraviolet, these disappear.

So Seiberg duality applies exactly in the ultraviolet, and therefore,
from our previous point, it applies at all scales.  Properly defined,
the duality cascade is a single flow, with multiple (indeed, an infinite
number of) descriptions, each of which has a limited range of usefulness.

If you don't like this argument (after all, it does give us an
infinite number of colors in the ultraviolet) I'm happy to regularize
it.  Put an ultraviolet cutoff at $k=k_0$ large but finite.  Seiberg
duality applies approximately in the ultraviolet, with $1/k_0$
corrections. But as is clear from the central
diagram of \reffig{Seidual3}, the duality applies better and better as
the flow approaches the infrared; the two flows shown there become
closer and closer to identical as the infrared is approached.  This is
because their difference is described by an irrelevant operator, whose
effect decreases to zero in the infrared.  So I can make the two flows
arbitrarily close to one flow by 
only asking questions in the physical regime where
$k\ll k_0$.  Then the two flows, and their two descriptions,
can be made identical, for fixed $k$, by taking
$k_0\to\infty$.  Again, this is because the difference between the
two flows is an irrelevant operator, whose effect must vanish at any
fixed energy scale as the cutoff at $k_0$ is removed.

In short, the duality cascade is about a single flow, with an infinite
number of descriptions, one for every $k$, obtained
from one another by application of an arbitrary
number of exact Seiberg duality transformations.
The flow given by the dashed curve in \reffig{KSbdy3} and
\reffig{cascade2} moves from one approximate fixed point to the
next, each with its own useful variables.  
We might wish to think about this flow as a ``cascade of
dualities,'' but in retrospect, this a rather poor choice of
terminology.  The dualities are not physical, while the cascade {\it
is} physical.  The cascade of the theory from one fixed point to
another involves a journey from a region where {\it one} description
of the theory is useful to a region where {\it another} is useful.
Each description is valid at all scales, but is {\it useful} only in
the region where the theory flows close to the $SU(kM)\times
SU[(k\pm1)M]$ Seiberg fixed points.\footnote{Strictly speaking, it is
possible that the $k$th description breaks down above some
$k$-dependent ultraviolet scale.  However, each choice of variables
can be used, in principle, for a simulation of the theory below this
scale.  Such a simulation will be valid to arbitrarily low energies,
unless the renormalization group, which is unambiguous heading into
the infrared, itself breaks down.  Thus there is certainly an
infinite number of descriptions in the infrared.}

We might be tempted to ask whether this theory really has an infinite
number of colors.  But we already know that this is not a meaningful
question. We know that $SU(20)$ SQCD with $N_f=28$ is exactly dual in
some contexts to $SU(8)$ SQCD+M with $N_f=28$, so it must be that the
number of colors cannot be computed from the answer to any physical
question.  Of course, the conformal anomalies of the theory are
physically meaningful, and it is true that for theories of free
particles, they do count the number of fields, and thus the number of
colors.  For interacting theories, however, the conformal anomalies do
not tell us the number of colors.  They give us only an estimate of
how many colors a convenient (weakly-coupled!)  description of the
theory is likely to have.  For the flow near the boundary of the
valley of the duality cascade, the conformal anomalies themselves gradually
cascade, decreasing sharply as the flow moves from the vicinity of one
fixed point to the vicinity of another, then becoming nearly constant
once the neighborhood of the new fixed point is reached, before
jumping again as the flow moves onward.  They do indeed scale as
$k^2M^2$, but again, that doesn't tell us how many colors our {\it
description} has to have.  At best, they can only suggest to us that
perhaps a $SU(kM)$ SQCD fixed point, with an extra weakly-coupled
$SU([k\pm1]M)$ gauge group, might lie nearby.  Certainly they never
count the number of fields in the model in a straightforward way,
because at no energy scale are both gauge groups simultaneously weakly
coupled.

\subsection{Far from the Boundary of the Valley}

All the statements made above apply when the flow lies near the
boundary of the valley.  But now look at the other curve in
\reffig{KSbdy3} and \reffig{cascade2}, represented by the dot-dashed
line. This curve never lies near any fixed points.  Nowhere are any of
the couplings particularly weak, or particularly strong.  We see that
the $\lambda$'s are always pretty large, and that $\eta\sim\tilde\eta$
is always of order one.  We are nowhere near any well-understood fixed
point, around which we might do perturbation theory.  How can we
analyze the theory when it flows along this curve?

\begin{figure}[th]
\begin{center} 
\centerline{\psfig{figure=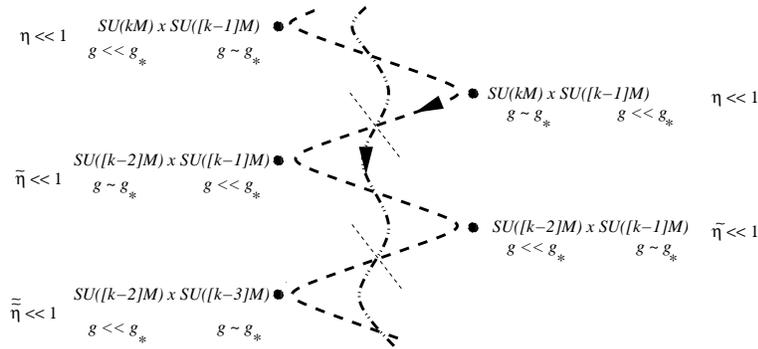,width=10cm,clip=}} 
\caption{\small As in \reffig{cascade}, showing the 
step-like character of flows (dashed) near the
boundary of the valley, and the smooth character of flows (dot-dashed)
far from the boundary.}  
\fig{cascade2}
\end{center}
\end{figure}

To appreciate what this curve means, let's represent the data in
another way, as we did for \reffig{tauflows},
by forgetting everything in the three-dimensional graph
except the valley and compressing the graph into the plane of the
gauge coupling constants.  In \reffig{KSbdy2}, it 
is shown where the
boundary of the valley lies; the right-hand diagram shows the
above-mentioned projection.  
In \reffig{bdycut} the boundary is shown again, extended to
include another two steps in the cascade: one sees not only the
gauge couplings for $SU(kM)$, $SU([k-1]M)$ and $SU([k-2]M)$ but also
$SU([k-3]M)$ and $SU([k-4]M)$.  {\it Don't be misled!}  The
valley does not actually close on itself, and the cascade is not an
Escher staircase.  Rather, think of this as a multi-sheeted graph,
with a branch cut off to the left.  The
boundary continues onto the next sheets above and below the one
shown.

\begin{figure}[th]
\begin{center} 
\centerline{\psfig{figure=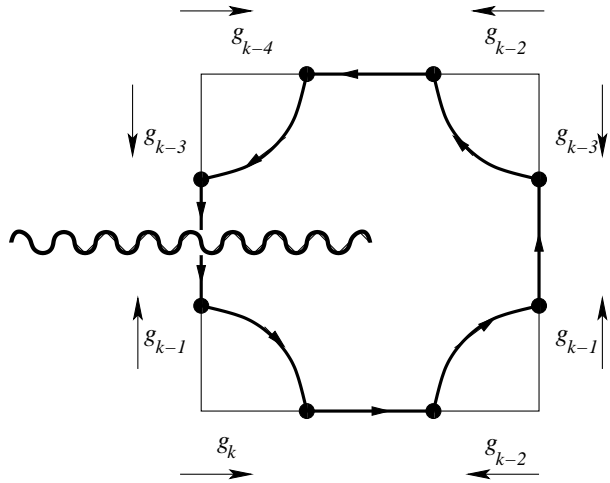,width=8cm,clip=}} 
\caption{\small The boundary of the valley, viewed from ``above'' (i.e.
with the $\eta$ coordinate(s) suppressed), shown for four steps
of the cascade.  Each dot represents a Seiberg SQCD fixed point.
The cut at the left edge indicates that the space
of couplings is viewed here as multi-sheeted, with flows emerging
from the sheet above the one shown, and continuing on the sheet below.}
\fig{bdycut}
\end{center}
\end{figure}

In \reffig{twoflows}, I have extended 
the two continuous RG flows of \reffig{KSbdy3}, so you
can see where they lie on the multi-sheeted graph of the
valley.  The outer one closely follows the boundary,
and has multiple quasi-fixed points where two of the three couplings
are small and the third one is almost fixed. The inner one blithely
circles the origin without much structure.  To see even more
dramatically what is happening, I have drawn in the next graph,
\reffig{eta}, the
lines of constant $\eta$, which I don't know precisely but can
estimate to order $1/k$.  (In this graph I have written the dual of
$\tilde \eta$ as $\eta$, rather than stacking 
tildes {\it ad infinitum} as I have 
started to do in Figs.~\ref{fig:cascade} and \ref{fig:cascade2}. 
This notation is quite reasonable conceptually,
however, as is clear from the form of
the figure.)
The outer curve flows from $\eta\ll1 $ to
$\eta\gg 1$ and back again, again and again during each step of the
cascade.  The inner curve, by contrast, sits at $\eta\sim 1$, sliding
back and forth between $\eta>1$ and $\eta<1$ but never deviating far
from it.  {\it Thus, the inner curve represents $\eta$ at a
quasi-fixed point.}  Its beta function is zero on average and is
always much less than one.

\begin{figure}[th]
\begin{center} 
\centerline{\psfig{figure=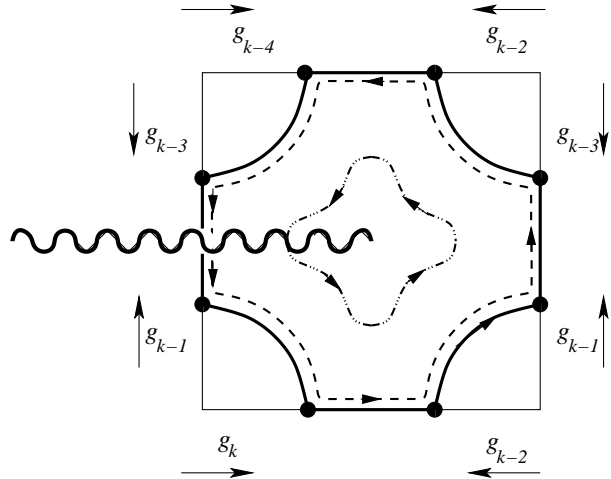,width=8cm,clip=}} 
\caption{\small An extension of \reffig{KSbdy3}: two flows, near to (dashed)
and far
from (dot-dashed) the boundary (solid) 
of the valley, viewed from ``above,'' with notation
as in \reffig{bdycut}.}
\fig{twoflows}
\end{center}
\end{figure}

Let us also consider what happens to the anomalous dimension
$\gamma_0$.  Lines of constant $\gamma_0$ will look similar to those
of constant $\eta$.  However, $\gamma_0$ never
deviates much from $-\half$ anywhere in the valley.  Furthermore,
on the inner
curve, it is even better than that.  The small $\beta_\eta$, combined
with the fact that $\eta\sim 1$ on this curve, implies $\gamma_0\sim
-\half$ to very good accuracy --- presumably much better accuracy than
on the outer curve, where it deviates from $-\half$ by order ${1/k}$.

\begin{figure}[th]
\begin{center} 
\centerline{\psfig{figure=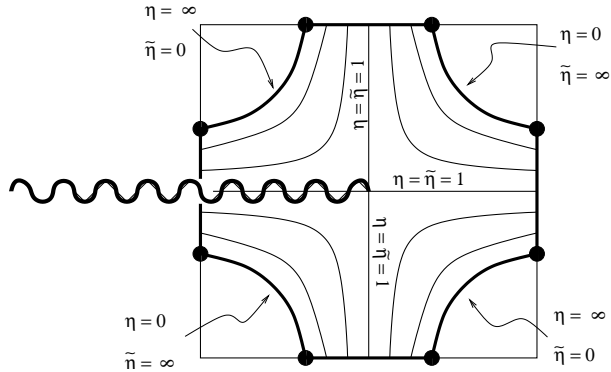,width=8cm,clip=}} 
\caption{\small Lines of constant $\eta$ and $\tilde \eta$
in the valley; notation as in \reffig{bdycut}. }
\fig{eta}
\end{center}
\end{figure}

Finally, let's look at the lines of constant $\tau_\pm$, the
parameters of the space of theories in the Klebanov-Witten case.  Comparing
\reffig{tauplus} and \reffig{tauminus} with \reffig{twoflows},  
we see that the flow lines tend to lie at nearly constant
$\tau_+$ and that the flow involves a continuously decreasing
$\tau_-$.   In fact, we can imagine that near the center
of the diagram there might exist
a scheme where the different flows in the valley
would be indexed by $\tau_+$, and
the flow along each line would be parametrized by $\tau_-$.

\begin{figure}[th]
\begin{center} 
\centerline{\psfig{figure=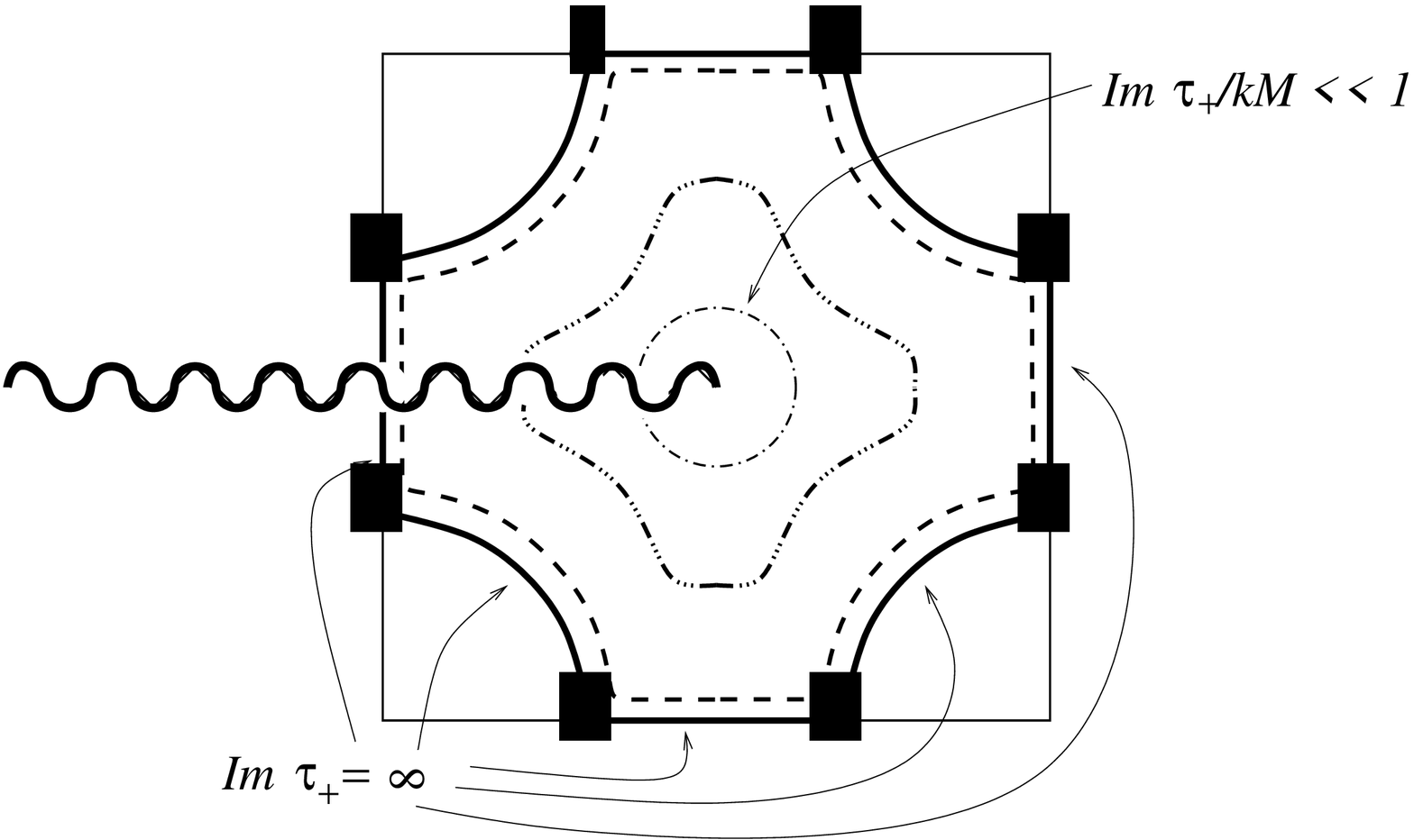,width=10cm,clip=}} 
\caption{\small Lines of constant $\tau_+$ in the valley; notation as in \reffig{bdycut}. }
\fig{tauplus}
\end{center}
\end{figure}

\begin{figure}[th]
\begin{center} 
\centerline{\psfig{figure=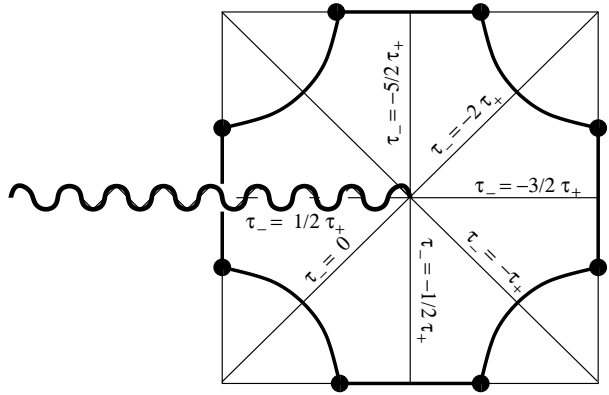,width=8cm,clip=}} 
\caption{\small Lines of constant $\tau_-$; reflection symmetries
under exchanging the two gauge groups or under $\eta\to 1/\eta$
fix the ratio $\tau_-/\tau_+$ on the lines shown, up to a shift
of $\tau_-$ by $\tau_+$.
Notation is as in \reffig{bdycut}. }
\fig{tauminus}
\end{center}
\end{figure}

Now, it should be clear where we are heading.  In the region around the
dot-dashed curve, none of the field
theory descriptions is particularly useful, but the string theory
description of the flow is particularly simple there.
The quantity $\tau_+$ is related to
the string coupling, given by the dilaton, which is constant.  The
center region is where $\tau_+\sim 1/g_s$ satisfies
$g_skM\gg 1$, which is the condition that the string theory be 
propagating on a space of small curvature.  This in turn assures
that a supergravity approximation to the string theory is valid.
For $g_skM\ll 1$, as is true (for any $k$) near the boundary
of the valley ($\tau_+\to i\infty$), 
the RG flow is choppy, involving a
stepping from one quasi-fixed point to another; this is the regime
where the calculations \eref{ksbetas}--\eref{ksbetasB} and the ensuing
conclusions are valid.  But in the region where
supergravity is a good approximation to the string description 
of the flow, the flow is smooth, never
approaching an approximate fixed point, though still flowing slowly,
at least for large $k$.  

Again, when the flow is smooth and lies far from the boundary of the
valley, none of the $SU(kM)\times SU[(k-1)M]$ descriptions of the
theory are particularly useful.  But they are still present.  There
are two direct ways to see this.  First, we can describe the smooth
flow, in any particular energy range, as follows: find a choppy flow
nearby (by increasing Im $\tau_+$, holding other couplings fixed),
describe the choppy flow by using any of the simple variables
inherited from the nearest Seiberg fixed point, and then go back to
the smooth flow by reducing Im $\tau_+$ again.  If Seiberg duality
applies to the choppy flow, and tells us what operator corresponds to
the coupling $\tau_+$ in two different descriptions, then it tells us,
in two different descriptions, how to get from the choppy flow back to
the smooth one, bringing the two different descriptions with it.
Second, all flows except those with $\tau_+\to i\infty$ reach the
supergravity regime in the ultraviolet, since $k\to\infty$ there.  Any
flow with $g_sM\ll 1$ will have a critical value of $k=k_c\sim 1/g_sM$
where for $k\gg k_c$ the supergravity description is useful, while for
$k\ll k_c$ the field theory descriptions become useful.  Essentially,
all flows gradually spiral out toward the boundary, and those with
$g_sM\ll 1$ actually reach its vicinity at $k\sim 1/g_sM$.  So for any
finite $\tau_+$ with $M/Im(\tau_+)\ll 1$, the field theory
descriptions are eventually going to be necessary for describing the
flow.  By adjusting $\tau_+$, we adjust the critical $k_c$ where this
occurs, and thus we can continuously slide from a flow for which a
particular field-theory description is useful to one in which it is
not, without in any way reducing the validity of that description.

Meanwhile, the question ``how many colors does the theory
have?'' is even less  meaningful here than on a choppy flow, since
no field-theory description, with {\it any}
number of colors, is useful for curves in the supergravity
regime.  But as before, we can turn to the conformal anomalies;
these provide a measure of  the
number of colors that nearby choppy flows would need if they were
to be described usefully using a choice of field-theory variables.
We will see below that these decrease smoothly, without jumps, and are of
order $k^2M^2$.

In summary, the entire valley, from its choppy boundary to its smooth
interior, can be described in an infinite number of ways, using an
infinite number of choices of variables, each involving an
$SU(kM)\times SU[(k-1)M]$ gauge group with bifundamental matter, along
with an additional string theory description to boot.  This statement
applies to all the flows within the valley, which form a one-parameter
family indexed by $\tau_+$, with $\tau_-$ the coordinate along the
flow.  Each of these flows gradually descends the valley, beginning
with a smooth flow in the regime where supergravity is valid.  The
flows gradually move out toward the boundary as they descend.  A flow
with $g_sM\gg 1$ remains always within the supergravity regime, and
its flow is always smooth; but a flow with $g_sM\ll 1$ will transition
out of the supergravity regime at $k\sim 1/g_sM$, and thereafter will
begin to cascade in a choppy fashion, passing through regimes in which
one or another field theory description will be useful.

\subsection{SUGRA and the Flow at Large $k$}

We will now construct the
supergravity description of the flow at large $g_skM$.  
For small $g_sM$, the description
will be good only for $k\gg k_c\sim 1/g_sM$, but for large $g_sM$ it
will describe the entire flow, start to finish, ultraviolet
to infrared.

Since the flow lies far from the boundary of the
valley, it flows smoothly, without jumps.
How can we describe it using supergravity?  What might we guess, if
we were naive?  The fact that the conifold is deformed plays 
no role here, since we are far from the region of deformation.
For $M=0$ we have 
$$
\cFfive\int_{T^{11}}F_5 = N = {4R^4\over 27\pi g_s} \ .
$$ 
Here we would expect this integral to equal the {\it local}
value of $N$, namely $k(r)M$, where I have indicated explicitly that
$k$ is a function of the momentum scale in the gauge theory, and thus
of $r$ in the supergravity.  In other words,
$$
\cFfive\int_{T^{11}}F_5 = k(r)M 
$$
which implies the $AdS$ radius gradually must change as $k(r)$ changes,:
\bel{R4vskM}
R^4(r)/\alpha'^2\approx {27\pi\over 4} g_s k(r)M .
\ee
We might guess this captures everything, and that we need
merely substitute $R(r)$ into the $AdS_5\times T^{11}$ metric 
\begin{eqnarray}
ds^2 &=& R^2(r) ds_{AdS_5}^2 + R^2(r) ds_{T^{11}}^2 \nonumber \\
&=&{r^2\over R^2(r)}dx^2 + {R^2(r)\over r^2} (dr^2 + r^2 ds^2_{T^{11}})
\ .
\end{eqnarray}
This space is said to be a {\it warped} version of the Klebanov-Witten
metric.

Now, what is $k(r)$?  
The beta function for $\tau_-$ near $\tau_-=0$, where
$g_k\approx g_{k-1}\equiv \bar g$, is  
$$
\beta_{\tau_-} \approx \beta_{4\pi\over g_k^2} - \beta_{4\pi\over g_{k-1}^2}
\approx {1\over 2\pi}{6M -2M\delta_0\over 1-{\bar g^2\over 8\pi^2}}
\ .
$$ 
Since $\delta_0$ is always of order $1/k$ in the valley, it can be
neglected.  We see, then, that $\tau_-$ has a beta function of the
same form as the beta function for pure \none\ $SU(M)$ supersymmetric
Yang-Mills theory itself: it is proportional to $M$, not $kM$.  The
solution to this equation is a logarithmically running $\tau_-$.  For
the string theory this implies, using \Eref{Bint},
$$
\cBtwo\int B_2\sim {\tau_-\over 2\tau_+} = {3g_sM\over 2\pi} \ln(r/r_s)
$$
where $r_s$ is where $k$ reaches 0.  
For every step down the cascade,
$k$ decreases by 1, $\tau_-$ decreases by $2\tau_+$, and thus
$\cBtwo\int B_2$ also decreases by 1.  We may therefore guess that 
\bel{kofr}
k(r) = \cBtwo\int_{S^2} B_2= {3g_s M\over 2\pi} \ln(r/r_s)  \ .
\ee
As $dk/dr \propto 1/r$, the speed of the cascade increases in 
the infrared; note also its speed is proportional to the 't Hooft coupling
$g_sM$.

Meanwhile, $F_3\propto  M\omega_3$, such that
$$
\cFthree \int_{S^3}F_3 = M \ .
$$ 
This should be constant, so we can use $F_5= B_2\wedge F_3$ (plus
another term to make it self-dual) to infer that $\cFfive\int F_5 =
3g_s M^2 \ln(r/r_s)$.  Consequently, from Eqs.~\eref{R4vskM} and
\eref{kofr},  $R^4(r) = (81/8) (g_s M)^2
\alpha'^2\ln(r/r_s)$.  Interestingly, this goes as $(g_sM)^2$, not as
$g_sM$.  This difference from non-cascading models impacts a number of
scaling relations and leads to confusion if one forgets about it.

Finally, from the curves we have been drawing we expect we should
require that the dilaton be constant, which can only be true if its
source, $F_3^2-H_3^2/g_s^2$, is zero.  You can easily check this is
true!  And in fact, this naive argument turns out to be almost
precisely correct, as shown in Klebanov and Tseytlin \mycite{ikat}, as
long as both $k\gg 1$ and $g_skM\gg 1$.  The only difference from what
I have suggested here is a small (order-$1/k$) constant shift in $R^4$
relative to $F_5$.

We can now check that, as claimed earlier, the conformal anomalies are
decreasing as $k^2M^2$.  This is essentially trivial.  The conformal
anomaly for \nfour\ $SU(N)$ Yang-Mills is a constant times $N^2$.  The
constant is independent of the gauge coupling, so the conformal
anomaly counts the number of degrees of freedom in the free theory.
We can compute the anomaly from the $AdS_5$ metric \mycite{HenSken} or
more easily from the volume of the $S^5$ that appears in the
gravitational metric \mycite{GubserT11}. For the Klebanov-Witten model
the anomaly is larger by a factor of $27/16$; this can be computed in
the field theory at the Seiberg fixed points, or computed far from the
valley's boundary using the fact that the volume of the $T^{11}$ space
is smaller by $27/16$ than that of the $S^5$ \mycite{GubserT11}. Note
this does not mean the number of fields in the theory is equal to
$27/16$ times the number of fields in \nfour\ $SU(N)$ Yang-Mills, as
you can easily check; the anomaly is not directly counting the number
of fields.  Finally, we have seen that all we have to do to write the
supergravity description of the duality cascade at large $r$ is to
replace $N$ in the Klebanov-Witten model with $N(r)=k(r)M$.  Thus the
anomaly is proportional to $[k(r)M]^2$, as we claimed, times the above
mentioned factor of $27/16$.  Again, one cannot read off, from this
formula, exactly how many colors would be useful in a description of
this or nearby flows.  One can only expect that the number of colors
in a useful description is probably decreasing roughly as $k(r)M$.

We should also address a nagging issue that may have bothered the reader.
Is it really sensible to extrapolate the flows under discussion up to
the extreme ultraviolet, where $k\to\infty$ and the number of colors
diverges?  A field theory is defined normally by holding the number of
degrees of freedom fixed and taking an ultraviolet cutoff to infinity.
Can we really take the number of degrees of freedom to be infinite?
Well, there is no law which forbids us to take the number of degrees
of freedom to gradually increase as the cutoff scale is raised. As
long as there are well-defined Green functions for operators that are
sensible in the limit, the theory is consistent.  Now that we have the
supergravity description of the large-$k$ regime, we can check that
the theory, though it has an infinite number of degrees of freedom,
still has many finite and computable amplitudes satisfying the usual
constraints of local field theory \mycite{Krasnitz2001}. In particular,
in the ultraviolet many amplitudes (properly normalized) gradually
approach those of the Klebanov-Witten model, with only very slow
logarithmic running to violate the power-law Green functions expected
in conformal theories.  These issues deserve further investigation,
but there is plenty of evidence that these theories are
well-behaved.\footnote{Other classes of theories of similar type are
trivial to construct.  For instance, consider an $SU(N)$ \nfour\ gauge
theory broken to $U(1)^{N-1}$ by a scalar whose expectation value is
$\vev{\Phi_1}= {\rm diag}\{v_1,v_2, \dots, v_N\}$, where
$v_1<v_2<\cdots<v_N$.  The $v_n$ are free parameters and can be chosen
arbitrarily.  Now take the limit $N\to\infty$.  By adjusting the
dependence of the $v_n$ on $n$, many different $SU(\infty)$ gauge
theories, with correspondingly different metrics in their supergravity
dual descriptions, may be obtained.}

In the infrared, on the other hand, something drastic must happen
as $k\to 0$.  If $g_s M\ll 1$, then supergravity breaks
down at $k\sim 1/g_sM$; 
we have little hope of describing the transition regime, though
at lower energies our field-theory methods will apply.  But
for $g_sM\gg 1$, supergravity should apply all the way to the infrared.
In this case, we should complete our metric, accounting for the
fact that the conifold is deformed as well as warped in the infrared.
We will round out our story with some general words about
this computation.

\subsection{The End of the Cascade}

Now we must match our complete understanding of the large $k$ region
onto the end of the flow as $k\to 0$.  We've already understood what
will happen for $k=3,2,1,0$ from our earlier field theory analysis.
I've drawn a graph using the technique I introduced earlier, though
I've made some changes.  Note the boundary of the valley is quite
different.  This is partly to account for the fact that $1/k$
corrections are not small.  More importantly, for $k\leq 4$ the
$SU(kM)$ theory with $2(k-1)M$ flavors is no longer in the conformal
window.  Instead it lies in the free magnetic phase, so there are no
Seiberg fixed points in this regime.  Instead, the infrared fixed
points for $k=4,3$ lie at $g_{k-2}\to zero$, and for $k=2$ the dual
theory has no gauge coupling (though its low-energy confined bound
states have other infrared-free couplings which I have indicated
schematically as $y_0$.)  At the top of the graph, the only remaining
light degrees of freedom of the theory are those of pure $SU(M)$
Yang-Mills.

\begin{figure}[th]
\begin{center} 
\centerline{\psfig{figure=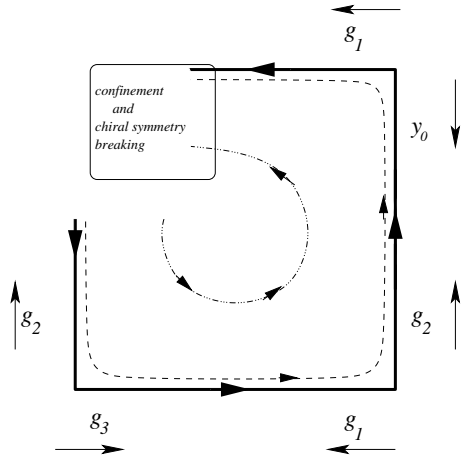,width=6cm,clip=}} 
\caption{\small The last steps in the cascade.  The flow on the outer
curve is choppy, remaining for an extended period in each corner.  The
flow on the inner curve is smooth and steady.  The endpoints of the
flows both exhibit confinement and chiral symmetry breaking, but in
other respects they are very different.}
\fig{finalflo}
\end{center}
\end{figure}

Our field theory calculations for $k\leq 3$ were always performed in
regions where one of the gauge couplings was small.  This was necessary
to make the calculations easy.\footnote{The calculations should be
tractable, though difficult, even without taking this limit.  To my
knowledge no one has looked seriously at this problem.}  Thus we
actually computed the RG flow appropriate to the {\it outer} curve in
\reffig{finalflo}, the choppy flow between quasi-fixed-points.  In
short, we were working at $g_sM\ll 1$.

But we now want to know what happens in the supergravity regime.  The
inner curve in \reffig{finalflo} indicates 
what we should expect.  Both curves in \reffig{finalflo} end in a
confining and chiral-symmetry breaking phase, and their paths look
reasonably similar, but actually they are very different. The issue is
the {\it rate} of flow.  Before flowing into the confining region, the
outer curve gets stuck for an extended period in the preceding corner.
In fact, the range of energy $\mu_H>\mu> \mu_L$ during which it
remains in this corner is {\it exponentially} large: ${\mu_H/ \mu_L}
\sim \exp({8\pi^2/g_1^2M})$, where $g_1$ is the minimum value of $g_1$
in this corner.  \rem{???!!!***MUST CHECK $\eta$ BEHAVIOR} This follows simply
from one-loop perturbation theory in $g_1$; the theory is essentially
pure Yang-Mills, once the $SU(2M)$ group confines and only $SU(M)$ remains
(along with a neutral Goldstone multiplet.)
But the inner curve flows smoothly into the confining region, with no
exponential hierarchies.  At no point does the theory resemble pure
Yang-Mills, even though it is in the same ``universality class.''  Its
extreme infrared structure is the same, but above the infrared lie
many massive partices.  For instance, the $Z^{adj}$ fields which
appeared in the $SU(2M)\times SU(M)$ analysis --- massive adjoint
scalars and fermions of $SU(M)$, which are not present in pure \none\
Yang-Mills theory --- lie close to the confinement scale.  So on the
inner curve we do not ever find, at any scale, pure \none\ Yang-Mills;
all we have is a theory with some similar features.  We can study
confinement and chiral symmetry breaking using supergravity, but this
magical ability doesn't come for free --- we don't get to study it in
the theory of greatest interest.\footnote{And indeed it is clear why
we cannot: real-world QCD lies outside the supergravity regime, and
any theory which resembles it, such as \none\ Yang-Mills, will also
lie outside the supergravity regime.  Only in very limited
circumstances might we gain some deep insight into QCD or \none\
Yang-Mills from supergravity.  The trick is to identify those
circumstances!}

How should we find the supergravity?  We will begin with a simple
ansatz: let us assume that the metric continues to be a warping of a
simple metric, of the form
$$
ds^2 =h(r)^{-1/2} (dx^\mu)^2 + h(r)^{1/2}ds^2_{6}
$$ 
where $ds^2_6$ is the metric of the {\it deformed} conifold.  In
the absence of the deformation, or equivalently at large $r$, we found
earlier that $h(r) = R^2(r)/r^2$, where $R^4=(81/8) (g_sM\alpha')^2
\ln(r/r_s)$; now presumably it will get some correction at $r\sim r_s$
and will not go negative.  We will also assume the dilaton is still
constant.  Now, this turns to be a good guess!  The ensuing
computation is too laborious to present here, and it is well-described
in the original paper, and more precisely in the Les Houches
Lectures written up by Herzog, Klebanov and Ouyang \mycite{HKO}.  But
here are some facts.  The metric is indeed sensible, with $h(r)$
nonsingular and approaching a nonzero constant as $r$ goes to a
minimum value $r_{min}$.  The resulting space is everywhere
weakly-curved and nonsingular, has no horizon, and is fully described
within supergravity.\footnote{This is in contrast to the \none$^*$
theory investigated by Polchinski and Strassler \mycite{nonestar}.  The
confining vacuum found there is a supergravity solution with an NS5
brane in it.  While this is a perfectly reasonable solution, it is
difficult to work with string theory in the presence of a single NS5
brane, at least for modes close to the 5-brane.  Since low-lying
hadrons are of this type, there would be some problems calculating
low-energy processes in the \none$^*$ model.  Also, the exact
metric for this theory remains unknown.}

We expect that the theory has the same qualitative features as \none\ \susic\
Yang-Mills, modulo the presence of a single massless
Goldstone supermultiplet.  Let's go through the list.

\begin{figure}[th]
\begin{center} 
\centerline{\psfig{figure=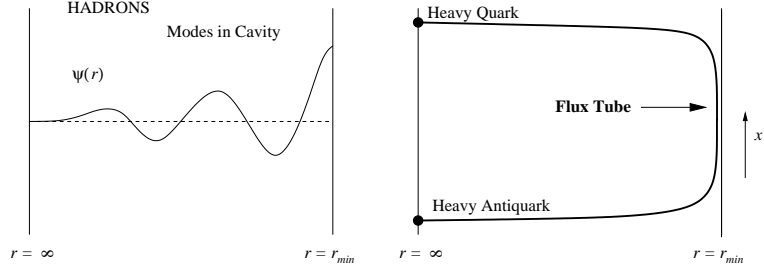,width=10cm,clip=}} 
\caption{\small On the left, the wave function for a 
cavity mode in five dimensions that
represents a typical four-dimensional hadron. On the right, 
a string with both ends on the boundary and
its middle lying at $r=r_{min}$ represents an infinitely
heavy quark and antiquark joined by an object of constant
energy per unit length: a confining flux tube.}
\fig{hadflux}
\end{center}
\end{figure}

\begin{itemize}
\item {\it A discrete spectrum, with spacing of order $m$.} \newline \ \newline
When the space is cut off at $r_{min}$, and has no horizon, it is effectively
made compact (since gravity tends to pull things in from the boundary.)
A five-dimensional space with one compact direction (or more
appropriately, a ten-dimensional
space with six compact dimensions) will give a discrete
four-dimensional spectrum, with modes carrying wave
functions such as shown in \reffig{hadflux}.  Let's call the typical
spacing between masses of low-lying states $m$.
\newline \ \newline
\item {\it Confinement (and associated strings carrying
flux)
} 
\newline \ \newline 
A Wilson loop corresponds to a string with its
ends on the boundary, as shown in \reffig{hadflux}.  
As long as its tension does not vanish
anywhere, it will give an area law.  And the tension cannot vanish,
because it is proportional to $h(r)$, which is nowhere zero (in
contrast to $AdS_5$ where it vanishes on the horizon at $r=0$.)  The
string tension is larger than $m^2$ by the ratio $R^2/\ell^2_s=gM$,
not $\sqrt{gM}$ as in \none$^*$.  (Here we see the effect of the
unusual dependence of $R$ on $gM$.)
\newline \ \newline
\item {\it Chiral symmetry breaking (a $\vev{\lambda\lambda}$
condensate spontaneously breaks $\ZZ_{2M}$ to $\ZZ_2$, since now only
$\alpha=0,\pi$ is an unbroken symmetry)} \newline \ \newline This
follows directly from the appearance of the {\it deformed} conifold
metric, which depends on the parameter $\epsilon$.  This parameter is
left unchanged only by the $\ZZ_2$ subgroup of the classical
$U(1)_{\mathcal R}$.
\newline \ \newline
\item {\it A moduli space with $M$ identical degenerate branches, with
$\vev{\lambda\lambda}=e^{2\pi i n/M} (\Lambda_0^{3M})^{1/M}$,
$n=0,\dots,M-1$ (since as with any spontaneously broken symmetry $G\to
H$, the vacua must form a representation of $G/H \approx
\ZZ_{2M}/\ZZ_{2}\approx\ZZ_M$.)} \newline \ \newline The fact that
$U(1)_{\mathcal R}$ is broken to $\ZZ_{2M}$ can be derived by studying
the anomaly of the theory, as worked out carefully in Klebanov, Ouyang
and Witten \mycite{OKW}.  Once this is established, the $\ZZ_{2M}$ rotations
of the phase of the parameter $\epsilon$, modulo the $\ZZ_2$ under
which it doesn't change, give $M$ different possible choices of
$\epsilon$ for a fixed $|\epsilon|$.
\newline\ \newline
\item {\it Domain walls which separate one vacuum
from the next (which are required by the previous condition); 
it turns out these are BPS saturated
(Dvali and Shifman, 1997).} \newline \ \newline
These are D5-branes wrapped on the finite-radius $S^3$ at the tip
of the deformed conifold.  With three dimensions on the $S^3$, these
branes have a 2+1 dimensional world-volume appropriate to a 
field theory domain wall.  D5-branes are magnetic sources for
$C_2$, so $dC_2$ is nonzero, leading to a shift in $C_2$ 
as we cross a D5-brane.  This is equivalent \mycite{OKW,HKO} to a shift in the
phase of the deformation parameter $\epsilon$. 
\end{itemize}
Thus our solution reproduces the qualitative physics of \none\
Yang-Mills theory, with the appropriate adjustments for the
extra Goldstone mode.\footnote{One 
can also look at dibaryons here, using
D3-branes wrapped on the $S^3$.  In field theory one can see that in
$SU(N)\times SU(N-M)$ these objects carry $M$ hanging indices, which
can be joined via confining flux tubes to $M$ external quark sources.
In short, they act as baryon vertices in the pure $SU(M)$ theory.
This is also seen in the supergravity. The $F_3$ flux from the $M$
D5-branes generates $M$ units of string charge on the wrapped
D3-brane, which must then be joined to $M$ external strings in order
that the wrapped brane have finite energy.}

\subsection{Goldstone modes}

One might ask whether the massless Goldstone mode (and its scalar
superpartner) makes this theory
profoundly different from pure \none\ Yang-Mills theory. This issue
has not been fully explored, but to my mind it has two answers.

On the one hand, important details of phenomenology are
different in certain processes; certain particles that 
are stable in \none\ Yang-Mills will be able to decay here, new
annihilation channels open up in hadron-antihadron scattering, etc.
Certain Green functions and scattering processes will
be quite different. 
This is very important if one wishes to use the duality cascade
for realistic model building.  (However, in realistic cases one
can expect the $U(1)_B$ to be gauged, which replaces the
massless Goldstone boson [and its superpartners] with a massive vector boson
[and its superpartners], alleviating
some of the phenomenological problems.  It also allows
for finite tension cosmic strings \mycite{cosmicstrings}.)
 
On the other hand, 
the Goldstone mode is odd under charge conjugation ($A_i\leftrightarrow
B_i$) which is an exact symmetry of the duality cascade.
For Green functions of operators which are 
even under this $\ZZ_2$ transformation,
or if we consider the scattering of hadrons for which both all initial
state hadrons and all final state hadrons are even, then
the Goldstone modes cannot appear in any tree-level string theory
diagram.  In other words, they cannot appear in any planar graph in
the field theory with even external operators or states.  In such
processes, the
Goldstone modes must be created in pairs, so they
can only appear in loops, giving $1/M^2$ corrections.

In short, if one's interest is in learning about the properties of matrix
elements of \none\ Yang-Mills, or of confining gauge theories in
general, it is straightforward to restrict oneself to matrix elements
where the Goldstone modes make no contribution at leading
non-vanishing order.  However, more care must be taken in 
building models of particle physics with this particular theory,
as this Goldstone mode could be more problematic (or useful) there.  This
remains to be explored further.

Finally, let us recall that we earlier encountered another simple
method by which we could obtain more interesting Goldstone modes, with
the same quantum numbers as pions.  In Sec.~\ref{subsec:extraD3} we
added an integer D3-brane to probe the space.  Looking back, we see
this extra brane makes the cascade of the form $SU(kM+1)\times
SU([k-1]M+1)$, with $k$ still shifting by 1 in each step of the
cascade.  The extra brane has two main effects. First, confining flux
tubes will now be able to break, as in QCD, because at the last step
($k=1$) there is matter in the fundamental representation of
$SU(M+1)$.  Also, as we saw following \Eref{solnAdSp}, the angular
coordinates of the position of the extra D3-brane are pions from the
coset $[SU(2)\times SU(2)]/SU(2)$ or $[SU(2)\times SU(2)]/U(1)$,
depending on the D3-brane location.  (It also appears that there
is no breaking of baryon number and no Goldstone mode, but this
remains to be verified.)  If we add the electroweak theory
to this model by gauging an $SU(2)\times U(1)$ subgroup of the
$SU(2)\times SU(2)$ global symmetry, we can obtain the standard model
symmetry-breaking pattern through strong-coupling dynamics.  This is a
version of technicolor, and it is natural to use it for
model-building, with application to particle physics and the Large
Hadron Collider.  Research of this sort is ongoing, and is closely
related to the work being done on Higgsless models \mycite{Higgsless}.

\subsection{Final Thoughts}

Now, what is this all good for?

Any calculable toy model of four-dimensional confinement and chiral
symmetry breaking (both discrete and continuous) is a very good thing.
To my knowledge, this is really the first example of its kind.  In
principle, everything in this theory that resides in the supergravity
regime (and many stringy effects too) can be calculated
semi-analytically.  Hadron spectra, scattering amplitudes, matrix
elements --- you want it, we can compute it.

Can we use this to learn something about QCD?  Many QCD processes are
already calculable, using the bona fide QCD Lagrangian, and these new
methods aren't needed.  These include processes truly dominated {\it
only} by ultraviolet physics, which are approachable using
perturbation theory at weak QCD coupling, and processes in which the
only long-distance ({\it i.e.} infrared, and therefore
nonperturbative) physics can be related to Euclidean Green functions
(which can be calculated using a computer.)  However, there are many
processes which are both nonperturbative and fundamentally
Minkowskian, including for example certain types of Regge physics,
quark-antiquark fragmentation into hadrons, quark-hadron duality, and
semi-inclusive or exclusive hadron scattering.
For these processes, no calculational methods exist, and a calculable
toy model could prove useful.

But I certainly do not expect the models described above to generate
numbers that we can compare with experiment.  These are toy models of
QCD: for one thing, they are supersymmetric, and we can only calculate
their properties for $M$ and $g^2M$ large.  Generally, toy models
don't give predictions; at best they give insights and new methodology
which can then be applied to the real model, hopefully leading
eventually to predictions.

Could we do better if we could find a version of the duality cascade
with real-world QCD as its low-energy limit?  We would need to break
supersymmetry and add quarks; see
for example \mycite{KK,ouyang,Kruc,Evans,Erdmeng,etaprime}.
Suppose we succeeded; would that improve our situation?  Somewhat, but
not too much.  At small $g^2M$, a theory such as the duality cascade
would be similar in the infrared to QCD, but would not be under
theoretical control.  Even in a supersymmetric theory, the field
theory methods of Seiberg {\it et al.} \mycite{NAD,kins,powerholo}
just aren't quite powerful enough to compute quantities of greatest
interest, such as the hadron spectrum, scattering amplitudes,
fragmentation, etc.  To compute anything we must be at large $g^2M$,
where supergravity methods are accurate, but this is a bad place to
look for real-world QCD.  Even if we could calculably embed QCD itself
into a model which could be continued to large $g^2M$, the theory in
the (super)gravity regime, by construction, would be very different
from the real world.  In particular, the small five-dimensional
curvature, required for supergravity to be useful, also {\it
unavoidably} implies all sorts of non-QCD-like behavior
\mycite{ewconfine,glueballs,nonestar,jpms,SonSteph}: (1) low-spin
hadrons with $m^2$ much less than the confining string tension; (2) a
large hierarchy between the masses of spin $\leq 2$ hadrons and spin
$\geq 3$ hadrons; (3) suppression of anomalous magnetic moments of
hadrons; (4) hadrons with large numbers of very soft partons and no
hard partons; (5) suppression of radiated high-transverse-momentum
gluon jets in high-energy scattering --- just to name a few.  So this
would not work well.

But despite this, it is worth investigating these models, because any
insights whatsoever into the difficult problems of QCD would be very
valuable.  It is in this limited but nonetheless important context
that I suggest progress is possible.  This kind of work is well
underway, and you may want to look at the more recent work that
Polchinski and I have been doing \mycite{jpms}, as well as the work of
Karch and Katz, Kruczenski et al., Belitsky et al., and Erlich, Katz,
Son and Stephanov, among others \mycite{KK,Kruc,Bel,SonSteph,EKSS}.

A second application is to physics of a very different kind.  One of
the greatest ``aw, shucks'' moments of my career came with the
realization that Randall and Sundrum got here first.  If they had not
discovered their five-dimensional method of stabilizing the weak-scale
hierarchy \mycite{rshier}, Klebanov and I would have discovered something
analogous in this model.  Take this cascading field theory, or one
very much like it.  Cut it off in the ultraviolet at some scale
corresponding to the Planck scale, and couple it to four-dimensional
gravity.  Then the dual string theoretic description is
five-dimensional $AdS$-like supergravity, plus a four-dimensional
graviton zero-mode localized near the boundary, with an infrared
cutoff at the value of the $AdS$ radius where $k\to 0$.  Unlike the
original Randall-Sundrum models, in which the infrared cutoff is
imposed by hand on a strictly-$AdS$ metric, here the $AdS$ metric is
warped and cuts itself off dynamically at small $r$.  This infrared
cutoff is
nothing more than the confinement and the generation of a mass gap
that occurs at an exponentially low scale in this field theory.  In
particular, if the number of colors at the Planck scale is $kM$, the
low-energy confinement scale is the Planck mass times $e^{-\# k}$.
Imagine, then, that the theory is coupled somehow to the
electroweak model.  At high energies it has many colors, but it
cascades down until the number of colors is small (and it could be
either that $g_sM\ll 1$, so that supergravity breaks down before the
infrared scale, or that $g_sM\gg 1$, so that supergravity is valid at
all scales.)  At the confinement scale, nonperturbative dynamics could
easily be imagined to break electroweak symmetry and/or supersymmetry,
leading perhaps to a realistic standard model. Essentially, this
provides a model for technicolor-like physics (in which the Higgs
mechanism is generated dynamically by strong coupling, and the Higgs
boson is a composite field.)  String-theoretic realizations of these
ideas along the lines suggested by H.~Verlinde \mycite{HVerlinde} were
later considered by Giddings, Kachru and Polchinski \mycite{GKP} (see
Kachru's lectures) who embedded the duality cascade into a fully
consistent string background.  They thereby provided the first
complete string-theoretic realization of the Randall-Sundrum approach
to the hierarchy, which in these models simply appears as a dual
description of dimensional transmutation.  (For realistic
model-building, one must deal successfully with the massless Goldstone
supermultiplet, or with the massive, weakly-interacting gauge boson 
supermultiplet by
which it is eaten.)

Finally, as I alluded to a moment ago, there is another possible
application.  The standard model itself, being a model with few colors
at weak coupling, could easily (with the addition of some massive
matter) be at the base of a similar duality cascade.  I toyed with
such ideas back in 1995, but I found that the generic extension of the
standard model with an ultraviolet chain of Seiberg dualities hits a
``duality wall'': a finite energy scale at which the number of colors
diverges \mycite{dualitywall}.  Unfortunately I didn't come across models
where this energy scale is infinite, as in the duality cascade of
these lectures.  Well, I didn't try too hard, honestly.  After all, in
the days before supergravity duals, it was completely crazy to discuss
field theories which had an ever- and rapidly-increasing number of
colors in the ultraviolet.  What an ugly idea! especially compared to
the simplicity of grand unification, with 5 or 10 colors.  But now we
know differently: this steadily increasing number of colors translates under
duality into a perfectly reasonable weakly-curved string-theoretic
background.  What previously appeared insane now
looks surprisingly beautiful.  Perhaps it is even true.

\end{document}